\RequirePackage{fix-cm}
\documentclass[%
    twoside, openright, titlepage, numbers=noenddot,%
    cleardoublepage=empty,%
    BCOR=5.5mm, paper=a5, fontsize=10pt,
]{scrreprt}
\usepackage{etex}
\reserveinserts{10}

\usepackage{paralist}
\usepackage{amsmath}
\usepackage{mathtools}
\usepackage{dsfont}
\usepackage{amssymb}
\usepackage{enumerate}
\usepackage[shortlabels]{enumitem}
\usepackage{bbm}
\usepackage{subfiles}

\usepackage[utf8]{inputenc}

\usepackage[main=english,ngerman]{babel}

\PassOptionsToPackage{%
  eulerchapternumbers,
  floatperchapter, pdfspacing,%
  beramono,%
  a5paper,%
}{classicthesis}

\newcommand{\myTitle}{%
 User Agency and System Automation in Interactive Intelligent Systems.
\xspace%
}

\newcommand{\myPlainTitle}{%
  user agency and system automation in  interactive intelligent systems.
\xspace%
}

\newcommand{\myDissNumber}{30445}
\newcommand{\myName}{Thomas Langerak\xspace}

\newcommand{\myTime}{2024\xspace}

\newcommand{\myDOI}{10.3929/ethz-b-000705078}

\usepackage{scrtime}

\newcounter{dummy}

\newcommand{\ie}{i.\,e.\xspace}

\newcommand{\eg}{e.\,g.\xspace}

\newcommand{\etal}{et al.\xspace}

\usepackage{csquotes}  
\usepackage[T1]{fontenc}
\usepackage{xspace}
\usepackage{textcomp}
\usepackage{mparhack}
\usepackage{relsize}
\usepackage{amssymb}

\usepackage[
  style=nature,%
  natbib=true,%
  clearlang=true,%
  backend=biber,%
]{biblatex}

\AtEveryBibitem{%
  \clearfield{day}%
  \clearfield{month}%
  \clearfield{endday}%
  \clearfield{endmonth}%
}

\DeclareRedundantLanguages{en,EN,English}{english}

\DeclareFieldFormat{pages}{\mkfirstpage{#1}}

\let\origcite\cite%
\def\cite#1{\unskip~\origcite{#1}}

\usepackage{amsmath}
\usepackage{mathtools}
\usepackage{isomath}

\usepackage{tabularx}
\usepackage{ltablex}
\usepackage{floatrow}
\usepackage{caption}
\usepackage{subcaption}
\expandafter\def\csname ver@subfig.sty\endcsname{}

\usepackage{algorithmic}
\usepackage[ruled,vlined]{algorithm2e} 

\usepackage{blindtext}

\usepackage[dvipsnames]{xcolor}
\usepackage[%
  hyperfootnotes=false,%
  pdfpagelabels,%
]{hyperref}
\usepackage{hyperxmp}
\hypersetup{%
  colorlinks=true, linktocpage=true,%
  breaklinks=true, pdfpagemode=UseNone, pageanchor=true, pdfpagemode=UseOutlines,%
  plainpages=false, bookmarksnumbered, bookmarksopen=true, bookmarksopenlevel=1,%
  hypertexnames=true, pdfhighlight=/O,
  pdftitle={\myPlainTitle},%
  pdfauthor={\myName},%
  pdfcopyright={Copyright (C) \myTime, \myName},%
  pdfsubject={},%
  pdfkeywords={},%
  pdflang={en},%
}

\usepackage{graphicx}
\usepackage{rotating}
\usepackage{tikz}
\usepackage{pgfplots}
\pgfplotsset{compat=newest}
\usetikzlibrary{shapes.geometric}

\usepackage{cleveref}

\PassOptionsToPackage{printonlyused}{acronym}
\usepackage{acronym}

\usepackage{enumitem}

\usepackage{scrhack}
\usepackage{classicthesis}

\KOMAoptions{headinclude=true,footinclude=false}
\areaset[current]{\textwidth}{1.618034\textwidth}

\clearplainofpairofpagestyles
\ofoot[\pagemark]{}
\definecolor{chapter-color}{cmyk}{1, 0.50, 0, 0.25}
\definecolor{link-color}{cmyk}{1, 0.50, 0, 0.25}
\definecolor{cite-color}{cmyk}{0, 0.7, 0.9, 0.2}

\usepackage{bookmark}
\hypersetup{
  urlcolor=Black, linkcolor=Black, citecolor=Black, 
}

\let\chapterNumber\undefined%
\newfont{\chapterNumber}{eurb10 scaled 5500}

\usepackage{units}

\usepackage{etoolbox}
\makeatletter
\pretocmd{\chapter}{\addtocontents{toc}{\protect\addvspace{15\p@}}}{}{}
\makeatother

\titleformat{\chapter}[display]%
  {\relax}{\vspace*{-3\baselineskip}\makebox[\linewidth][r]{\color{halfgray}\chapterNumber\thechapter}}{10pt}%
  {\raggedright\spacedallcaps}[\normalsize\vspace*{.8\baselineskip}\titlerule]%

\def\chapterabstract#1{%
  \begingroup
  \baselineskip1.3em
  \leftskip1em
  \rightskip\leftskip\itshape#1
  \par
  \endgroup
}

\setkomafont{dictumtext}{\itshape\small}%
\setkomafont{dictumauthor}{\normalfont}

\usepackage{balance}       
\usepackage{graphics}      
\usepackage[T1]{fontenc}   
\usepackage{txfonts}
\usepackage{mathptmx}
\usepackage{booktabs}
\usepackage{textcomp}
\usepackage{graphicx,import}

\usepackage{multirow}
\usepackage{makecell}
\usepackage{array}
\usepackage{bbm}
\usepackage{xcolor,colortbl}
\usepackage{mdframed}
\usepackage{ifthen}
\usepackage{mathtools}
\usepackage[export]{adjustbox}
\usepackage{xspace}

\usepackage[normalem]{ulem}

\definecolor{darkgreen}{RGB}{1, 128, 1}
\definecolor{darkred}{RGB}{128, 1, 1}
\definecolor{teal}{RGB}{1, 160, 160}

\newcommand{\newparagraph}[1]{\\\mbox{}\\}

\newcommand{\figref}[1]{Figure~\ref{#1}}
\newcommand{\tabref}[1]{Table~\ref{#1}}

\newcommand{\secref}[1]{Section~\ref{#1}}
\newcommand{\chapref}[1]{Chapter~\ref{#1}}

\newcommand{\appref}[1]{Appendix~\ref{#1}}

\newcommand{\contribution}[1]{
\begin{mdframed}[backgroundcolor=gray!20, hidealllines=true]
\vspace{1em}
\chapterabstract{#1}
\vspace{1em}
\end{mdframed}
\clearpage}

\DeclarePairedDelimiterX{\norm}[1]{\lVert}{\rVert}{#1}

\newcommand{\Fig}[1]{Fig.~\ref{fig:#1}}
\newcommand{\Figure}[1]{Figure~\ref{fig:#1}}

\newcommand{\Table}[1]{Table~\ref{tab:#1}}

\newcommand{\Eq}[1]{Eq.~\ref{eq:#1}}

\newcommand{\Sec}[1]{Sec.~\ref{sec:#1}}

\newcommand{\Appendix}[1]{Appendix~\ref{app:#1}}
\newcommand{\useragent}{{user agent}\xspace}
\newcommand{\interfaceagent}{{interface agent}\xspace}
\newcommand{\add}[1]{#1}
\newcommand{\addiui}[1]{#1}
\newcommand{\deliui}[1]{}
\newcommand{\del}[1]{}

\makeatletter
\DeclareCiteCommand{\fullcite}
  {\defcounter{maxnames}{\blx@maxbibnames}%
    \usebibmacro{prenote}}
  {\usedriver
     {\DeclareNameAlias{sortname}{default}}
     {\thefield{entrytype}}}
  {\multicitedelim}
  {\usebibmacro{postnote}}
\makeatother

\newcommand{\omniHap}{\emph{Omni-v1}\xspace}
\newcommand{\omniUIST}{\emph{Omni-v2}\xspace}
\newcommand{\magpen}{\emph{MagPen}\xspace}
\newcommand{\marlui}{\emph{MARLUI}\xspace}

\newcommand{\omniHapTitle}{Contact-free Non-Planar Haptics with a Spherical Electromagnet}
\newcommand{\omniUISTTitle}{Volumetric Sensing and Actuation of Passive Magnetic Tools for Dynamic Haptic Feedback}
\newcommand{\magpenTitle}{Optimal Control for Electromagnetic Haptic Guidance Systems}
\newcommand{\marluiTitle}{Multi-Agent Reinforcement Learning for Point-and-Click Adaptive User
Interfaces}

\newcommand{\interfaces}{Shared Variables Interfaces\xspace}
\newcommand{\control}{Joint Control of Shared Variables\xspace}

\newcommand{\interfacesLower}{\MakeLowercase {\interfaces}\xspace}
\newcommand{\controlLower}{\MakeLowercase {\control}\xspace}

\usepackage{bibentry}
\usepackage{tabularx}
\usepackage[final]{pdfpages}

\newcommand{\SetOfStates}{\mathcal{S}}
\newcommand{\StatePerPolicy}{S}
\newcommand{\SetOfObservations}{\mathcal{O}}
\newcommand{\ObservationPerPolicy}{O}

\newcommand{\SetOfActions}{\mathcal{A}}
\newcommand{\ActionPerPolicy}{A}
\newcommand{\Transitions}{T}

\newcommand{\ObservationTransitions}{{F}}
\newcommand{\SetOfObservationTransitions}{\mathcal{F}}

\newcommand{\SetOfRewards}{\mathcal{R}}
\newcommand{\RewardPerPolicy}{R}
\newcommand{\SetOfPolicies}{\Pi}
\newcommand{\discount}{\gamma}
\newcommand{\belief}{b}

\newcommand{\stack}{\mathbf{o}}
\newcommand{\error}{\mathcal{E}}
\newcommand{\action}{\mathbf{a}}
\newcommand{\inputu}{\mathbf{u}}
\newcommand{\statex}{\mathbf{x}}
\newcommand{\valuev}{v}

\newcommand{\observation}{\mathbf{o}}
\newcommand{\state}{\mathbf{s}}
\newcommand{\policy}{\pi}

\newcommand{\nitems}{n_{i}}
\newcommand{\nslots}{n_{s}}
\newcommand{\timet}{t}

\newcommand{\pos}{\mathbf{p}}
\newcommand{\tools}{\mathbf{x}}
\newcommand{\gattr}{\mathbf{g}}
\newcommand{\menu}{\mathbf{m}}

\newcommand{\target}{\mathbf{t}}
\newcommand{\satweight}{\lambda}
\newcommand{\miss}{\lnot h}

\newcommand{\mt}{T_M}
\newcommand{\dect}{T_D}

\newcommand{\RpenBold}{\mathbf{r_{p}}}
\newcommand{\rpen}{r_{p}}
\newcommand{\rsBold}{\mathbf{r_{0}}}

\newcommand{\meBold}{\mathbf{m_{e}}}
\newcommand{\me}{m_{e}}
\newcommand{\mpp}{m_{p}}
\newcommand{\FrBold}{\mathbf{F_{r}}}

\newcommand{\FtBold}{\mathbf{F_{t}}}

\newcommand{\er}{\mathbf{e_r}}

\newcommand{\RmagtopenBold}{\mathbf{r_{p}}}

\newcommand{\RmagtopenBoldt}{\mathbf{\Tilde{r}_{p}}}
\newcommand{\Rmagtopen}{r_{p}}

\newcommand{\mpBold}{\mathbf{m_p}}
\newcommand{\mpBoldt}{\mathbf{\Tilde{m}_{p}}}
\newcommand{\mmBold}{\mathbf{m_e}}
\newcommand{\posm}{\mathbf{p_e}}
\newcommand{\posp}{\mathbf{p_p}}

\newcommand{\ed}{\mathbf{e_d}}
\newcommand{\et}{\mathbf{e_t}}
\newcommand{\ez}{\mathbf{e_z}}
\newcommand{\Rtheta}{\mathbf{r_{\theta}}}
\newcommand{\Cost}{\mathcal{C}}
\newcommand{\stheta}{\sin{\angt}}
\newcommand{\ctheta}{\cos{\angt}}
\newcommand{\sphi}{\sin{\angp}}
\newcommand{\cphi}{\cos{\angp}}
\newcommand{\angt}{\theta}
\newcommand{\angp}{\varphi}

\newcommand{\BBold}{\mathbf{B}}
\newcommand{\BiBold}{\mathbf{B_i}}
\newcommand{\BeBold}{\mathbf{B_e}}
\newcommand{\BpBold}{\mathbf{B_p}}
\newcommand{\BnBold}{\mathbf{B_n}}
\newcommand{\BiTildeBold}{\mathbf{\Tilde{B_i}}}

\newcommand{\mBold}{\mathbf{m}}

\newcommand{\siBold}{\mathbf{s_i}}
\newcommand{\SBold}{\mathbf{S}}
\newcommand{\oBold}{\mathbf{o}}

\newcommand{\ReBold}{\mathbf{r_{e}}}
\newcommand{\rBold}{\mathbf{r}}
\newcommand{\rpBold}{\mathbf{r_{p}}}

\newcommand{\RsiBold}{\mathbf{r_{s_i}}}

\newcommand{\HatrBold}{\mathbf{\hat{r}}}

\begin{document}
\frenchspacing
\raggedbottom%
\selectlanguage{english}
\pagenumbering{roman}
\pagestyle{scrplain}

\definecolor{darkgreen}{RGB}{1, 128, 1}

%
%
%

%
%


\begin{titlepage}
    \begin{center}
        \large
        \begingroup
            \spacedlowsmallcaps{Diss. ETH No. \myDissNumber}
        \endgroup

        \hfill

        \vfill

        \begingroup
            \spacedallcaps{\myTitle}
        \endgroup

        \vfill

        \begingroup
            A thesis submitted to attain the degree of\\
            \vspace{0.5em}
            \spacedlowsmallcaps{Doctor of Sciences}\\
            (Dr.\ sc.\ ETH Zurich)
        \endgroup

        \vfill

        \begingroup
            presented by\\
            \vspace{0.5em}
            \spacedlowsmallcaps{\myName}\\
            M.Sc., Aalto University\\
            M.Sc., University of Twente\\
            \vspace{0.5em}
            born on 13 January 1994
            \endgroup

        \vfill

        \begingroup
            accepted on the recommendation of\\
            \vspace{0.5em}
            Prof.\ Dr.\ C. Holz\\
            Prof.\ Dr.\ A. Oulasvirta\\
            Prof.\ Dr.\ R. Murray-Smith\\
        \endgroup

        \vfill

        \myTime%

        \vfill
    \end{center}
\end{titlepage}

\thispagestyle{empty}

\hfill

\vfill

\noindent\myName: \textit{\myTitle} 
\textcopyright\ \myTime

\bigskip

\noindent\spacedlowsmallcaps{DOI}: \myDOI

%
%
%
%
%

\cleardoublepage

\pdfbookmark[1]{Abstract}{Abstract}
\begingroup
\let\clearpage\relax
\let\cleardoublepage\relax
\let\cleardoublepage\relax

\chapter*{Abstract}
The balance between user agency and system automation in interactive intelligent systems is crucial for intuitive and efficient interactions. While fully automated systems could potentially offer greater efficiency and demonstrably improved performance, making them perfect is notoriously hard. The inevitable shortcomings of automated systems diminish usability and overall experience, thereby compromising users' perceived self-determination. Conversely, tools, systems that rely entirely on user agency and have no level of automation, though offering full control to the user, can be inefficient and fail to enhance the user's capabilities. Hence, for effective human-AI interactions, we need to find a balance between user agency and system automation. The question we address in this dissertation is "How can we balance user agency and system automation for the interaction with intelligent systems?"

We approach this challenge through four main contributions. First, we introduce a novel spherical electromagnet capable of generating adjustable forces on an untethered tool, allowing users to feel grounded forces while maintaining full agency. Second, we develop an integrated sensing and actuation system that tracks a passive magnetic tool in 3D space while simultaneously delivering haptic feedback, eliminating the need for external tracking. Third, we propose an optimal control method for electromagnetic haptic guidance systems that balances user input and system control, allowing users to adjust trajectories and speed as needed. Finally, we present a model-free reinforcement learning approach for adaptive user interfaces that learns interface adaptations without relying on heuristics or real user data.

Our findings, based on simulations and user studies, suggest that the shared control of intelligent systems has the potential to significantly outperform naive control strategies. Thus, we contribute methodologies that find an agency-automation trade-off and pave the way for more interaction with intelligent systems. Our research demonstrates that integrating models of human behavior, either explicitly or implicitly, into control strategies enables intelligent systems to better account for user agency. We show that the trade-off between user agency and system automation is not solely an algorithmic problem but must also be considered in the engineering of physical devices and interface design. We advocate for an integrated end-to-end approach to interaction with intelligent systems that incorporates algorithmic, engineering, and design perspectives.
\endgroup

\cleardoublepage%

\begingroup
\let\clearpage\relax
\let\cleardoublepage\relax
\let\cleardoublepage\relax

\begin{otherlanguage}{ngerman}
\pdfbookmark[1]{Zusammenfassung}{Zusammenfassung}
\chapter*{Zusammenfassung}
Das Gleichgewicht zwischen der Handlungsfreiheit des Benutzers und der Systemautomatisierung in interaktiven intelligenten Systemen ist entscheidend für intuitive und effiziente Interaktionen. Während vollautomatisierte Systeme potenziell größere Effizienz und nachweislich verbesserte Leistung bieten könnten, ist es bekanntermaßen schwierig, sie perfekt zu machen. Die unvermeidbaren Mängel automatisierter Systeme verringern die Benutzerfreundlichkeit und das Gesamterlebnis und beeinträchtigen dadurch die wahrgenommene Selbstbestimmung der Benutzer. Umgekehrt können Werkzeuge, d.h. Systeme, die sich vollständig auf die Benutzerautonomie stützen und kein Automatisierungsniveau aufweisen, zwar die volle Kontrolle für den Benutzer bieten, aber ineffizient sein und die Fähigkeiten des Benutzers nicht erweitern. Daher müssen wir für effektive Mensch-KI-Interaktionen ein Gleichgewicht zwischen Benutzerautonomie und Systemautomatisierung finden. Die Frage, die wir in dieser Dissertation behandeln, lautet: "Wie können wir Benutzerautonomie und Systemautomatisierung für die Interaktion mit intelligenten Systemen ausbalancieren?"

Wir nähern uns dieser Herausforderung durch vier Hauptbeiträge. Erstens stellen wir einen neuartigen sphärischen Elektromagneten vor, der in der Lage ist, einstellbare Kräfte auf ein kabelloses Werkzeug auszuüben und es den Benutzern ermöglicht, geerdete Kräfte zu spüren, während sie volle Autonomie behalten. Zweitens entwickeln wir ein integriertes Erfassungs- und Betätigungssystem, das ein passives magnetisches Werkzeug im 3D-Raum verfolgt und gleichzeitig haptisches Feedback liefert, wodurch die Notwendigkeit einer externen Verfolgung entfällt. Drittens schlagen wir eine optimale Kontrollmethode für elektromagnetische haptische Führungssysteme vor, die Benutzereingaben und Systemsteuerung ausbalanciert und es den Benutzern ermöglicht, Trajektorien und Geschwindigkeit nach Bedarf anzupassen. Schließlich präsentieren wir einen modellfreien Ansatz des verstärkenden Lernens für adaptive Benutzeroberflächen, der Schnittstellenanpassungen erlernt, ohne sich auf Heuristiken oder reale Benutzerdaten zu stützen.

Unsere Ergebnisse, basierend auf Simulationen und Benutzerstudien, deuten darauf hin, dass die geteilte Kontrolle intelligenter Systeme das Potenzial hat, naive Kontrollstrategien deutlich zu übertreffen. Somit tragen wir Methoden bei, die einen Kompromiss zwischen Autonomie und Automatisierung finden und den Weg für mehr Interaktion mit intelligenten Systemen ebnen. Unsere Forschung zeigt, dass die Integration von Modellen menschlichen Verhaltens, sei es explizit oder implizit, in Kontrollstrategien es intelligenten Systemen ermöglicht, die Benutzerautonomie besser zu berücksichtigen. Wir zeigen, dass der Kompromiss zwischen Benutzerautonomie und Systemautomatisierung nicht nur ein algorithmisches Problem ist, sondern auch bei der Entwicklung physischer Geräte und dem Interface-Design berücksichtigt werden muss. Wir befürworten einen integrierten End-to-End-Ansatz für die Interaktion mit intelligenten Systemen, der algorithmische, technische und Design-Perspektiven einbezieht.
\end{otherlanguage}

\endgroup

\vfill

\cleardoublepage

\pdfbookmark[1]{Acknowledgements}{acknowledgements}

\bigskip

\begingroup
\let\clearpage\relax
\let\cleardoublepage\relax
\let\cleardoublepage\relax
\chapter*{Acknowledgements}

\def\thanks#1{%
\begingroup
\leftskip1em
\noindent #1
\par
\endgroup
}

First and foremost, I would like to express my gratitude to Prof. Otmar Hilliges. He believed in me, despite my unconventional background, and gave me the opportunity to explore this  academic world. Under Otmar's supervision and mentorship, I have grown into the researcher I am today. My PhD journey was a long learning experience, filled with both ups and downs. I am immensely grateful for his patience and trust throughout this process. The weekly one-on-one meetings with him were invaluable to my work, and I only came to appreciate them too late.

I would also like to extend my sincere thanks to my committee. Prof. Christian Holz, who was not only a close collaborator but also stepped in as my thesis supervisor when it became necessary, has my sincere gratitude. I am also deeply thankful to Prof. Antti Oulasvirta, who introduced me to the field of computational interaction and whose opinions I value. Antti ignited my initial passion for HCI and research. Finally, I would like to thank Prof. Roderick Murray-Smith, whose work was instrumental in shaping my understanding of interaction, and who graciously agreed to join my committee despite our limited prior interaction.

Furthermore, I want to thank all the external collaborators who played a part in this journey, including Prof. Daniele Panozzo, Dr. Bernhard Thomaszewski, Dr. Tanya Jonker, and Dr. Kashyap Todi. Their fresh perspectives were always invaluable.

A very special thanks goes to all my colleagues and friends at the AIT Lab. Together, we truly achieved more than we could have as individuals. You ensured that the lab was a place of great times, fun, and an enjoyable atmosphere. I had countless discussions with each of you that shaped my research. I want to specifically thank Dr. Juan Z\'arate for always engaging in conversations about science, magnets, and life. I am also grateful to Sammy Christen for patiently listening to my frustrations and for teaching me about Reinforcement Learning. My thanks also go to Dr. Velko Vechev and Dr. Christoph Gebhardt for their unwavering support, whether it was meeting deadlines, answering HCI-related questions, or help TAing. Marcel Buehler deserves thanks for making the office a great place to be, encouraging me to go bouldering, and for bringing me El Tony or a Fr\"uchtbrotli. I also want to thank Prof. David Lindlbauer for our beer-fueled discussions, his selfless can-do attitude during deadlines, and for enabling me to join the UIST Organization Committee. Finally, I am grateful to Prof. Xucong Zhang, Dr. Emre Aksan, Dr. Manuel Kaufmann, Yufeng Zheng, Prof. Jie Song, Prof. Anna Feit, Alex Fan, Mert Albaba, and all the others in the lab for being part of this journey.

I also want to express my gratitude to my friends for patiently listening to my frequent complaints, helping me switch off from work, and ensuring I stayed grounded in reality. A special thanks goes to my friends back in the Netherlands; I know it’s not always easy to stay in touch with someone who has moved away, and I appreciate the effort they  made. I also want to thank Salma Thalji for always being there, for providing me with pragmatic advice, for teaching me that motivation isn't a requirement to accomplish something, and for her understanding. Meeting you on that New York rooftop during the summer school is a memory I will always cherish.

Last but certainly not least, I want to thank my parents for supporting me in countless ways throughout my studies and for offering (sometimes unsolicited) advice when I needed it, even when I didn’t realize I did.

\endgroup

\pagestyle{scrheadings}
\cleardoublepage

\refstepcounter{dummy}
\pdfbookmark[1]{\contentsname}{tableofcontents}
\setcounter{tocdepth}{1} 
\setcounter{secnumdepth}{3} 
\manualmark%
\markboth{\spacedlowsmallcaps{\contentsname}}{\spacedlowsmallcaps{\contentsname}}
\makeatletter
\renewcommand{\toclevel@part}{10}
\makeatother
\bookmarksetup{level=part}
\tableofcontents
\automark[section]{chapter}
\renewcommand{\chaptermark}[1]{\markboth{\spacedlowsmallcaps{#1}}{\spacedlowsmallcaps{#1}}}
\renewcommand{\sectionmark}[1]{\markright{\thesection\enspace\spacedlowsmallcaps{#1}}}

\cleardoublepage%
\part{Introduction}
\cleardoublepage\pagenumbering{arabic}%

\def\dir{chapters/01_introduction}

\chapter{Introduction}
\label{ch:introduction}

\section{Motivation}
Although human-computer interaction may appear relatively unchanged since the post-war era, as we still mainly rely on the mouse and keyboard, today's intelligent systems depend on much more than just explicit user input. Autonomous vehicles utilize computer vision to anticipate human intent and movement \cite{janai2020computer, maqueda2018event, hee2013motion}. GitHub Copilot automatically adapts and generates code using large language models \cite{wermelinger2023using, imai2022github}. Smart home devices, like Nest thermostats, learn user preferences to automatically optimize comfort and energy efficiency \cite{yang2013learning, pisharoty2015thermocoach}, and advanced software like photo editing tools amplify human creativity \cite{liu20233dall}. These modern systems leverage contextual understanding to adapt to user needs. However, the shift from explicit to implicit user input, which comes from contextual understanding, is not without challenges. Systems that take action without explicit inputs reduce user agency. Thus, despite significant advances in the contextual understanding that these intelligent systems leverage, interaction with these systems remains challenging. In this dissertation, we focus on algorithmic methods to balance the automation provided by implicit contextual understanding with user agency achieved through explicit input in the area of Human-Computer Interaction (HCI).

In intelligent systems, a user interacts with an artificial agent that has contextual knowledge. The incorporation of contextual understanding aims to make these systems more natural and beneficial for real-world human tasks \cite{xu2023transitioning}. The goal of the artificial agent is to assist users in completing tasks, perform tasks on the users' behalf, or otherwise help them. In contrast to non-context-aware devices (e.g., \cite{engelbart1962augmenting, engelbart1968research}) that rely on explicit and precise user input, intelligent systems are always available. This crucial property allows intelligent systems to proactively engage with their environment and users. In turn, this property enables implicit and imprecise input from the user. Humans leverage these kinds of interactions on a daily basis -- e.g., saying "can you hand me \emph{this}?" \cite{lee2024gazepointar}. Furthermore, the proactive capabilities enable intelligent systems to act on behalf of the user by inferring their intent -- e.g., a robot handing you a glass of water without being told to do so \cite{christen2023learning}. Yet, intelligent systems are still capable of acting based on explicit user commands, seemingly losing no ground to classical forms of human-machine interaction. However, as intelligent systems are capable of both implicit (system automation) and explicit (user agency) interactions, they need to strike a balance between user agency and the automation of tasks.

This agency-automation trade-off becomes even more crucial in interactions where the system and user operate on a shared variable, that is, a variable that can be manipulated simultaneously by both the user and the system. For instance, consider the control of a semi-autonomous car. The user changes the car's acceleration, while the intelligent system simultaneously adjusts it. The artificial agent might increase safety and driving efficiency. However, it might not include passenger preferences. Similarly, a drawing assistance system might take control of a pen, thereby increasing accuracy yet constraining the user's creative expression. Also consider a virtual reality intelligent adaptive user interface. The interface may show optimal adaptations and decrease task completion time, but it does not necessarily align with user expectations. In all these scenarios, the ownership of the variable is ambiguous, and therefore, the control is ambiguous.

This ambiguity and agency-automation balance is where the primary challenge for interaction with intelligent systems lies. Fully automated systems could potentially provide greater efficiency and measurably better performance but at the cost of compromising user agency and diminishing perceived usability and overall experience. On the other hand, systems that rely solely on user agency provide complete control but may be sub-optimal, inefficient, and fail to augment the user's abilities. This dissertation explores the question: \emph{How can we algorithmically control intelligent systems with shared variables to balance user agency and system automation?}

Researchers have explored the balance between agency and automation in intelligent systems. In the 1960s, \citeauthor{BarHillel1960} \cite{BarHillel1960} addressed optimizing human-computer labor division in translation tasks. The 1990s saw debates over user agency versus system automation \cite{Shneiderman1997}, leading to a consensus on automation that enhances productivity while maintaining control. Additionally, research on human-AI collaboration emphasizes the importance of bidirectional communication, trust, and shared goals to enhance team effectiveness \cite{demir2017team, shively2017human}. To facilitate bidirectional communication, both models of human behavior and intelligent control strategies are necessary. Traditional methods to model human-computer interaction have focused on manual control theory \cite{McRuer1967, Costello1968} using heuristics \cite{card1986model, card1980klm, card1983the, kieras1997overview, anderson1997act} and low-parameter mathematical models \cite{fitts1954information, hick1952rate}. More recent approaches leverage reinforcement learning (RL) due to its ability to predict user behavior \cite{jokinen2021multitasking, jokinen2021touchscreen, gebhardt2020hierarchical}. These models of human behavior need to be integrated into the control strategies of systems. However, most haptic devices use a form of open-loop control \cite{yamaoka2013depend}, and thus do not take the user into account. Alternatively, some devices use proportional-integral-derivative control (PID) \cite{abut2018interface, ramos2016wavenet, pothi2014design}. However, PID, among other shortcomings, optimizes for a fixed setpoint, thereby limiting user freedom. This dissertation builds upon these foundations by integrating simple heuristics and RL to model human behavior in both the control of pen-based haptic feedback systems and the policy of adaptive user interfaces. We introduce novel kinesthetic haptic devices, essential for exploring the agency-automation spectrum, and employ model-based control strategies and multi-agent RL to dynamically adjust to user interactions, providing real-time optimal adaptations and enhancing user experience.

In summary, AI-driven systems are becoming increasingly ubiquitous in people's work and lives. These systems are no longer just "conventional" computing systems, but intelligent systems that exhibit unique characteristics that bring new interaction paradigms. This shift presents a challenge in balancing user agency and system automation, particularly when the user and system share control over a variable. This dissertation aims to explore how to strike an optimal balance between user control and system efficiency in interactions with intelligent systems and shared variables. We approach this question by introducing novel haptic devices, model-based control, and reinforcement learning approaches.
\section{Approach}
\label{sec:Approach}
This dissertation aims to bridge the gap between user agency and system automation through four projects divided into two primary directions. First, we discuss the design of shared variable interfaces. Second, we explore the control strategies for such interfaces.
\subsection{The Design of Shared Variable Interfaces}

In shared variable interfaces, both the user and the system act on a shared variable, which can take many forms, such as the acceleration of a car or a graphical menu in an operating system. A special instance of this shared variable is kinesthetic haptic feedback. Here, not only is the variable itself shared (e.g., the position of a joystick), but the action and perception of the variable are in the same modality from the user's perspective. For example, to change or perceive a joystick's position, the user uses their hands. This singular interaction modality contrasts sharply with, for example, touch interfaces, where users perceive the UI through their eyes but interact with it using their hands. Using an interface with a single interaction modality potentially reduces confounding factors when investigating interactions with shared variable interfaces. This shared domain allows for simultaneous interaction between a user and an intelligent system.

Previous work in kinesthetic haptic feedback generally revolves around complex mechanical contraptions (\eg \cite{Massie94, Stamper1997, VanDerLinde2002, Araujo2016, zoller2019assessment, Sinclair2019Capstan}). Further extensions of such systems include exoskeletons~\cite{Gu2016, Choi2016}, gloves~\cite{Cybergrasp, hinchet2018dextres}, and tilt-platforms~\cite{Prattichizzo2013, Kim2016}. However, these often require user instrumentation and inherently involve system friction, which users will always perceive. This friction leads to a system that, even when off, does not allow for full user agency.

To overcome the limitations of not covering the full spectrum from user agency to system automation and to enable our research, we introduce novel haptic interfaces. These haptic devices sense and actuate a]passive tool. We introduce a novel spherical electromagnet that acts on an embedded permanent magnet, enabling an untethered tool that allows for full user agency without physical constraints while simultaneously providing large grounded forces, which enables system automation.

In Part II, we introduce i) a novel spherical electromagnet (\chapref{ch:shared:contact}, published in \cite{zarate2020contact}), and ii) a gradient-based tool tracking algorithm (\chapref{ch:shared:volumetric}, published in \cite{Langerak:2020:Omni}). This combination allows us to deliver dynamic haptic feedback. We evaluate our system both technically and with users. We find that interacting with a shared variable in the same modality allows for natural interactions that increase user agency. However, our findings also suggest that more intelligent system control is necessary to enable more complex dynamics.

\subsection{The Control of Shared Variable Interfaces}
Previous systems typically use open-loop control \cite{yamaoka2013depend} to control intelligent systems. This approach fails to consider the user, eliminating the capability to trade off user agency with system automation. Alternatively, some systems use heuristics \cite{Lopes16, Browne1990, Smith2010, Stephanidis1997}, which require tediously crafted rules by experts. Furthermore, supervised learning \cite{Maes1995, Lashkari1997, McCreath2006, Faulring2010, Shen2009a, Shen2009b, Berry2011, Pejovic2014, Mehrotra2015} and multi-armed bandits \cite{glowacka2019bandit, lomas2016interface, koch2019may, kangas2022scalable, Koyama2014, Koyama2016} are popular approaches for controlling intelligent systems. However, these approaches optimize for a myopic decision, failing to consider future states of both the system and the user. Optimizing purely for the next timestep limits the system's capabilities to intelligently trade off user agency and system automation. In contrast, our work minimizes a cost function over a receding horizon[,] taking into account expected user and system behavior. This enables an explicitly optimized balance between agency and automation, taking into account possible future states.

In Part III, we first discuss Model Predictive Contour Control (MPCC), an optimization-based control strategy that uses a supplied model of the system dynamics in combination with a cost function (\chapref{ch:control:optimal}, published in \cite{langerak2020magpen}). We apply this in the context of a prototypical pen-based electromagnetic haptic-feedback system. Our approach allows users to easily override the system, adapt their input spontaneously, and draw at their own speed. While beneficial, crafting a system dynamic that includes user behavior is challenging. Secondly, we focus on UI adaptation as a use case of an interface with a shared variable (\chapref{ch:control:multi}, published in \cite{Langerak:2024:MARLUI}). Moving from haptics to a digital-only interface offers several advantages. Adaptive User Interfaces allow us to focus on control strategies, overcoming any sim-to-real gap confounding factors. Furthermore, existing cognitive models (such as Fitts' Law \cite{fitts1954information}) can be utilized in this context. To overcome the challenge of crafting system dynamics that include user behavior, we turn to Reinforcement Learning (RL) as a control strategy and a means to learn both user and system dynamics models that are intertwined. By formulating UI adaptation as a multi-agent reinforcement learning problem, we introduce a user agent that mimics a real user and learns to interact with an interface, while simultaneously introducing an agent that learns a control strategy to maximize the user agent's performance. The control agent learns the task structure from the user agent's behavior and, based on that knowledge, can support the user agent in completing its task.

\clearpage
\chapter{Contributions}
We summarize the four main contributions of our dissertation.
\section*{1. \omniHapTitle}
Many kinesthetic haptic feedback devices rely on mechanical contraptions to provide forces. However, even when unactuated, these devices limit user agency, as the mechanical components restrict movement. To investigate the agency-automation trade-off, we need to design an \emph{untethered} haptic feedback device capable of delivering grounded forces. To that end, \textbf{we introduce \omniHap (\chapref{ch:shared:contact}), a novel contact-free volumetric kinesthetic haptic feedback device.} This system combines a spherical electromagnet with a dipole magnet model and a simple control law to deliver dynamically adjustable forces onto a handheld tool. We conducted a user experiment with 6 participants to characterize the force delivery aspects and perceived precision of our system.
\section*{2. \omniUISTTitle}
Many haptic devices (including \omniHap) require external optical tracking, which is often expensive, cumbersome, and requires a clear line of sight. To overcome these limitations, we developed \omniUIST (\chapref{ch:shared:volumetric}). \omniUIST is a haptic feedback device with integrated spatial tracking. The spatial tracking capabilities of \omniUIST are enabled by \textbf{a novel gradient-based method, which reconstructs the 3D position of the permanent magnet in mid-air using measurements from eight off-the-shelf Hall sensors integrated into the base.} Furthermore, following improvements to the actuator design, \omniUIST delivers over twice the forces compared to \omniHap. We detail \omniUIST's hardware implementation and our 3D reconstruction algorithm, providing an in-depth evaluation of its tracking performance. \omniUIST shows how integrating sensing and actuation provides natural interaction that encourages user agency. Both \omniHap and \omniUIST open up new applications in AR/VR, particularly in design, gaming, and object exploration.
\section*{3. \magpenTitle}
Existing haptic devices typically employ open-loop control, which does not account for user feedback. Alternatively, they use proportional–integral–derivative (PID) control or heuristics, which are usually based on timed references, limiting user agency. \textbf{We propose \magpen (\chapref{ch:control:optimal}), a time-independent closed-loop control strategy that allows users to retain agency while receiving haptic guidance.} Our real-time approach assists in pen-based tasks such as drawing, sketching, or designing. By iteratively predicting the motion of an input device, such as a pen, and adjusting the position and strength of an underlying dynamic electromagnetic actuator, our method provides flexible guidance without diminishing user control. Experimental results demonstrate that our approach is more accurate and preferred by users compared to open-loop and time-dependent closed-loop methods.
\section*{4. \marluiTitle}
Most control strategies rely on known system dynamics. However, in many HCI scenarios, users are integral to the system, and their behavior is complex and not easily modeled. To address this, we treat HCI as a multi-agent problem where an agent learns system, task, and user dynamics through interaction with a synthetic user agent. Specifically, \textbf{we introduce \marlui (\chapref{ch:control:multi}), a multi-agent reinforcement learning approach for adaptive user interfaces (AUIs).} In our formulation, a \useragent mimics real user interactions with the UI, while an \interfaceagent learns to adapt the UI to maximize the \useragent's performance. Our method captures the underlying task structure, system dynamics, and user behavior. Experiments show that the learned policies generalize well to real users and achieve performance comparable to data-driven supervised learning baselines.
\clearpage
\section{Structure of Dissertation}
We describe the structure of this doctoral dissertation in the following. After establishing the state of the art and background knowledge on the relevant methods, we outline our main contributions. More specifically:

\begin{tabularx}{\textwidth}{lX}
\multicolumn{2}{l}{\color{CTtitle}\textsc{\MakeLowercase{I. Introduction}}} \\
\textsc{\MakeLowercase{\chapref{ch:related_work}}} & introduces the related work on haptic devices, adaptive interfaces, user modelling and control strategies. \\
\textsc{\MakeLowercase{\chapref{ch:background}}} & provides theoretical and mathematical background in electromagnetism and control theory. \\
\multicolumn{2}{l}{\color{CTtitle}\textsc{\MakeLowercase{II. The Design of Shared Variable Interfaces}}} \\ 
\textsc{\MakeLowercase{\chapref{ch:shared:contact}}} & \textbf{Contribution 1: }\emph{\omniHapTitle.}\\
\textsc{\MakeLowercase{\chapref{ch:shared:volumetric}}}& \textbf{Contribution 2: }\emph{\omniUISTTitle.}\\
\textsc{\MakeLowercase{\chapref{ch:shared:conclusion}}} & summarizes, discusses the implications, and highlights the limitations of our work on \interfacesLower. \\
\multicolumn{2}{l}{\color{CTtitle}\textsc{\MakeLowercase{III. The Control of Shared Variable Interfaces}}} \\ 
\textsc{\MakeLowercase{\chapref{ch:control:optimal}}} & \textbf{Contribution 3: }\emph{\marluiTitle.}\\
\textsc{\MakeLowercase{\chapref{ch:control:multi}}}& \textbf{Contribution 4: }\emph{\magpenTitle.}\\
\textsc{\MakeLowercase{\chapref{ch:control:conclusion}}} & summarizes, discusses the implications, and highlights the limitations of our work on \controlLower. \\
\multicolumn{2}{l}{\color{CTtitle}\textsc{\MakeLowercase{IV. Conclusion}}} \\ 
\textsc{\MakeLowercase{\chapref{ch:summary}}}& summarizes our contributions.  \\
\textsc{\MakeLowercase{\chapref{ch:outlook}}} & highlights possible future research directions. and control theory.
\end{tabularx}
\clearpage
\pdfbookmark[1]{Publications}{publications}
\section{Publications}

\noindent
The contributions of this thesis are based on the following publications, in order of appearance:

\begin{enumerate}
    \item \begin{refsection}\fullcite{Langerak:2020:Contact:Bold}\end{refsection}
    \footnote{My specific contribution in this work is the ideation, fabrication, control strategy, technical evaluation and user study.}
    \item \begin{refsection}\fullcite{Langerak:2020:Omni:Bold} \end{refsection}
    \footnote{My contribution in this work is the ideation, fabrication, method and evaluation.}
    \item \begin{refsection}\fullcite{Langerak:2020:Optimal:Bold}\end{refsection}
    \footnote{I have been fully responsible for this work and contributed to all components, with exception of the derivation of the electromagnetic model we use.}
    \item \begin{refsection}\fullcite{Langerak:2024:MARLUI:Bold}\end{refsection}
    \footnote{I have been fully responsible for this work and contributed to all components.}

\end{enumerate}

\noindent
Further publications that were conducted during the course of my PhD research but are out of scope of this thesis are listed below, in order of publication:
\begin{enumerate}
    \item \begin{refsection}\fullcite{Langerak:2020:Demonstration:Bold}\end{refsection}
    \footnote{I have been fully responsible for this work and contributed to all components.}

    \item \begin{refsection}\fullcite{Langerak:2021:Hedgehog:Bold}\end{refsection}
    \footnote{I have contributed with ideation, discussions and to the writing of the paper. The first author was a Master Thesis student under my supervision}

    \item \begin{refsection}\fullcite{langerak2021generalizing}\end{refsection}
    \footnote{I have been fully responsible for this work and contributed to all components.}

    \item \begin{refsection}\fullcite{Langerak:2022:Robust:Bold}\end{refsection}
    \footnote{I have contributed with ideation, discussions and to the writing of the paper. The first author was a Master Thesis student under my supervision}

    \item \begin{refsection}\fullcite{Langerak:2024:xaiui:Bold}\end{refsection}
    \footnote{I have been fully responsible for this work and contributed to all components.}

    \item \begin{refsection} \fullcite{Langerak:2024:rile:Bold}\end{refsection}
    \footnote{I have contributed with fruitful discussions and to the writing of the paper.}

\end{enumerate}


\cleardoublepage%

\def\dir{chapters/related_work}

\chapter{Related Work}
\label{ch:related_work}

We provide an overview of related work. We start by discussing human-AI interaction from a high-level perspective. Then, we will discuss control theory in the context of HCI, including, a brief overview of computational user modeling and its role in the control loop. Finally, we focus on two types of interfaces with shared variables, discussing what they are and specific examples of control. First, we discuss haptic interfaces, and secondly, we discuss adaptive user interfaces.

\section{Human-AI Interaction}
Research on human-AI collaboration has revealed that integrated teams of humans and AI systems are more effective than either working independently \cite{bansal2019beyond, bansal2019updates, demir2018conceptual}. Research explores various aspects of this collaboration, such as conceptual architectures and frameworks \cite{johnson2019ai, madni2018architectural, oneill2020human, prada2014human}, and performance metrics \cite{bansal2019beyond}.

To gain insights into human-machine collaboration, researchers have applied psychological theories of human teamwork \cite{devisser2018automation, mou2017media}. These theories highlight core principles for bidirectional communication, trust, goals, situational awareness, language, intentions, and decision-making between humans and AI systems \cite{demir2017team, shively2017human}. This approach marks a shift from the traditional one-way communication model in conventional HCI contexts \cite{xu2023transitioning}, emphasizing a reciprocal relationship. In this dissertation, we embed this reciprocal relationship by modeling human-AI interaction as a multi-agent problem. Furthermore, having a shared variable inherently forms a bidirectional communication channel.

There are foundational concepts in interpersonal teaming that require adaptation for application to human-machine teams. These include Autonomy (compared to Automation), Trust (compared to Reliability), and Teaming (compared to the Use of Automation) \cite{greenberg2023foundational}. This adaptation has become particularly important in the context of autonomous car research \cite{li2016trolley, awad2018moral} and military applications \cite{chen2014human}. \citet{sycara2004integrating} identified three general roles that machines can support within teams: assisting individuals with their tasks, acting as equal team members, or supporting the team as a whole. This dissertation focuses on the first role, supporting individuals. Usually, this role is investigated in the context of decision support systems \cite{lyons2021humanautonomy}. However, in contrast to decision support systems, our focus is on optimizing the control of interfaces to best support user interactions.

\section{Control Theory in Interaction}
\begin{figure}
\centering
\includegraphics[width=0.75\textwidth]{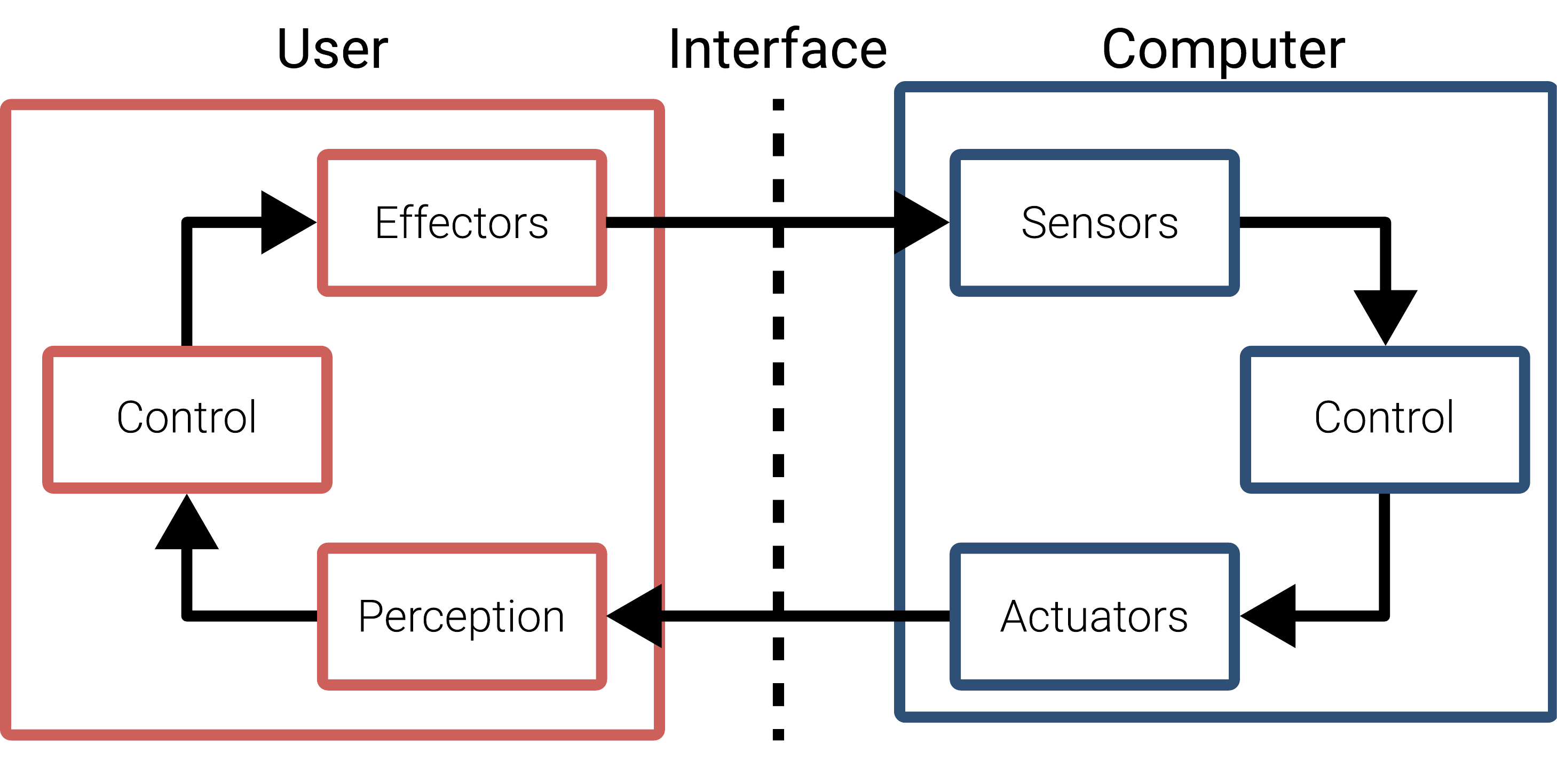}
\caption{Human-Computer Interaction as a closed-loop system, adapted from \cite{murray2018control}}
\label{fig:control_loop}
\end{figure}

Control theory deals with the behavior of dynamic systems with inputs and how their behavior is modified by feedback \cite{thummala2007, seal2019}. The goal of control theory is to develop methods for influencing the behavior of dynamical systems to achieve desired outcomes \cite{seal2019}. The key idea is to measure the system's output, compare it to a reference or desired state, and adjust the system's input accordingly to minimize the difference between the actual and desired outputs \cite{grigas2023}. This feedback loop allows the system to adapt and maintain stability in the presence of disturbances or uncertainties \cite{seal2019, vizvarova2021}. Control theory provides a mathematical framework for modeling, analyzing, and optimizing such feedback systems \cite{babblenewt2023, seal2019}.
In an HCI context, a human interacts with a machine. While the human interaction can be modeled as a control system, the machine uses control theory to achieve desired outcomes given the human input. Together they form a closed-loop system (\figref{fig:control_loop}) \cite{murray2018control}. First, we will discuss control theory from a machine perspective. Then, we will discuss modeling human behavior from a control theory perspective. For a mathematical introduction, see the background chapter (\secref{sec:ocp}).
\subsection{Control of Machines}
There has been extensive research on optimal control techniques and their applications to various domains. Proportional-integral-derivative (PID) control is one of the most widely used approaches due to its simplicity and effectiveness. For example, PID control has been used to control haptic devices \cite{abut2018interface, ramos2016wavenet, pothi2014design}. However, PID control has some shortcomings, such as poor adjustment and follow-up when the system has strong interference or high nonlinearity and uncertainty \cite{krstic2017applicability}. It can also struggle with precise control and overshoot in some applications \cite{somefun2021dilemma}.

Model Predictive Control (MPC) is another optimal control technique that has gained popularity in recent years. MPC addresses some of the limitations of PID control by explicitly taking into account the nature of the actuators and system dynamics over a certain time horizon before deciding the control action to be applied \cite{dasilva:2008:mpc, qin1997overview}. This allows MPC to handle constraints, nonlinearities, and uncertainties more effectively. MPC has been used, for instance, in computer numerical controlled machines \cite{lam2010model, lam2013model}, the control of drones \cite{Gebhardt:2018}, or autonomous vehicles \cite{liu2017path}. However, MPC also has some shortcomings, such as the need for an accurate system model and the computational cost associated with solving the optimization problem in real-time \cite{holkar2010overview}. Furthermore, MPC is a time-dependent control strategy, in which the setpoint moves at a fixed pace. This limits user creativity and freedom.

Reinforcement Learning (RL) is an increasingly popular approach for optimal control in intelligent systems that interact with humans. RL addresses some of the limitations of MPC by learning control policies directly from interaction with the environment, without requiring an explicit system model \cite{sutton1998introduction}. This allows RL to handle complex, uncertain, and nonlinear systems more effectively. RL has been applied to robotics, where it has been used for learning complex behaviors such as manipulation and locomotion \cite{christen2019guided, christen2023synh2r}. In self-driving cars, RL has been employed for learning driving policies that can handle diverse traffic scenarios and road conditions \cite{kiran2021deep}.

In this dissertation, we look at a time-independent MPC for haptic devices and RL for adaptive user interfaces. In both scenarios, we embedded user dynamics into the control loop. To that end, we must have models of human behavior.

\subsection{Human Behavior as Control Strategy}
The specific area of control systems relating to human users became a major focus starting in the 1950s \cite{McRuer1967, Costello1968}. Our discussion of the relevant work is based on \cite{murray2018control}; we refer to that for a more extensive discussion. In general, it is accepted that models of human behavior enable simulation. Simulating human behavior, in turn, has many advantages in theory crafting, design, novel interaction paradigms, and safety \cite{murray2022simulation}.

Manual control theory \cite{McRuer1967, Costello1968}, which seeks to model the interaction of humans with machines like aircraft or cars, grew out of Craik's early, war-related work \cite{Craik1947}. It became more well-known in the broader framing of Wiener's Cybernetics \cite{Wiener1948}. According to Wickens and Hollands \cite{Wickens1999}, modeling human control behavior emerged from two main schools: the skills researchers, who focused on learning and acquisition in undisturbed environments, and the dynamic systems or manual control theory researchers \cite{Kelley1968, Sheridan1974}, who modeled the interaction of humans with machines in vehicle control and complex industrial process control, driven by engineering motivations to eliminate error in closed-loop systems. Poulton \cite{Poulton1974} reviews the early tracking literature, while Jagacinski and Flach \cite{Jagacinski2003} provide an accessible textbook review of manual control approaches.

Early work relies on heuristics \cite{card1986model, card1980klm, card1983the, kieras1997overview, anderson1997act} and on low-parameter mathematical models \cite{fitts1954information, hick1952rate}. More recent work extends these models and, for instance, predicts the operating time for a linear menu \cite{10.1145/1240624.1240723}, gaze patterns \cite{salvucci2001integrated}, pointing \cite{muller2017control, martin2021intermittent, murray2021forward, ikkala2022breathing}, or cognitive load \cite{duchowski2018index}.

Recently, reinforcement learning gained popularity within the research area of computational user models. This popularity is due to its neurological plausibility \cite{botvinick2012hierarchical, frank2012mechanisms}, allowing it to serve as a model of human cognitive functioning. The underlying assumption of RL in HCI is that users behave rationally within their bounded resources \cite{gershman2015computational, oulasvirta2022computational}. There is evidence that humans use such a strategy across domains, such as in causal reasoning \cite{denison2013rational} or perception \cite{gershman2012multistability}. In human-computer interaction, researchers have leveraged RL to automate the sequence of user actions in a keystroke-level-model framework \cite{leino2019computer} or to predict fatigue in volumetric movements \cite{cheema2020predicting}. It was also used to explain search behavior in user interfaces \cite{yang2020predicting} or menus \cite{chen2015emergence} and as a model for multitasking \cite{jokinen2021multitasking}. Most similar to our work is research on hierarchical reinforcement learning for user modeling. \citeauthor{jokinen2021touchscreen} \cite{jokinen2021touchscreen} show that human-like typing can emerge with the help of Fitts' Law and a gaze model. Other works show that HRL can elicit human-like behavior in task interleaving \cite{gebhardt2020hierarchical} or touch interactions \cite{jokinen2021touchscreen}.

Specifically, Multi-Agent Reinforcement Learning (MARL) is of interest \cite{debard2020multiagent}. MARL is interesting as it captures the implicit closed-loop iterative nature of HCI. We can see both the human and computer as two agents that interact together in an environment. MARL is popular in robotics \cite{yang2004multiagent}. Furthermore, MARL is closely related to alignment theory as MARL deals with training multiple AI agents to behave in desired ways. The challenge of aligning multiple agents' objectives with each other and with human values relates closely to fundamental alignment problems \cite{ma2022elign, rodriguez2022instilling}. MARL is also highly relevant to cooperative AI, as it provides frameworks for training multiple agents to work together toward common goals. Techniques from MARL, like centralized training with decentralized execution \cite{chen2019new}, can be used to develop AI systems that cooperate effectively \cite{tan1993multi}.

In this dissertation, we first use simple heuristics to model human behavior in a pen-based haptic feedback guidance system. We embedded this human behavioral model into an MPC that controls the haptic system. Afterwards, we use RL to model human behavior in adaptive interfaces. This is used to implicitly learn a behavioral model by the adaptive interface, to show optimal adaptations to the user.

\subsection{Shared Control}
Traditionally, there is a clear distinction between assistance systems, where the machine only supports the human, and automation, where the machine takes over the main task, replacing the human. Sheridan recognized that this distinction should not be so black and white, and proposed levels of automation \cite{sheridan1978human}. There are many situations where both the human and the machine should act together simultaneously, and where authority and tasks need to be shifted or adapted \cite{sheridan2011adaptive, miller2003beyond}.
Human-machine cooperation is a broader concept that includes shared control but also extends to cooperation at higher tactical and strategic levels like guidance and navigation. Shared control, specifically, is an approach where control is shared between human users and automated components, such as robots or computer systems. It involves cooperation at the control level between humans and machines \cite{flemisch2016shared}. Shared control is often discussed in the context of physical systems, such as (semi-)autonomous vehicles \cite{marcano2020review} and human-robot interaction \cite{losey2018review}.

\section{Haptic Interfaces}
\begin{figure}[h]
 \center
  \includegraphics[width=0.8\columnwidth]{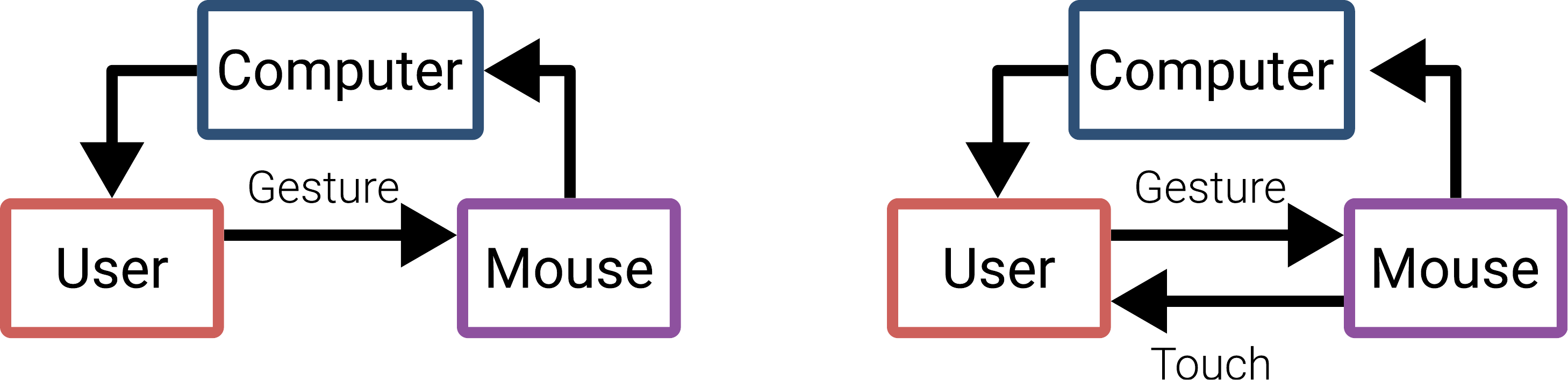}
  \caption{The Haptic-Loop allows for bi-directional exchange of information between machine and user (\citet{hayward2004haptic}).}
  \label{fig:haptic-loop} 
\end{figure}
Human haptic perception, one of our primary senses, is spread throughout the body in the skin, muscles, and joints. It allows us to perceive our environment, both natural and simulated, through two distinct modes: tactile and kinesthetic feedback \cite{Culbertson18}. Tactile sensing provides information about pressure, shear forces, and vibrations via mechanoreceptors in the skin. Kinesthetic feedback, closely tied to proprioception, enables us to sense the position of our limbs, body posture, and large-scale forces acting on our hands and body through muscle and joint activations. Haptic perception is unique in that it actively involves mechanical interaction - we manipulate the environment through touch and force, using the resulting feedback to continuously adjust our actions.

A haptic interface facilitates a two-way flow of information, allowing the user and machine to exchange data simultaneously, setting it apart from haptic displays and related graphical displays that only provide one-way information transfer. The haptic interface provides haptic feedback based on the user's input, creating a "haptic loop." For example, a standard mouse (left in \figref{fig:haptic-loop}) can be enhanced to provide haptic feedback (right in \figref{fig:haptic-loop}) that resists movement when the user approaches a disabled on-screen action. \citet{hayward2004haptic} further describe this information exchange as an exchange of energy, a concept that can be applied when using physics-based optimization techniques to develop haptic interfaces.

Haptic interfaces are especially interesting as a use case in our settings of shared variables. First, haptic technologies have found renewed interest in the human-computer interaction community due to the recent reemergence of Augmented and Virtual Reality systems. Second, the physical nature of haptic interaction makes variable ownership more impactful.

\subsection{Devices}
\subsubsection{Mechanical haptic feedback}
The most common form of rendering haptic feedback has been vibrotactile actuators.
They are popular in commercial devices, such as the controllers of game consoles, AR and VR controllers, and mobile phones. They can also be embedded directly into displays \cite{Wellman1995} or in clothing \cite{Cybertouch, Gloveone}.

Vibrotactile actuators usually produce coarse and global feedback, especially when used in handheld controllers (\eg rendering touch contact in VR~\cite{Benko2016,Choi2018}). Researchers have investigated how to overcome this limitation, such as by rendering interpolations between several such motors~\cite{israr2011tactile} or strategically distributing them across a controller (\eg placing them under the fingers to render local grasp feedback~\cite{Lee2019Torc}).

Alternatives to vibrotactile feedback often involve more complex articulated haptic elements, such as arms and braking mechanisms (\eg \cite{Massie94, Stamper1997, VanDerLinde2002, Araujo2016, zoller2019assessment, Sinclair2019Capstan}). In general, these types of systems can provide local haptic feedback at higher levels of fidelity and can render both tactile and kinesthetic feedback. Further extensions of such systems are exoskeletons~\cite{Gu2016, Choi2016}, gloves~\cite{Cybergrasp, hinchet2018dextres}, and tilt-platforms~\cite{Prattichizzo2013, Kim2016}. These platforms are usually (rigidly) anchored and can therefore supply large forces. These approaches, except DextrES \cite{hinchet2018dextres}, rely on mechanical structures and anchoring to the environment.
Therefore, their use is mostly limited to high-end applications such as teleoperation.

\subsubsection{Contact-free haptic feedback}
A second line of research focuses on contact-free haptics, which provide rich and strong feedback, and overcome the need for expensive and complex mechanical setups \cite{brink2014factors}. 

Within the contact-free domain, many different actuation devices have been explored, \eg ultra-sound pressure waves \cite{hoshi2010noncontact}, active control of stylus motions \cite{kianzad2018harold}, and drones \cite{heo2018thor}. The most popular and practical actuators in this domain use magnetism. The simplest form of magnetism is delivered by passive magnets that are embedded into interactive objects (\eg~\cite{yamaoka2013depend}). The recent advance of consumer 3D printing has allowed this approach to actuate objects with arbitrary shape and function~\cite{zheng2019mechamagnets, ogata2018magneto}. A big shortcoming of passive magnets, however, is the lack of dynamic control over them and thus the forces users perceive during interaction.

\subsubsection{Electromagnetic haptic feedback}
The shortcoming of passive magnets can be addressed using electromagnetism and computational control of magnetic forces. Two-dimensional arrays of electromagnets can be combined with passive magnets that are worn \cite{weiss2011fingerflux, zhang2016magnetic, yamaoka2013depend, adel2019magnetic, berkelman2012co, berkelman2013interactive, berkelman2018electromagnetic} or embedded in tools and interactive objects \cite{ju2002origami, weiss2010madgets, Langerak:2021:Hedgehog}. Pangaro et al.\cite{pangaro2002actuated} model the force-field of each electromagnet and combine these using standard aliasing techniques, allowing directed movement of multiple objects on the surface. Similarly, Yoshida et al.\cite{yoshida2006proactive} use linear induction motors to control objects on a tabletop. Strasnick et al.\cite{strasnick2017shiftio} use electromagnets to control an object on a mobile phone case. Suzuki et al.\cite{suzuki2018reactile} combine these two works and use a grid of electromagnetic cores to move objects on a tabletop. The actuation area can be increased by attaching an electromagnet to a biaxial linear stage \cite{langerak2020magpen, langerak2019demonstration}.

Similarly, by leveraging the electromagnetic forces in a coil between two permanent magnets, large and grounded forces can be delivered onto a joystick \cite{berkelman2009extending}. The main drawback of this approach is the requirement for a mechanical connection, limiting the range of motion and impeding contact-free haptics. Senkal et al.\cite{senkal2009spherical, senkal2011haptic} and Li et al.\cite{li20072} propose to use magnetorheological fluids in joysticks. With the help of an electromagnetic field, the internal friction can be significantly increased. This allows for a large breaking force; however, it does not allow adding energy to the system.

Perhaps the most well-known haptic interface that builds on Lorentz forces is the Butterfly Haptics Maglev \cite{berkelman1996design, berkelman2000lorentz}. The design of this system consists of a flotor bowl with six integrated coils to which an interaction handle is rigidly attached. This flotor bowl is levitated between magnets assemblies that are part of a stator bowl. Due to the Lorentz levitation, haptic feedback can be achieved in a degree of rotary movement and also a small translation. Due to the small movements allowed, the device is mainly suitable for small-scale manipulations, e.g., where only the fingers are used.

The design of our system is fundamentally different. In contrast, \omniHap and \omniUIST use a single omnidirectional electromagnet, thereby largely increasing the rotary capabilities. Furthermore, by omitting the levitating flotor bowl, our devices greatly reduce the complexity and therefore make it easier to fabricate, thereby more likely to foster adaptation. Finally, our designs allow for contact-free haptics, which is not possible in the Butterfly system.

Closely related work to \omniHap and \omniUIST also includes Omnimagnet by Petruska et al. \cite{petruska2014omnimagnet} and its variants \cite{iqbal2019design}. Similar to our devices, their system generates an omnidirectional magnetic field. Their design differs by using three nested cuboid-shaped coils, causing force decay as the user moves along the surface of the device as well as an obstruction of heat dissipation. This limits the maximal strength and duration of actuation~\cite{esmailie2017thermal}, and also makes the device only suitable for rendering vibrotactile stimuli \cite{zhang2018six}. Due to the cuboid shape, the center-to-center distance between the electromagnet and the permanent magnet is not constant among the surface, which causes high variance in the forces perceived. In contrast, \textit{Omni}'s design is spherical, symmetrical, and has intertwined coils. This results in better heat dissipation and less variance in the force, thereby arguably improving the user experience.

\subsubsection{Haptic Guidance Devices}
Providing haptic guidance to users can provide benefits for learning \cite{teranishi2018combining} and short-term performance (cf. Abbink et al.\cite{Abbink2012}). Teranishi et al.\cite{teranishi2018combining} demonstrate that participants showed improved learning for handwriting skills when receiving guidance through a 3-DOF Phantom Omni device. Mullin et al.~\cite{mullins2005haptic} use a similar device as a handwriting aid for rehabilitation. Forsyth and MacLean \cite{Forsyth2006} show that force cues are beneficial in navigation tasks. The focus of these works is that users receive tight guidance (i.e., they are supposed to follow the system as closely as possible).

There exists a large range of devices and systems that aim at providing guidance to users. Comp*Pass \cite{Nakagaki14} uses pantograph-like devices to assist users in drawing, while I-Draw \cite{Fernando14} is a motorized drawing assistant. Lin et al. \cite{Lin16} use a magnet mounted on a small robotic arm to retain the correspondence between the pen and a portable base. Digital rubbing employs a comparable system using a solenoid for tracing over digital images on real paper \cite{kim2008digital}. While users handle larger-scale motions of the devices, they generally aim at having full control over the resulting drawing. Users can take back this control; however, these systems do not provide a way to guide users back to the target trajectory. Besides the aforementioned systems, several works aid users in the process of crafting and manufacturing (cf. Zoran et al.\cite{zoran2014wise}). Free-D \cite{zoran2013freed} and D-Coil \cite{peng2015d} assist users in sculpting physical artifacts by guiding them on a predefined 3D shape. Shilkrot et al.\cite{Shilkrot2014} propose an augmented paintbrush to assist users in painting. While users can override these systems to deviate from the target shape, they have no mechanism that guides users back to the target.

\subsection{Control of Haptic Devices}
To effectively control haptic devices, a closed loop needs to be formed. This means that the user needs to be sensed, and based on this inference, an actuation decision needs to be made. We first discuss magnetic sensing and then the control for haptic guidance devices.

\subsubsection{Magnetic Sensing}
Permanent magnets have been used for tracking objects in 3D, ranging from styli and other interactive objects \cite{liang2012gausssense, kuo2016gaussmarbles}, jewelry \cite{ashbrook2011nenya} all the way to fingers \cite{han2007wearable}, joints, and other biological tissues \cite{bhadra2002implementation, tarantino2017myokinetic}.

Ample research exists on tracking permanent magnets. Most of the existing literature uses isotropic (i.e., spherically) shaped permanent magnets, because the dipole model most accurately resembles these \cite{jackson2007classical}. However, some work also uses electromagnets attached to fingertips \cite{chen2016finexus}. Due to this, the fingers can be tracked individually. One of the biggest challenges is that a closed-form solution of the magnet states (e.g., position) is unlikely to exist in most scenarios. Therefore, optimization-based methods are commonly used \cite{schlageter2001tracking, taylor2019low} and more recently also neural networks \cite{russel2017neural}. However, these methods were often employed offline, suffered from large latency, or converged to local minima. 

A key related work is Magnetips by McIntosh et al.\cite{McIntosh2019}. They use a permanent magnet attached to a fingertip to interact with a watch. The permanent magnet is tracked around the watch and also used to provide vibrotactile haptic feedback. Magnetips multiplexes actuation and sensing. However, this causes significant delays (2 ms for every swap between tracking and actuating), which may pose an issue for scenarios that require continuous interactions.

The work that most resembles our work from an algorithmic point of view is \cite{taylor2019low}. They track multiple spherical magnets online using an analytical Jacobian, allowing solving with a quasi-Newton method. In contrast, \omniUIST tracks a single magnet while compensating for drastically changing electromagnetic fields, rather than tracking multiple permanent magnets in a static environment. We do this by adjusting the dipole model so that it is suitable for electromagnetism. We also propose a novel formulation of the position reconstruction problem and an implementation in PyTorch that can leverage the framework's auto-grad capabilities, thus avoiding the need for an analytical Jacobian.

\subsubsection{Haptic Guidance Control}
Closest to \magpen in terms of hardware is dePENd by Yamaoka et al.~\cite{yamaoka2013depend}.
They move a permanent neodymium magnet on a two-axis setup to control the metal tip of a ballpoint pen. The neodymium magnet "drags" the input pen around a predefined path, similar to a plotter. dePENd employs an open-loop strategy to control the magnet, which means users cannot deviate from the predefined path without risking losing haptic guidance. In contrast, in \magpen we propose a mathematical model and optimal control strategy that allows users to move at their own pace through a drawing, for example, and reacts in real-time to user input by altering the position and strength of the magnet. We show that our approach provides better results than their open-loop approach, as well as existing closed-loop approaches.

Kianzad et al.~\cite{Kianzad20} use a ballpoint drive to assist users in sketching. They employ a proportional-derivative (PD) control loop, which allows users to deviate from the target to a certain extent. We show in our experiments that our optimization scheme outperforms such existing closed-loop approaches. Muscle-Plotter \cite{Lopes16} proposes active guidance for users based on electrical muscle stimulation. Their control strategy is based on heuristics for users to share control with the system. Our approach could be applied to their work if the electromagnetic force model is replaced by a model of the interaction between the muscle stimulation and the force users produce.

Optimal reference following given real-world influences is studied in depth in the control theory literature. Methods like Model Predictive Control (MPC) \cite{Faulwasser:2009} optimize the reference path and the actuator inputs simultaneously based on the system state.
MPC is widely applied in many robotics (e.g., to control quadcopters, Mueller et al.\cite{Mueller2013}) and graphics applications (e.g., for human motion prediction, Da Silva et al.\cite{dasilva:2008:mpc}). However, Aguiar et al.~\cite{AGUIAR2008} show that the tracking error for following timed trajectories can be larger than if following a geometric path only.

To address this issue, Lam et al. propose Model Predictive Contouring Control (MPCC) \cite{lam2013model} to follow a time-free reference, optimizing system control inputs for time-optimal progress. MPCC has been successfully applied in industrial contouring \cite{lam2013model}, RC racing \cite{Liniger2014}, and in drone cinematography \cite{Naegeli:2017:MultiDroneCine}.

We also pose our optimization problem in this well-established framework. However, to the best of our knowledge, we are the first to apply it for haptic guidance systems where one has to consider both a controllable (i.e., the electromagnetic force) and non-controllable (i.e., the user) system. We contribute a formulation of the problem including models and control algorithms to enable employing MPCC in this context.

\section{Adaptive Interfaces}
UI adaptation can either be offline, to computationally design an interface, or online, to adapt the UI according to users' goals. We will focus on online adaptive UIs and refer readers to \cite{combinatorialoptimizationdesign2020oulasvirta, functionalityselection2017oulasvirta} for an overview of computational UI design.
\subsection{Control of Adaptive Interfaces} We introduce an overview of different control strategies for adaptive interfaces. Note that we use control strategy and policy in the context of adaptive interfaces interchangeably.

\subsubsection{Heuristics, Bayesian Networks \& Combinatorial Optimization}
In early works, heuristic- or knowledge-based approaches are used to adapt the UI \cite{Browne1990,Stephanidis1997,Smith2010}. Similarly, multi-agent systems employ rule-based and message-passing approaches \citep{Rich1998,Rich2005,Yorke2012}. Another popular technique for AUIs is domain-expert-designed Bayesian networks \citep{Horvitz1998, Bosma2004}. More recently, combinatorial optimization was used to adapt interfaces dynamically \cite{park2018adam, lindlbauer2019context}. The downside of these approaches is that experts need to specify user goals using complex rule-based systems or mathematical formulations.
Creating them comprehensively and accurately requires developers to foresee all possible user states, which is tedious and requires expert knowledge.Commonly, these approaches also get into conflict when multiple rules or objectives apply. This conflict often results in unintuitive adaptations. In contrast, \marlui only requires the layout of the UI.
From its representation as an RL environment, we learn policies that meaningfully adapt the UI and realistically reproduce user behavior.

\subsubsection{Supervised Learning}
Leveraging machine learning can overcome the limitations of heuristic-, network-, and optimization-based systems by learning appropriate UI adaptations from user data.
Traditional machine learning approaches commonly learn a mapping from user input to UI adaptation.

Algorithms like nearest neighbor \cite{Maes1995, Lashkari1997}, Naïve Bayes \cite{McCreath2006,Faulring2010}, perceptron \cite{Shen2009a, Shen2009b}, support vector machines \cite{Berry2011}, or random forests \cite{Pejovic2014, Mehrotra2015} are used and models are learned offline \cite{Berry2011} and online \cite{Shen2009a}.
Due to the problem setting, these approaches require users' input to be highly predictive of the most appropriate adaptation. Furthermore, it restricts the methods to work in use cases where myopic planning is sufficient, i.e., a single UI adaptation leads users to their goal.
In contrast, \marlui considers multiple goals when selecting an adaptation and can lead users to their goal using sequences of adaptations.

More recent work overcomes the limitations stemming from simple input-to-adaptation mapping by following a two-step approach. They (1) infer users' intention based on observations and (2) choose an appropriate adaptation based on the inferred intent \cite{oulasvirta2018computational}. Such work uses neural networks, and user intention is modeled either explicitly \cite{kolekar2010learning, soh2017deep} or as a low-dimensional latent representation \cite{RIZZOGLIO2021}.

However, these approaches are still highly dependent on the quality of the training data, which may not even be available for emerging technologies. In contrast, \marlui can learn supportive policies without pre-collected user data by just observing simulated user behavior.

\subsubsection{Bandits \& Bayesian Optimization}
Bandit systems are a probabilistic approach often used in recommender systems \cite{glowacka2019bandit}. In a multi-armed bandit setting, each adaptation is modeled as an arm with a probability distribution describing the expected reward. The Bayes theorem updates the expectation, given a new observation and prior data. Related work leverages this approach for AUIs \cite{lomas2016interface,koch2019may,kangas2022scalable}.

Bayesian optimization is a sample-efficient global optimization method that finds optimal solutions in multi-dimensional spaces by probing a black box function \cite{shahriari2015taking}. In the case of AUIs, it is used to find optimal UI adaptations by sampling users' preferences \cite{Koyama2014, Koyama2016}. Both approaches trade off exploration and exploitation when searching for appropriate adaptations (i.e., exploration finds entirely new solutions, and exploitation improves existing solutions), rendering them suitable approaches to the AUI problem. However, such methods are not able to plan adaptations over a sequence of interaction steps, i.e., they plan myopic strategies.

In addition, these approaches need to sample user feedback to learn or optimize for meaningful adaptations and, hence, also rely on high-fidelity user data.
Furthermore, as users themselves learn during training or optimization, solutions can converge to sub-optimal user behavior as such methods reduce exploration with convergence.
In contrast, \marlui can plan adaptations over a sequence of interaction steps learned from realistic, simulated user data.

\subsubsection{Reinforcement Learning}
Reinforcement learning is a natural approach to solving the AUI problem, as its underlying decision-making formalism implicitly captures the closed-loop iterative nature of HCI \cite{howes2018interaction}. It is a generalization of bandits and learns policies for longer horizons, where current actions can influence future states. This generalization enables selecting UI adaptations according to user goals that require multiple interaction steps. Its capability makes RL a powerful approach for AUIs with applications in dialog systems \cite{Gasic2014, Su2017}, crowdsourcing \cite{PengCrowdsourcing2013, Hu2018}, sequential recommendations \cite{Chen2019, Liu2018, Liebman2015}, information filtering \cite{seo2000reinforcement}, personalized web page design \cite{ferretti2014exploiting}, and mixed reality \cite{gebhardt2019learning}. Similar to our work is a model-based RL method that optimizes menu adaptations \cite{todi2021adapting}.

Current RL methods sample predictive models \cite{todi2021adapting, Gasic2014, Hu2018} or logged user traces \cite{gebhardt2019learning}. However, these predictive models and offline traces represent user interactions with non-adaptive interfaces. Introducing an adaptive interface will change user behavior; so-called co-adaptation \cite{mackay2000responding}. Hence, it is unclear if the learned model can choose meaningful adaptations when user behavior changes significantly due to the model's introduction. In contrast, our user agent learns to interact with the adapted UI; hence, our interface agent learns on behavioral traces from the adapted setting.

\subsubsection{Multi-Agent Reinforcement Learning}
MARL is a generalization of RL in which multiple agents act, competitively or cooperatively, in a shared environment \cite{zhang2021multi}.Multi-agent systems are common in games \cite{baker2019emergent, jaderberg2019quake}, robotics \cite{ota2006multiagent,sariff2018multiagent}, or modeling of social dilemmas \cite{chao2015social,leibo2017multi}. MARL is challenging since multiple agents change their behavior as training progresses, making the learning problem non-stationary. Common techniques to address this issue are via implicit \cite{tian2020implicit} or explicit \cite{foerster2016learning} communication, centralized critic functions \cite{lowe2017multi}, or curricula \cite{epciclr2020,wang2020curriculum}.

\cleardoublepage%

\def\dir{chapters/background}

\chapter{Background}
\label{ch:background}

We introduce background in electromagnetic models, specifically the dipole and dipole-dipole model. Afterwards, we introduce optimal control theory. 

\section{Magnetic Model}
\label{sc:back.mag}

\subsection{Dipole Model}
\label{sc:back.dipole}
In our haptic applications, we employ electromagnetic feedback. To effectively provide this feedback, it is crucial to understand the relationships between the electrical current, the magnetic field, and the force experienced by a permanent magnet embedded in a tool.

\begin{figure}
    \centering
    \includegraphics[width=0.5\textwidth]{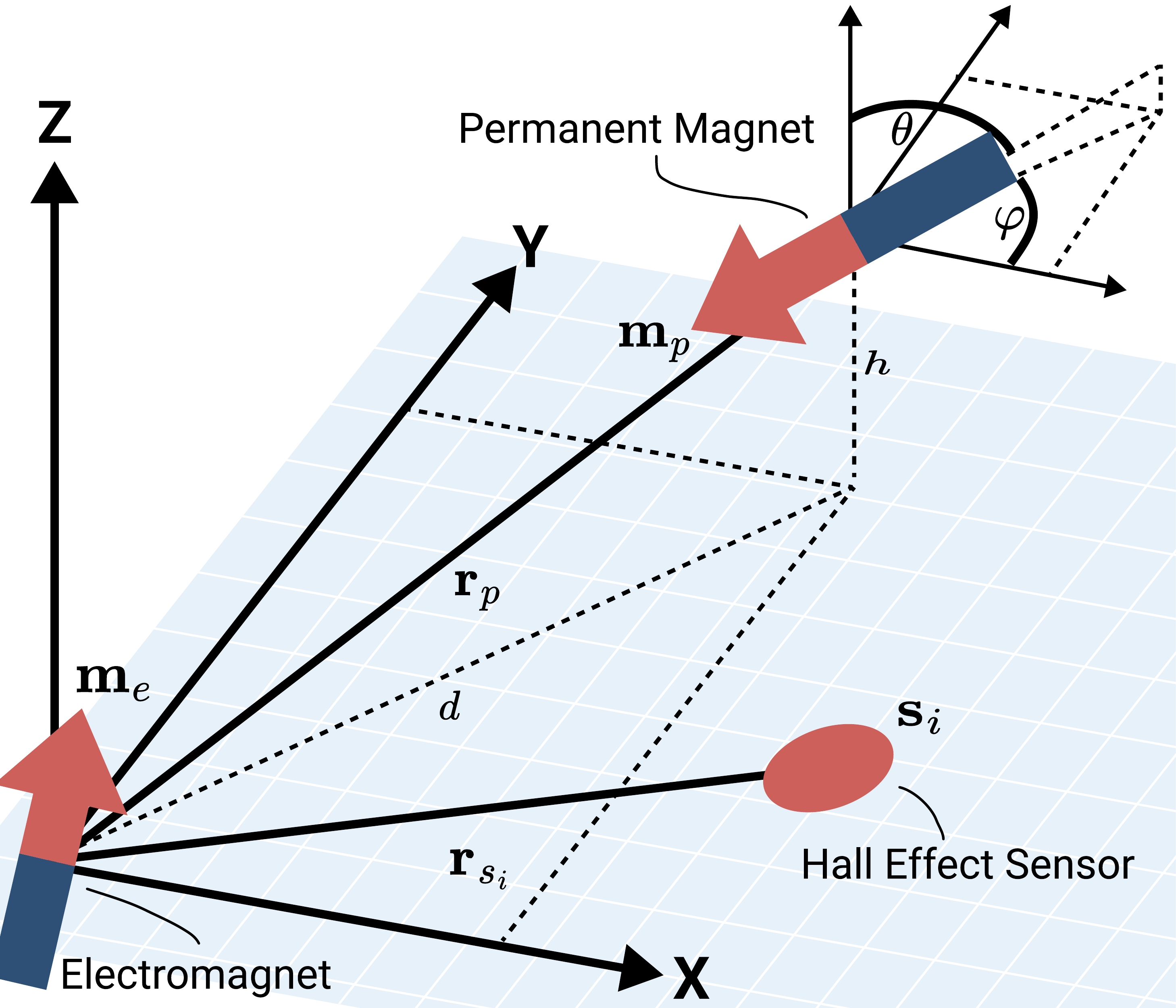}
    \caption{Diagram of the coordinate system used in the models}
    \label{fig:background:em_coordinates}
\end{figure}

We utilize simplified magnetic field models from previous research \cite{thomaszewski2008magnets}, which allow us to simulate the magnetic field in real-time. These models require only the magnitude, orientation, and position of each dipole magnet. We apply similar simplifications in our approach.

The formula for the magnetic field created by a dipole is:
\begin{equation}
    \BBold (\rBold, \mBold)=\frac {\mu _{0}}{4\pi }\frac {3 \ \HatrBold \ (\mBold \cdot \HatrBold  )-\mBold}{|\rBold |^3} \ ,
    \label{eq:basic_B}
\end{equation}        
where $\mu_0$ represents the permeability of air, $\rBold$ is the vector from the center of the magnetic source to the testing point, and $\mBold$ symbolizes the magnetic moment containing the \emph{strength} and orientation of the magnetic source. The magnetic moment varies depending on the magnet type: $\meBold$ for an electromagnet and $\mpBold$ for a permanent magnet. For the permanent magnet embedded in the tool, angles $\theta$ and $\varphi$ are measured from the positive z-axis, as depicted in \figref{fig:background:em_coordinates}.

The three components of the magnetic moment for the tool’s embedded magnet are described by:
\begin{equation}\label{eq:mpBold}
    \mpBold =  \frac{1}{\mu_0}*Br*V * \begin{bmatrix}\sin(\theta) \ \cos(\varphi)\\\sin(\theta) \ \sin(\varphi)\\\cos(\theta)\end{bmatrix}\ ,
\end{equation}
where $V$ is the volume and $Br$ is the residual magnetic flux of the permanent magnet, expressed in Tesla units.

\subsubsection{Electromagnet Dipole Equivalent} 
\label{sc:back.dipole.cyl}
Our designs employ electromagnets to generate haptic feedback. We extend the dipole model to include cylindrical and spherical electromagnets used in different devices.

The magnetic field generated by an electromagnet dipole at a point influenced by another dipole or sensor is given by:
\begin{equation}
  \mathbf{B_e} (\rBold,\mmBold) =  \BBold (\rBold, \mmBold),
  \label{eq:ap.B2}
\end{equation}

\noindent where $\rBold$ represents the vector from $\mmBold$ to either $\mpBold$ or $\siBold$, as shown in \figref{fig:background:em_coordinates}.

\paragraph{Cylinderical}
    \magpen uses a cylindrical electromagnetic. Hence we need a dipole model for cylindrical electromagnets used in our applications. The magnetic field $\mathbf{B_e}$ is modeled using a cylindrical coordinate system aligned with the magnet. The magnet's magnetic moment, $\mmBold$, can be expressed as:
    \begin{equation}
        \mmBold = \alpha \ m_e \ [0,0,1]^T,\\
    \end{equation}
    \noindent where $\alpha$ represents the normalized input to the electromagnet and $m_e$ the calibrated strength of the magnet.

\paragraph{Spherical}
    \omniHap and \omniUIST consit of custom spherical electromagnets. These operate under different constraints compared to cylindrical electromagnets. However, we can approximate them as the superposition of three orthogonal cylindrical electromagnets. However, in practice the coils are not perfectly orthogonal. Hence we use a calibration matrix $\mathbb{C}$ (cf. a scalar $\alpha$ in a cylindrical EM). Thus, the magnetic moment of these electromagnets is proportional to the actuation current:
    \begin{equation}\label{eq:me_from_c_i}
    \meBold = \mathbb{C} * \mathbf{I}^T \ ,
    \end{equation}
    \noindent where $\mathbf{I}$ is the current vector for each coil, and $\mathbb{C}$ is a calibration matrix derived from measurements.

    
    
    
\subsubsection{Magnetic Field at a Sensor}
To compute the magnetic field at each sensor location, we consider the combined influence of the permanent magnet, the electromagnet, and any background interference:
\begin{equation} \label{eq:b_sum}
\BiBold =\underbrace{\BBold(\RsiBold-\RpenBold, \mpBold)}_{\BpBold} + \underbrace{\BBold(\RsiBold, \meBold)}_{\BeBold} + \BnBold \ .
\end{equation}
This equation allows us to predict the total magnetic field experienced at each sensor location accurately.

\subsection{Dipole-dipole Model} 
\label{ap:em_model}

We have introduced the dipole model to calculate the magnetic fields produced by both permanent and electromagnets.

Primarily, our objective is to determine the actuation force on the pen, $\mathbf{F_p}$, which can be computed by integrating the gradient of the magnetic potential over the volume of the pen's permanent magnet:
\begin{equation}
\mathbf{F_p} = \iiint \nabla \left( \mathbf{M_p} \cdot \mathbf{B_m}(\cdot)\right) dxdydz , \label{eq:gradB2}
\end{equation}

The dipole-dipole interaction model, as described analytically by \citet{yung1998analytic}, provides an equation for the force between two magnetic dipoles. This model is crucial for determining the necessary magnetic dipole moment of the electromagnet, given the magnetic dipole and position of the permanent magnet to achieve a specific force:

\begin{multline}
   \mathbf{F_p} = {\dfrac  {3\mu _{0}}{4\pi \Rmagtopen^{5}}}
   \left [ \left(\langle\mpBold,\RmagtopenBold\rangle \right) \mmBold + 
   \left(\langle\mmBold,\RmagtopenBold\rangle\right) \mpBold \right . +
   \\
   \left(\langle\mpBold,\mmBold\rangle\right) \RmagtopenBold - 
    \left . {\dfrac{5\left(\langle\mpBold,\RmagtopenBold\rangle\right)
    \left(\langle\mmBold,\RmagtopenBold\rangle\right)}{\Rmagtopen^{2}}} \RmagtopenBold \right ] \ , \label{eq:F21-dip}
\end{multline}
A detailed derivation for specifically planar forces, as used in \magpenTitle, is provided in \appref{app:dipoledipole}.

\section{Optimal Control}
\label{sec:ocp}

This chapter offers an overview to introduce the reader to the problem formulations and methods of optimal control for discrete-time dynamic systems, which are the main formulations and methods used in the second half of this thesis chapters. This discussion is partially is based on \citeauthor{gebhardt2020optimal}'s earlier work and we refer to their publication for an overview \cite{gebhardt2020optimal}. 
We start by discussing the optimal control problem (OCP) for scenarios where the dynamic system's model is known. We proceed to examine model predictive control, an optimization technique employed to address these scenarios. Next, we introduce Markov decision processes (MDPs), which serve to frame OCPs when the system model is not known. We extend MDPs to their partial-observable counterparts (POMDPs), these enable scenarios where the system does not have access to the full state space. From there, we explore reinforcement learning (RL), methodologies applied to solve (PO)MDPs. Finally, we review multi-agent reinforcement learning (MARL) for scenarios with multiple agents in a shared environment. 

\subsection{Optimal Control}

\begin{figure}
    \centering
    \includegraphics[width=\textwidth]{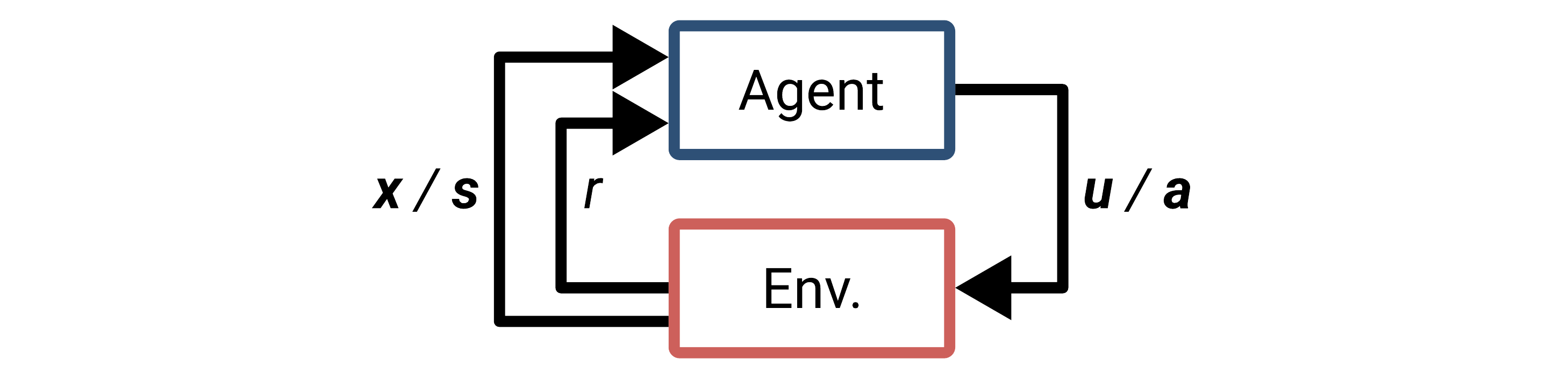}
    \caption{An overview of the control problem. An agent/system takes action $\action$ or input $\mathbf{u}$ based on the state $\mathbf{x}$ or $\mathbf{s}$. The action/input updates the state. Furthermore the agent receives a reward $r$.}
    \label{fig:ocp}
\end{figure}

 Control theory is a field that studies how to influence the behavior of dynamic systems using external inputs, with the goal of achieving a target state. Optimal control, a subfield of control theory, involves finding the best possible control inputs and state trajectories for a dynamic system. This is done by optimizing a measure of performance or cost. Optimal control can be applied to systems with either continuous or discrete dynamics. In this context, we concentrate on discrete dynamic systems.

First, we introduce the OCP formulation for the case of a discrete system model which inputs are optimized according to a cost function (\figref{fig:ocp}). Therefor, we consider the following discrete-time linear time-invariant system
\begin{equation}
\label{eq:linear_system}
    \mathbf{\statex}(\timet+1) = \mathbf{A\statex}(\timet) + \mathbf{B\inputu}(\timet),
\end{equation}
\noindent where $\statex \in \mathbb{R}^{n}$ is the state vector, $\inputu \in \mathbb{R}^{m}$ the input vector, $\mathbf{A} \in \mathbb{R}^{n \times n}$ the system matrix, $\mathbf{B} \in \mathbb{R}^{n \times m}$ the input matrix, $\timet \in \mathbb{N}_0$ the discrete time, and the pair $(\mathbf{A},\mathbf{B})$ is assumed stabilized.

We define $\statex(\timet)$ as the state vector measured at discrete time $\timet$ and $\statex_{\timet+k}$ be the state vector predicted at discrete time $\timet+k$ using state equation \ref{eq:linear_system} with initial condition $\statex_\timet = \statex(\timet)$.
We constraint the problem with:
\begin{equation}
\label{eq:constraints}
    \statex(\timet) \in \mathbb{X} \subseteq \mathbb{R}^{n}, \inputu(\timet) \in \mathbb{U} \subseteq \mathbb{R}^{M},
\end{equation}
\noindent where $\mathbb{X}$ and $\mathbb{U}$ are the sets of all possible states and inputs respecively.

Then, the a general cost function can be defined as:
\begin{equation}
\label{eq:cost_function}
    \valuev_N(\statex(\timet)) = r_T(\statex_{\timet+N}, \inputu_{\timet+N}) + \sum^{N-1}_{k=0} \gamma^k r(\statex_{\timet+k}, \inputu_{\timet+k}),
\end{equation}
\noindent with the stage cost:
\begin{equation}
\label{eq:stage_cost}
    r(\statex_{ime\timet+k}, \inputu_{\timet+k}) = \statex_{\timet+k}^T\mathbf{Q}\statex_{\timet+k} + \inputu_{\timet+k}^T\mathbf{R}\inputu_{\timet+k},
\end{equation}
\noindent and terminal cost:
\begin{equation}
\label{eq:terminal_cost}
    r_T(\statex_{\timet+N}, \inputu_{\timet+N}) = \statex_{\timet+N}^T\mathbf{P}\statex_{\timet+N} ,
\end{equation}
\noindent where $N$ is the prediction horizon, $\gamma \in (0,1]$ a discount factor to give less weight to errors and inputs further in the future, $\mathbf{Q} \in \mathbb{R}^{n \times n}$ and $\mathbf{R} \in \mathbb{R}^{m \times m}$, and $\mathbf{P} \in \mathbb{R}^{n \times n}$ are the state, input and terminal weighting matrices respectively. The weighting matrices are required to be symmetric and positive definite to ensure the objective function is convex. The terminal cost matrix $\mathbf{P}$ is obtained by solving an algebraic Riccati equation, which arises from the infinite-horizon linear quadratic regulator problem \citep{Camacho2013}.

Optimal control problems can be categorized as either finite-horizon (FHOCPs) or infinite-horizon (IHOCPs) based on the time period considered for optimization.
In FHOCPs, the control inputs are optimized over a fixed, finite number of future states or time steps. The goal is to find the best control sequence that minimizes the objective function from the initial state to a specified terminal state. We will use a FHOCP for haptic control.  
In IHOCPs, the control inputs are optimized assuming the system will continue to operate indefinitely. The objective is to find a control policy that minimizes the cost over the entire unbounded future, often expressed as an infinite sum of stage costs. The optimal control law is time-invariant - it provides the optimal input as a function of the current state, regardless of the time index. We will use an infinite horizon in our work on \marluiTitle. 

The FHOCP is defined as:
\begin{align}
\label{eq:finite_horizon}
    \valuev_N^*(\statex(\timet)) = \underset{\inputu(\timet)}{\min}~&r_T(\statex_{\timet+N}, \inputu_{\timet+N}) + \sum^{N-1}_{k=0} \gamma^k r(\statex_{\timet+k}, \inputu_{\timet+k}) \\ \nonumber
    \text{subject to} & \\ \nonumber
    &\statex_{\timet+k+1} = \mathbf{Ax}_{\timet+k} + \mathbf{Bu}_{\timet+k},~k = 0,..., N - 1 \\ \nonumber
    &\statex_{\timet+k} \in \mathbb{X},~k = 1,..., N  \\ \nonumber
    &\inputu_{\timet+k} \in \mathbb{U},~k = 0,..., N - 1  \\ \nonumber
    &\statex_{\timet} = \statex(\timet) \nonumber
\end{align}
with input sequence $\inputu(\timet) = (\inputu^{T}_\timet, ..., \inputu^{T}_{t+N-1})^T \in \mathbb{R}^{Nm}$.

Similarly, the IHOCP is defined as
\begin{align}
\label{eq:infinite_horizon}
    V_\infty^*(\statex(\timet)) = \underset{\inputu(\timet)}{\min}~&\sum^{\infty}_{k=0} \gamma^k r(\statex_{\timet+k}, \inputu_{\timet+k}) \\
    subject~to~ &\statex_{\timet+k+1} = \mathbf{Ax}_{\timet+k} + \mathbf{Bu}_{\timet+k},~k = 0, 1,... \\ \nonumber
    &\statex_{\timet+k} \in \mathbb{X},~k = 1, 2,...  \\ \nonumber
    &\inputu_{\timet+k} \in \mathbb{U},~k = 0, 1,... \\ \nonumber
    &\statex_\timet = \statex(\timet) \nonumber
\end{align}
with input sequence $\inputu(\timet) = (\inputu^{T}_\timet, \inputu^{T}_{t+1}, ...)^T$.

We introduced the OCP for a linear system so far. However, not all systems are linear or can be linearized; such as human behavior. In the nonlinear case, the right part of \eqref{eq:linear_system} is substituted by a nonlinear function that describes the evolution of the dynamic system:
\begin{equation}
\label{eq:nonlinear_system}
    \statex(t+1) = F(\statex(\timet), \inputu(\timet)).
\end{equation}
The terminal penalty for non-linear stytem is given by the Lypunov equation \citep{Johansen2004}, rather than the Ricatti equation.

\subsection{Model Predictive Control}
\label{sec:back:mpc}
Model predictive control (MPC) is a practical implementation of (in)finite horizon optimal control for the measured state vector $\statex(\timet)$ to attain the predicted optimal input sequence $\inputu^(\timet)$. Specifically, it is a closed-loop control method that measures the state ($\statex$) at every time step ($\timet$), solves the optimal control equations \eqref{eq:finite_horizon}, and then applies its first element ($\inputu^(\timet)$) to the system. This is repeated at each discrete time $\timet$ with a receding prediction horizon. Thus, MPC is also denoted as receding horizon control (RHC). 

Both the FHOCP and the IHOCP can be reformulated as a quadratic programming (QP) problem for linear systems with quadratic cost functions and linear constraints (\citep{Gorges2017}):

\begin{align}
	\label{eq:quadratic_program}
	\underset{\mathbf{X}}{\text{minimize}} \ & \frac{1}{2} \mathbf{X}^{T} \mathbf{H} \mathbf{X} + \mathbf{f}^{T} X \\
	\text{ subject to } &\mathbf{A}_{\mathit{ineq}} \statex \leq \mathbf{b}_{\mathit{ineq}} \nonumber \\
	\text{ and } & \mathbf{A}_{\mathit{eq}} \statex = \mathbf{b}_{\mathit{eq}} \nonumber~,
\end{align}

Here, $\mathbf{X}$ represents the stacked state vectors $\statex_{t}$ and inputs $\inputu_{t}$ at each time point. The matrices $\mathbf{H}$ and $\mathbf{f}$ contain the quadratic and linear cost coefficients, respectively, as defined in \eqref{eq:cost_function}. The matrices $\mathbf{A}{\mathit{ineq}}$ and $\mathbf{b}{\mathit{ineq}}$ describe the linear inequality constraints on states and inputs from \eqref{eq:constraints}, while $\mathbf{A}{\mathit{eq}}$ and $\mathbf{b}{\mathit{eq}}$ represent the linear equality constraints from our model in \eqref{eq:linear_system} for each time point $k \in {0,\ldots,N}$.

To solve nonlinear dynamic systems and cost functions, we can use nonlinear programming (NLP) or sequential quadratic programming (SQP) methods. NLP addresses the nonlinear optimization problem directly, whereas SQP iteratively solves local quadratic programming (QP) approximations of the nonlinear problem. The selection of horizon length, terminal cost, and constraints in these formulations is crucial for ensuring closed-loop stability and performance. We use numerical solvers to address MPC problems \cite{forcespro}.

\subsection{Markov Decision Processes}
\label{sec:mdp}
Model Predictive Control (MPC) addresses the OCP of a system assumed to behave deterministically according to a known dynamic model. However, many OCPs inherently have stochastic state transitions, making it impossible to model the underlying system dynamics accurately. Markov Decision Processes (MDPs) offer a mathematical framework for these situations, where a system model and suitable control policies are learned from the system's interactions with its environment. Generally, MDPs can be used to represent a wide range of sequential decision-making problems in stochastic settings \citep{kaelbling1998planning}.

An MDP satisfies the Markov property, which states that state transitions depend only on the current state and action, not on any prior states or actions. An MDP is defined by a five-tuple ($\StatePerPolicy$, $\ActionPerPolicy$, $\Transitions$, $\RewardPerPolicy$, $\discount$), where $\StatePerPolicy$ is the set of states, $\ActionPerPolicy$ is the set of actions, $\Transitions: \StatePerPolicy \times \ActionPerPolicy \times \StatePerPolicy \rightarrow [0,1]$ is the transition probability function, and $\Transitions (\state', \action, \state)$ represents the probability of transitioning from state $\state'$ to state $\state$ after taking action $\action$. $\RewardPerPolicy : \StatePerPolicy \times \ActionPerPolicy \rightarrow \mathbb{R}$ is the reward function, with rewards discounted by a factor $\discount$. The expected discounted reward for taking action $\action$ in state $\state$ under policy $\policy$ is called the Q value: 

\begin{equation}
\label{eq:q_value}
    Q^{\policy}(\state,\action) = \mathbb{E}_\policy\left[\sum_{\timet=0}^{\infty}\gamma^t \RewardPerPolicy(\state_\timet, \action_\timet) \mid | \state_\timet=\state, \action_\timet=\action \right ],
\end{equation}

The value function $V^\policy(s)$ is the expected return starting from state $s$ and following policy $\policy$: 
\begin{equation}
V^\policy(s) = \mathbb{E}_\policy\left[\sum_{t=0}^\infty \discount^t \RewardPerPolicy(\state_\timet,\action_\timet)\mid | \state_\timet = \state \right  ]
\end{equation} 

The Q values and value function are related via: 
\begin{equation}
V^\policy(\state) = \sum_\action \policy(\action|\state) Q^\policy(\state,\action)
\end{equation}

The Q values of following states are related via the Bellman equation:
\begin{equation}
\label{eq:bellman}
    Q^\policy(\state, \action) = \sum_{\state'} P(\state'|\state,\action)[\RewardPerPolicy(\state',\state,\action) + \discount Q^\policy(\state',\policy(\state'))].
\end{equation}
The optimal policy can then be computed as $\policy^* = \arg\max_a Q^\policy(s,a)$.
 
\subsection{Partially Observable Markov Decision Processes}
\label{sec:back:pomdp}

Partially Observable Markov Decision Processes (POMDP)  extent the MDP, and is a mathematical framework for single-agent decision-making in stochastic partially observable environments \cite{aastrom1965optimal}, which is a generalization over Markov Decision Processes \cite{howard1960dynamic}. A POMDP is a seven-tuple ($\StatePerPolicy, \ObservationPerPolicy, \ActionPerPolicy, \Transitions, \ObservationTransitions, \RewardPerPolicy, \gamma$). In POMDPs, the exact states ($s \in \StatePerPolicy$) of the evolving environment may or may not be captured fully. Therefore, observations ($\observation \in \ObservationPerPolicy$) represent the observable states, which may differ from the exact state. Similar to $\Transitions$, $\ObservationTransitions: \StatePerPolicy \times \ActionPerPolicy \times \ObservationPerPolicy \rightarrow [0,1]$ is an observation probability function, where $\ObservationTransitions (\observation, \action, \state')$ is the probability of observing $\observation$ while transitioning to $\state'$ after taking action $\action$. 

The Q-value function for a POMDP is defined over belief states and actions, rather than exact states and actions. Let $b$ denote a belief state, which is a probability distribution over states, so that $\belief'(\state')$ is the updated belief of being in state $\state'$. To compute the belief state in a Partially Observable Markov Decision Process (POMDP), we use a recursive state estimation process based on the previous belief state, the action taken, and the current observation received. Then, the belief state update equation is: 
\begin{equation}
    \belief'(\state')=\eta \ObservationTransitions(\observation | \state',\action)\sum_{\state \in \StatePerPolicy}\Transitions(\state'| \state, \action) \belief(\state)
\end{equation}
\noindent where $\eta = 1/P(\observation|\belief,\action)$ is a normalizing constant with:
\begin{equation}
    P(o|b,a) = \sum_{\state' \in \StatePerPolicy} \ObservationTransitions(\observation, \state',\action) \sum_{\state \in \StatePerPolicy} \Transitions(\state'|\state,\action) \belief(\state).
\end{equation}

This process is repeated after each action and observation to maintain a current belief state that summarizes all information received so far. The belief state forms a sufficient statistic for the entire action-observation history and allows the POMDP to be cast as a continuous-state MDP, called a belief MDP

The value function for a POMDP policy $\policy$ over belief states is defined as: \begin{equation}
V^\policy(b) = \mathbb{E}_\policy\left[\sum_{t=0}^\infty \gamma^t R(\state_\timet,\action_\timet) \mid \belief_0 = \belief\right]
\end{equation} 

Then the Q-value function for a POMDP policy $\policy$ (casted as belief MDP) is defined as:
\begin{equation}
Q^{\policy}(\belief,\action) = \sum_{\state \in \StatePerPolicy} \belief(\state) \left[\RewardPerPolicy(\state,\action) + \gamma \sum_{\observation \in \ObservationPerPolicy} \ObservationTransitions(\observation, \state,\action) V^{\policy}(\belief^\action_\observation) \right],
\end{equation}
\noindent where $\belief^\action_\observation$ is the updated belief state after taking action $\action$ and observing $\observation$ and $V^{\policy}(\belief^\action_\observation)$ is the value of the updated belief state $\belief^\action_\observation$ under policy $\policy$. The optimal policy can then be computed as $\policy^* = \arg \max_{\action} Q^{\policy}(\belief,\action)$

\subsection{Stochastic Games}
MDPs or POMDPs assume a single policy. Stochastic games generalize MDPs for multiple policies \cite{shapley1953stochastic}. When players do not have perfect information about the environment, stochastic games become partially observable stochastic games. A partially observable stochastic game is defined as an eight-tuple $\left(N, \SetOfStates, \SetOfObservations, \SetOfActions, T, \SetOfObservationTransitions, \SetOfRewards, \gamma \right )$, where $N$ is the number of policies. $\SetOfStates = \StatePerPolicy_1 \times ... \times \StatePerPolicy_N$ is a finite set of state sets, and $\SetOfObservations = \ObservationPerPolicy_1 \times ... \times \ObservationPerPolicy_N$ is a finite set of observation sets, with subscripts indicating different policies. $\SetOfActions = \ActionPerPolicy_1 \times ... \times \ActionPerPolicy_N$ is a finite set of action sets. $\SetOfObservationTransitions = \ObservationTransitions_1 \times ... \times \ObservationTransitions_N$ defines a set of observation probability functions of different players. A set of reward functions is defined as $\SetOfRewards = \RewardPerPolicy_1, ... \RewardPerPolicy_N $. Furthermore, we define a set of policies as $\SetOfPolicies = {\policy_1, ... \policy_N}$.

All policies have their individual actions, states, observations, and rewards. In this paper, we optimize each policy individually, while the observations are influenced by each other's actions, hence we can treat this as multiple distinct POMDPs as outlined in \secref{sec:back:pomdp}.

Interactive POMDPs (I-POMDPs) are an alternative framework for multi-agent decision making \cite{gmytrasiewicz2004interactive}. A core different with partially observable stochastic games is the I-POMDPs explicitly model belief states, and take actions base on them. Whereas stochastic games involve less cognitive modeling and have an implicit representation. 

\subsection{Reinforcement Learning}
\label{sec:back:rl}
Reinforcement learning (RL) solves MDPs by learning a state-action value function $Q(s,a)$ (or $Q(b,a)$) that approximates the Q value defined by the Bellman equation $Q^\policy(s,a)$. RL algorithms fall into two categories: model-based and model-free. In model-based algorithms, state transition probabilities are known, and policies are determined by enumerating possible state sequences following an initial state and action, summing the expected rewards along these sequences. This dissertation focuses on model-free RL algorithms for solving MDPs, so for an overview of model-based approaches, see \cite{moerland2023model, polydoros2017survey}. In model-free RL, the transition probability functions are unknown but can be sampled. These algorithms learn the approximate state-action value function $Q(s,a)$. In Deep RL, the policy $\policy_\theta$ is learned as a multi-layered perceptron (MLP), where $\theta$ represents the learnable parameters, or the weights of the MLP.

Trust Region Policy Optimization (TRPO) \cite{schulman2015trust} is a policy optimization algorithm designed to achieve stable and efficient training in reinforcement learning by maintaining a trust region around the current policy. The algorithm optimizes the policy by solving the following constrained optimization problem:
\begin{equation}
\label{eq:trpo}
\arg\max_{\theta} \mathbb{E}_{\state_\timet,\action_\timet \sim \policy_{\theta_{\text{old}}}} \left[ \frac{\policy_{\theta}(\action_\timet|\state_\timet)}{\policy_{\theta_{\text{old}}}(\action_\timet|\state_\timet)} A^{\policy_{\theta_{\text{old}}}}(\state_\timet,\action_\timet) \right],
\end{equation}

subject to a constraint on the Kullback-Leibler (KL) divergence between the new and old policies:
\begin{equation}
\label{eq:trpo_kl}
\mathbb{E}_{\state_\timet \sim \rho{\theta_{\text{old}}}} \left[ D_{\text{KL}} \left( \policy_{\theta_{\text{old}}}(\cdot|\state_\timet) | \policy_{\theta}(\cdot|\state_\timet) \right) \right] \leq \delta,
\end{equation}
where $\policy_{\theta}$ represents the policy parameterized by $\theta$, and $A^{\policy_{\theta_{\text{old}}}}(\state_\timet,\action_\timet)$ is the advantage function estimated under the old policy $\policy_{\theta_{\text{old}}}$:

\begin{equation}
A^\policy(\state, \action) = Q^\policy(\state, \action) - V^\policy(\state),
\end{equation}

The constraint ensures that each policy update is conservative, limiting the divergence between successive policies, thereby enhancing the stability and performance of the learning process. This approach mitigates the risk of drastic policy changes, which can lead to performance degradation or instability in the learning process.

In this dissertation we use Proximal Policy Optimization (PPO) \cite{schulman2017proximal}, which is a simplified variant of TRPO designed to retain its benefits while reducing computational complexity. PPO maximizes a clipped objective function:
\begin{equation}
\label{eq:ppo}
\arg\max_\theta \mathbb{E}_\timet \left[ \min \left( r_t(\theta) \action_\timet, \text{clip}(r_t(\theta), 1 - \epsilon, 1 + \epsilon) \action_\timet \right) \right],
\end{equation}
where $r_t(\theta) = \frac{\policy{\theta}(\action_\timet|\state_\timet)}{\policy_{\theta_{\text{old}}}(\action_\timet|\state_\timet)}$ is the probability ratio between the new and old policies, and $\action_\timet$ is the estimated advantage function. The clipping function constrains $r_t(\theta)$ within a range of $[1 - \epsilon, 1 + \epsilon]$, with $\epsilon$ being a small hyperparameter. This restriction prevents excessively large updates that could destabilize training, striking a balance between exploration and exploitation. PPO's design simplifies the implementation of policy updates by eliminating the need for complex constraints on KL divergence, thus making it a practical and widely adopted algorithm in reinforcement learning applications.

\cleardoublepage%
\part{\interfaces}
\cleardoublepage%

\def\dir{chapters/04_joint_interfaces}

\chapter{Contact-free non-planar haptics with a spherical electromagnet}
\label{ch:shared:contact}

\contribution{
We investigated a shared variable interface using a contact-free volumetric haptic feedback device. This device employs a symmetric electromagnet in combination with a dipole magnet model and a simple control law to deliver dynamically adjustable forces onto a handheld tool. The tool requires only an embedded permanent magnet, allowing it to remain completely untethered. Despite being contact-free, the force is grounded via the spherical electromagnet, enabling the user to feel relatively large forces (up to 1N at contact). The device can render both attracting and repulsive forces within a thin shell around the electromagnet. We conducted a user experiment with six participants to characterize the force delivery aspects and perceived precision of our system. Our findings indicate that users can discern at least 25 locations for repulsive forces. This research marks the first step towards understanding shared variable interfaces. By eliciting forces on the tool manipulated by the user, we enable haptic feedback in the same modality as the user's exertion, resulting in natural and intuitive interactions.
}

\section{Introduction}
A specific instance of a shared variable interface is a subset of haptic interfaces. Haptic interfaces are particularly interesting because they operate in the physical world, where the challenge of variable ownership is more impactful. Additionally, two agents exerting force on a single object intuitively exemplify shared variable interfaces. Furthermore, many emerging computing paradigms, such as virtual and augmented reality (VR/AR), rely on haptic feedback as an additional information channel to enhance the user experience. For example, in VR, haptic feedback increases the sense of presence and immersion by rendering collisions, shapes, and forces between the user and virtual objects.

For a haptic device to qualify as having a shared variable interface, it must function as both an input and output device. Many haptic devices primarily function as output devices, such as vibrotactile actuators embedded in handheld controllers \cite{whitmire2018haptic} or worn on the body \cite{hinchet2018dextres}. Vibrotactile feedback offers one-way communication and can only render coarse, non-localized haptic sensations. In contrast, our device allows for bi-directional communication on a shared variable. While complex setups such as articulated arms and exoskeletons can render large-force haptic feedback in three-dimensional space, they typically require force anchoring in the environment and involve cumbersome, bulky mechanisms, which hinder user uptake in walk-up-and-use scenarios.

To address the limitations of vibrotactile and large complex devices, we propose a contact-free, volumetric haptic feedback approach via an omnidirectional electromagnet. The device consists of a single 60 mm diameter spherical electromagnet capable of rendering attractive and repulsive forces onto permanent magnets embedded in pointing tools, such as a stylus or magnets worn on the user's fingertip. Leveraging a dipole-dipole approximation of the electromagnet-magnet interaction, our system can calculate and control the forces exerted on the permanent magnet in real-time, dynamically adjusting the force perceived by the user. The system can deliver perceptible forces up to 1N in a thin volume above the surface. Furthermore, we demonstrate that users can distinguish at least 25 different set-points separated by 18° on the surface of the sphere.

\begin{figure}[t]
\centering
\includegraphics[width=0.7\columnwidth]{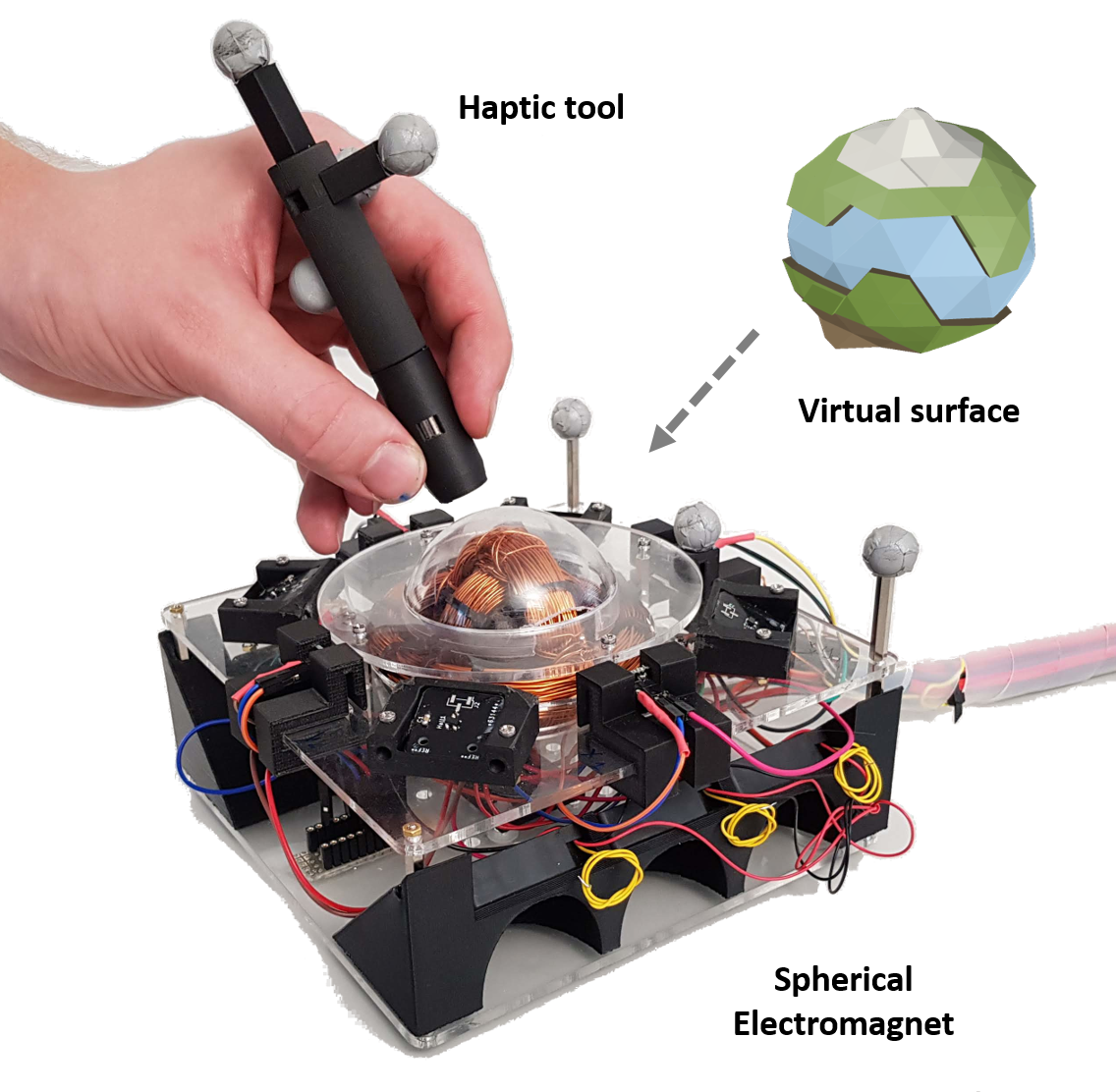}
\caption{We introduce a novel contact-free mechanism to render haptic feedback onto a tracked stylus via a hemispherical electromagnet. An approximate model of the magnet interaction and a computationally efficient control strategy allow for the dynamic rendering of attracting and repulsive forces, for example, allowing users to explore virtual surfaces in a thin shell surrounding the device (inset).}
\label{fig:syst_overview}
\end{figure}

To demonstrate the efficacy of our approach, we designed a functional prototype comprising an iron core and three custom-wound copper coils. The electromagnet is encased in a plastic dome upon which tools can come into contact and move about its surface (see \figref{fig:syst_overview}). The prototypical system can render radial (along the vector from the magnet to the tool) and tangential forces, both in attractive and repulsive polarities. The system can dynamically adjust the opening angle and steepness of the electromagnetic potential to gently guide the user towards a desired set-point in the thin volume above the device.

Modulating the magnetic field based on tool position opens the door to various interactive applications. For example, in virtual terrain exploration, the tool can be repelled when moved along mountains and attracted to valleys while descending (see \figref{fig:syst_overview}, inset). Another example is creating the sensation of stirring a viscous liquid by emulating the fluid's drag on the tool. To enable these interactive experiences, our device builds on three key components representing our contributions in this chapter:
\begin{itemize}
\item A computational model based on magnetic dipole-dipole interaction to produce force maps that allow for designing and generating location-dependent feedback,
\item The design and implementation of a 3 degree-of-freedom (DoF) spherical electromagnet prototype,
\item A control strategy that translates desired high-level forces into low-level input signals (currents/voltages) for the coils, fast enough for interactive use.
\end{itemize}
To assess the efficacy of the proposed design, we experimentally characterized the system properties and conducted a perceptual study exploring the thresholds for perception and localization capabilities of the electromagnetic actuation approach. Results from these early user tests indicate that users can perceive at least 25 different spatial locations with high precision.

\section{System Description}

We introduce a haptic feedback system that enables dynamic interactions with virtual surfaces through an untethered, contact-free tool.
Our device is a hemispherical shell. The core consists of three coils with mutually orthogonal axes. By controlling the current flow through the coils, we can shape the magnetic field around the device. This, in turn, enables the device to exert controlled electromagnetic forces on the permanent magnet located inside a handheld tool such as a stylus. Despite being contact-free, the forces perceived by the user are ultimately grounded to the support onto which the device is mounted, allowing for comparatively strong feedback.
We now detail the main components that make up our contribution: 1) a computational model of the electromagnet-magnet interactions; 2) the prototypical hardware design; and 3) a real-time control algorithm.

\subsection{Haptic force mapping}
\begin{figure}[t]
\centering
\medskip
\includegraphics[width=\columnwidth]{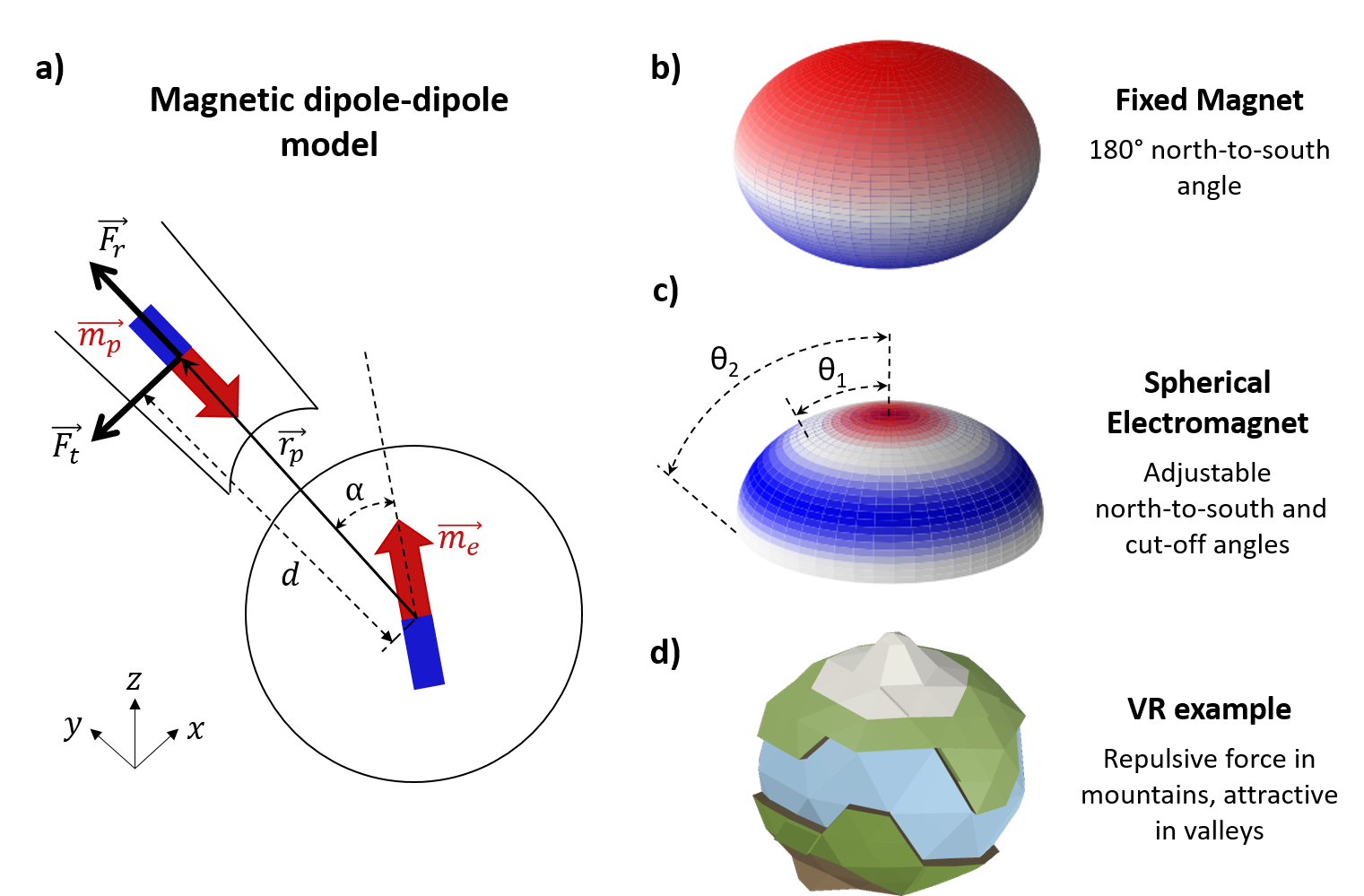}
\caption{Schematic of the main quantities necessary to compute desired radial and tangential forces (a). Insets show: force map of a permanent magnet (b). Adjustable force map generated by our approach (c). Here $\rsBold = d_{min} \ez$, $\theta_1 = \pi/10$ and $\theta_2 = 3\pi/10$. Example virtual surface that can be felt by the user (d).}
\label{fig:magnetic_model}
\end{figure}{}

To enable the envisioned interactive experiences, we must be able to dynamically adjust the haptic feedback. We therefore require a model for the magnetic interaction between device and tool that is \emph{1}) precise enough to predict forces with sufficient accuracy and \emph{2}) fast enough to run at the feedback rates required for haptic interaction.

Computing the magnetic field around, and resulting interaction between, arbitrarily-shaped objects is a challenging and computationally expensive task. However, even though the magnetic field can be very complex in the direct vicinity of an object, this complexity rapidly decays with increasing distance and approaches a simple dipole field. This fact has been exploited in previous work to construct fast, approximate models based on dipole-dipole interaction \cite{thomaszewski2008magnets}. Instead of solving the Maxwell equations on a discretization of ambient space, this approximate model only requires the magnitude and orientation of the magnetic moment of each dipole, leading to drastically reduced computation times.

In adopting this approach, we model both the electromagnet of the device and the tool as a single dipole (see Figure \ref{fig:magnetic_model}.a). Let $\mpBold, \meBold \in \mathbb{R}^3$ denote the magnetic moments of the permanent magnet in the tool and the electromagnet in the device, respectively. The forces exerted on the tool, expressed in local coordinates, are obtained as:
\begin{eqnarray}
    \FrBold &=& - \frac{3 \mu_0 \ \me \ \mpp}{2 \pi \ d^4} \cos(\alpha) \ \er \label{eq:Fr}\ ,\\ 
    \FtBold &=& - \frac{3 \mu_0 \ \me \ \mpp}{4 \pi \ d^4} \sin(\alpha) \ \et \label{eq:Ft}\ ,
\end{eqnarray}
\noindent where $\me=|\meBold|$, $\mpp=|\mpBold|$. In the above expression, $\FrBold$ is the force in the radial direction $\RpenBold = d \ \er$ from the center of the device to the tool. Likewise, $\FtBold$ is the force in the tangential direction $\et$ that tends to align the location of the two dipoles along $\er$. Assuming that the tool is in contact with the shell, both force components depend only on the relative angle $\alpha$ between the dipoles. Furthermore, $\FrBold$ and $\FtBold$ are attractive (negative) when the two dipoles have the same sign and $\alpha<\pi/2$. Conversely, the forces become repulsive (positive) when the dipoles have opposite orientations (see Figure \ref{fig:magnetic_model}.a).

The interaction forces decay quickly, as $1/d^4$, with increasing magnet-magnet distance. The maximum force $\mathbf{F}\mathrm{r,max}$ is obtained when the tool is in contact with the device ($d = d\mathrm{min}$). In our case, $d_{min} = 50$ mm, since the outer case radius is 30 mm and inside the tool, the magnet center is 20 mm away from the tool tip. Our proposed geometry ensures that the distance $d$ will remain constant across the working surface as long as the tool is kept in contact with the surface, allowing for much simpler control of the force. However, it is worth noting that moving the tool 1 cm away in the radial direction makes the force fall to approximately $\mathbf{F}\mathrm{r,max}/2$, another extra centimeter results in a force $\mathbf{F}\mathrm{r,max}/4$. This rapid decay of the interaction forces can, to some extent, be mitigated by increasing the intensity of the magnetic field. However, to maintain power consumption and thermal effects within reasonable bounds, we constrain our interactions to a volumetric shell ($d_{min} \leq d \lesssim d_{min} + 2cm$) above the device's surface.

Equations \ref{eq:Fr} and \ref{eq:Ft} also reveal the comparatively weak variation of force magnitude with respect to angle that one would expect when two magnets interact: switching from attractive to repulsive forces requires a change in orientation of $\alpha = \pi$; see Figure \ref{fig:magnetic_model}.b. This weak force variation is inherent to permanent magnets: whereas the far-field interaction is dominated by torque (which decays only as $1/d^3$), the near-field force interaction is governed by the location of the dipoles, not their orientation. In our setting, this property would translate into weak angular resolution with a permanent magnet. To address this problem, we introduce the concept of a \emph{force map} that uses magnetic pole transformation to take advantage of the spherical symmetry and that is compliant with the physics of the system. Our system can generate force maps equivalent to multiple alternating pole regions, having sharper repulsive domes and attractive valleys. The force map is defined by four parameters:
\begin{itemize}
\item The center $\rsBold$ of the potential. When rendering a mountain-like dome, for instance, $\rsBold$ is the summit.
\item The height of the dome is measured as the maximum magnetic moment intensity $m_{e0}$.
\item The angle $(\theta_1)$ (i.e., the location of the tool in polar coordinates with respect to $\rsBold$) where the radial force vanishes for the first time. In our example, $(\theta_1)$ is the angle from the summit to the base. 
\item The cut-off angle $\theta_2$ after which the potential is set to be zero. Having such a cut-off mechanism allows us to control how many individual potentials can be combined into one force map without mutual interference.
\end{itemize}

\begin{figure}[!t]
    \bigskip
    \textbf{Algorithm to calculate desired forces}\\
    \rule[5.0pt]{\columnwidth}{0.75pt}
    \vspace{-15pt}
    \begin{algorithmic}
    \STATE \% \emph{To compute} $\meBold$ \emph{given the tool position and the force map.}
    \REQUIRE $calc\_Me \ (\RpenBold, \rsBold, m_{e0}, \theta_1, \theta_2)$:
    \STATE $\RpenBold \rvert_{\rsBold} = \mathbb{T}_{\mathbf{r} \to \rsBold} \cdot  \RpenBold $
    \STATE $\mathbf{F} \rvert_{\rsBold} = calc\_F \ (\RpenBold\rvert_{\rsBold}, \rsBold, m_{e0}, \theta_1, \theta_2)$
    \STATE $\mathbf{F} = (\mathbb{T}_{\mathbf{r} \to \rsBold})^{-1} \cdot  \mathbf{F} \rvert_{\rsBold} $
    \STATE $\mathbf{F} \rvert_{\RpenBold} = \mathbb{T}_{\mathbf{r} \to \RpenBold} \cdot \mathbf{F}$
    \STATE $\meBold \rvert_{\RpenBold} = \frac{4 \pi d^4}{3 \mu_0} \ [1,1,-1/2] \cdot \mathbf{F} \rvert_{\RpenBold}$
    \STATE $\meBold = (\mathbb{T}_{\mathbf{r} \to \RpenBold})^{-1} \cdot \meBold \rvert_{\RpenBold}$
    \RETURN $\meBold$
    \end{algorithmic}
    \vspace{5pt}
    \begin{algorithmic}
    \STATE \% \emph{To compute the actuation force  in the} $\rvert_{\rsBold}$ \emph{coordinates.}
    \REQUIRE $calc\_F \ (\RpenBold\rvert_{\rsBold}, \rsBold, m_{e0}, \theta_1, \theta_2)$:
    \STATE $F_r = 0$
    \STATE $F_t = 0$
    \IF{$d < d_{max}$ \AND $\alpha < \theta_2$}
    \STATE $F_r = 2 F_0 \ cos(\alpha\frac{2\theta_1}{\pi}) \ \left(\frac{||\rsBold||}{||\RpenBold||}\right)^4$
    \STATE $F_t = F_0 \ sin(\alpha\frac{2\theta_1}{\pi}) \ \left(\frac{||\rsBold||}{||\RpenBold||}\right)^4$
    \ENDIF 
    \STATE $\mathbf{F} \rvert_{\rsBold} = [F_r, F_t, 0]$
    \RETURN $\mathbf{F} \rvert_{\rsBold}$
    \end{algorithmic}
    \rule[0.0pt]{\columnwidth}{0.75pt}
    \caption{Pseudo-code of our force calculation algorithm. Note that $\mathcal{T}_{r_i \to r_j}$ is the rotation matrix that maps from coordinate system $r_i$ to $r_j$, and that $\mathbb{T}_{r_j \to r_i} = (\mathbb{T}_{r_i \to r_j})^{-1} = (\mathbb{T}_{r_i \to r_j})^{T}$.}
    \label{fig:algorithm}
\end{figure}

Figure \ref{fig:algorithm} summarizes our algorithm to calculate the actuation vector $\meBold$ given the tool position and force map as input.
For simplicity and efficiency, we perform the different calculations in their natural coordinate system: the Cartesian system $\mathbf{r} = [x,y,z]$, the spherical system relative to the map's center $\rsBold$, and the spherical system centered around the tool position $\RpenBold$.

The force calculation incorporates the angular scaling by using $(\alpha\frac{2\theta_1}{\pi})$ as an argument for the trigonometric functions in Equations \ref{eq:Fr} and \ref{eq:Ft}. Note that if $\theta_1 = \pi/2$, we recover a permanent magnet. In Figure \ref{fig:magnetic_model}.c, we show an example where the center of the potential (\emph{red}) is on the north pole of the sphere, the first vanishing region (\emph{white}) appears at 18$^\circ$ and the forces are cut off at 54$^\circ$ (\emph{blue}).

Using the algorithm described in Figure \ref{fig:algorithm}, we obtain at each time step an actuation input $\meBold = (m_{e-x}, m_{e-y}, m_{e-z})^T$ given the tool position. Depending on the requirements of the application, the potential parameters (center position, intensity, angular variation, and cut-off) may also change as a function of tool position. For example, the force map for the terrain example can be dynamically adapted to emulate changes in landscape over time.

\subsection{Spherical electromagnetic actuator}
\begin{figure}[!t]
\centering
\medskip
\includegraphics[width=0.7\columnwidth]{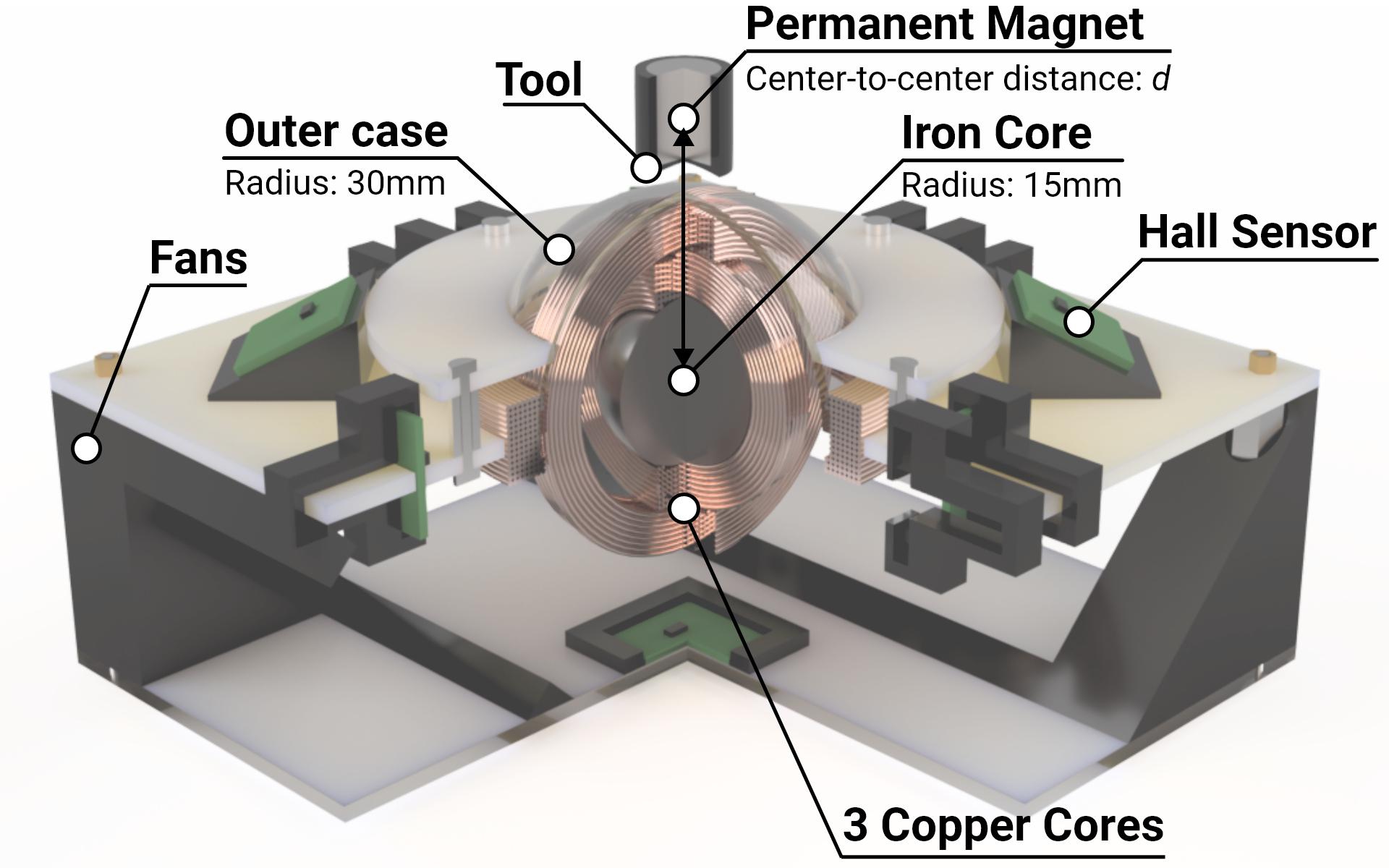}
\caption{3D cross-section of the proposed hardware setup. The device measures $15 \times 15$ cm across the base. Three coils are placed, orthogonal to each other and surrounding the iron core. Forces can be rendered onto a permanent magnet moving above the device. Hall sensors are used for calibration. A plastic cover isolates the coils from the user thermally and electrically. Active cooling is provided via several fans mounted in the base.}
\label{fig:hardware_labels}
\end{figure}

Having laid out the computational model for generating haptic feedback based on dipole interactions, we now describe hardware and implementation aspects for rendering these forces on our device (Fig. \ref{fig:syst_overview}).
Our device renders haptic forces by controlling the magnetic field generated by a spherical electromagnet. Compared to other alternatives, this approach has several advantages. First, there are no mechanically moving parts in the actuator, reducing complexity and eliminating wear. Changing the orientation of the resulting force on the tool is accomplished by adapting the currents in each coil such as to rotate the induced dipole in the core as desired; see also Figure \ref{fig:hardware_labels}. The underlying physical principle is that, in the presence of linear and isotropic materials, the magnetic field $\mathbf{B(\mathbf{r})}$ at any given point $\mathbf{r}$ can be calculated as the sum over all contributions of all magnetic sources \cite{petruska2014omnimagnet}. Under this linearity property of $\mathbf{B}$, the magnetic field produced by the three orthogonal coils is the superposition of the fields generated by each coil individually. Finally, we insert a magnetic core with isotropic (i.e., spherical) geometry and material at the center of the coils and operate it in the linear regime (i.e, $\me << m_{saturation}$), linearity is maintained such that $\mathbf{B(\mathbf{r})}$ can be computed by summing up each coil's contributions.
{}

In order for the previous statement to remain valid, two assumptions have to be made. First, hysteresis effects can be neglected: the lower the coercivity and remanence of the core material, the lower the effect of past states of the electromagnet on the current one. The second assumption is that the distance $d$ between dipoles is large enough such that the core magnetization due to $\mpp$ is small compared with the effect of the coils. This will not be true if, for example, the tool snaps to the sphere with no electrical current in the coils. In our setting, however, a spherical cap around the coils prevents too close approach of the tool and, at the same time, provides the grounding required for generating sufficiently strong interaction forces.

For the standard low-carbon steel core, we did not observe any hysteresis effects for the update of $\meBold$ at 50 Hz refresh rate. To avoid the undesired self-magnetization of the core due to the tool, we tuned the size of the permanent magnet and the coil parameters using FEM simulations, followed by minor design adjustments informed by real-world tests.

The design choices for the hardware of our prototype are motivated by our goal to develop a device that is affordable and easy to manufacture. In particular, we use off-the-shelf electronic components but custom-wound coils. FEM simulations in Comsol Multiphysics are used to assist in the exploration of the design space. In Figure \ref{fig:hardware_labels} we show a 3D CAD rendering of our device. The external dimensions are 150 mm by 150 mm by 95 mm. The structure is built out of laser-cut acrylic glass and 3D-printed parts. The three orthogonal coils are arranged around the 30 mm steel core. All coils have a resistance of roughly 0.6 
$\Omega$ at room temperature. We use the 12V line of a standard CPU power supply to drive the coils, meaning a maximum electrical current of 24A per coil at full strength. The electrical current in each coil is controlled by a high-power motor driver (Pololu 18v17). The PWM signals are generated by a 12-bit driver (PCA9685) that allows for easy tuning of the carrier frequency and the duty cycle with 12-bit resolution. To be able to accurately control the electrical current and compensate for thermal drifts, we use INA219 current sensors in each coil with a 0.01 $\Omega$ shunt resistor. Finally, an Arduino board creates the bridge between the I2C components and the PC. The hardware is completed by 9 Hall sensors arranged collinearly with the axes and diagonals of the coils. Six fan coolers below the coils provide active cooling.

\subsection{Control Strategy}
\begin{figure}[!t]
\centering
\medskip
\includegraphics[width=0.95\columnwidth]{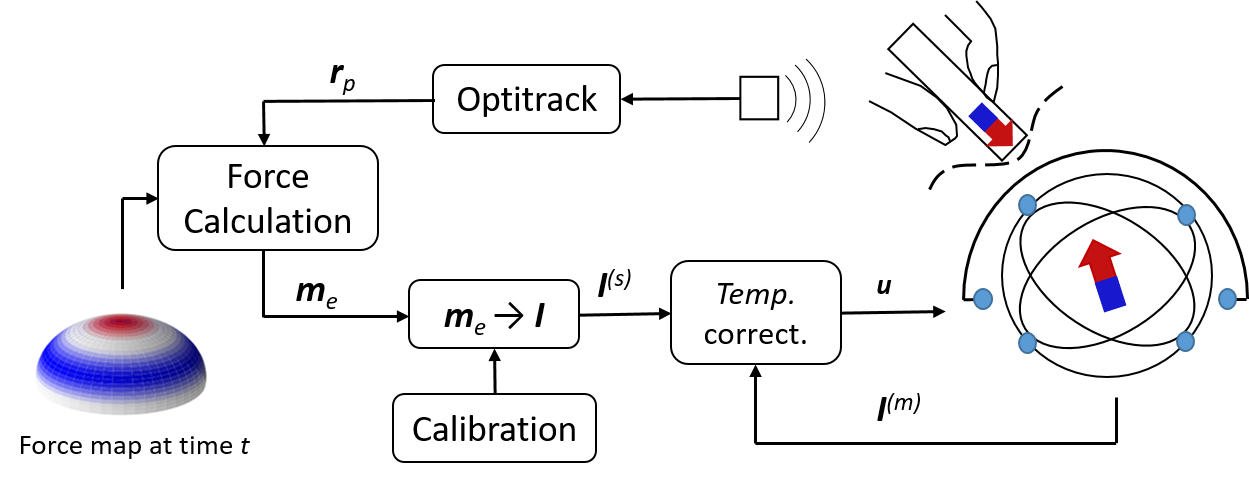}
\caption{Schematic overview of the software pipeline. Given the desired force map at time $t$, and the tool position provided by an external tracking system, we calculate the input value $\meBold$ using the algorithm of Fig. \ref{fig:algorithm}. Then the system inputs are computed using Eq. \ref{eq:control-law}, and finally a temperature compensation step corrects the system inputs.}
\label{fig:control}
\end{figure}

The main objective of the actuator control loop is to generate a stable and controllable force on the haptic tool. Although the mathematical principles are straightforward, the practical implementation poses some problems. Since the magnetic field is directly proportional to the current (Fig. \ref{fig:current_m}), controlling the latter is sufficient to determine the state of the system. If the resistance is known, controlling the voltage is equivalent to controlling the current via Ohm's law,
\begin{equation}\label{eq:ohms}
I = V/R \ ,
\end{equation}
\noindent and the voltage in turn can be controlled via Pulse-Width Modulation (PWM). Therefore, the input to our system is the PWM frequency. The complete control loop is shown in Figure \ref{fig:control}.
However, significant heating occurs due to the necessary power that in turn increases the resistance. Therefore, the PWM duty cycle (i.e., voltage) needs to be adjusted to maintain a constant current. Measuring the current allows determination of the resistance via inversion of Eq \ref{eq:ohms}. A simple controller then computes an input $\mathbf{u} \in [-1,1]$ at time $t$, corresponding to the PWM duty cycle. This depends on the desired current in Ampere ($\mathbf{I}^{(s)}t$), the resistance in Ohm ($\mathbf{R}t$), and the maximum voltage in the system, $V_0=12$:
\begin{equation}\label{eq:control-law}
\mathbf{u}{t} = \frac{\mathbf{I}^{(s)}{t}}{V_0}*\mathbf{R}t \quad ,
\end{equation}
\noindent where $I_t^{(s)}$ is based on the desired magnetization, $\mathbf{m}\text{e}$, computed via the algorithm presented in Fig. \ref{fig:algorithm} and can be determined via Biot-Savart Law (adapted for our purpose):
\begin{equation}\label{eq:I_vs_me}
I_t^{(s)} = \mathbf{c} * \frac{\mathbf{m}_e * \mu_0}{2*\pi*\mathbf{d}^3} \quad ,
\end{equation}
\noindent here $\mathbf{c}$ is a constant coming from a calibration procedure that, with the help of five Hall sensors, maps input current to $\mathbf{m}_e$ (Fig. \ref{fig:current_m}). $\mu_0=4*\pi*10^{-7}$ is the relative permeability of air and $d$ is the distance from the core to the Hall sensors used for calibration (0.055 meter).
Due to the thermal effects, $\mathbf{R_t}$ however is not a constant but depends on the measured current ($\mathbf{I}t^{(m)}$) computed and averaged over a sliding window:
\begin{equation}
\mathbf{R}t = \frac{V{0} * \frac{1}{N}\sum^N \mathbf{u}{t-i}}{\frac{1}{N}\sum^N\mathbf{I}^{(m)}_{t-i} } \quad .
\end{equation}


\section{System Evaluation}

\begin{figure}[!t]
    \centering
    \medskip
    \includegraphics[width=0.8\columnwidth]{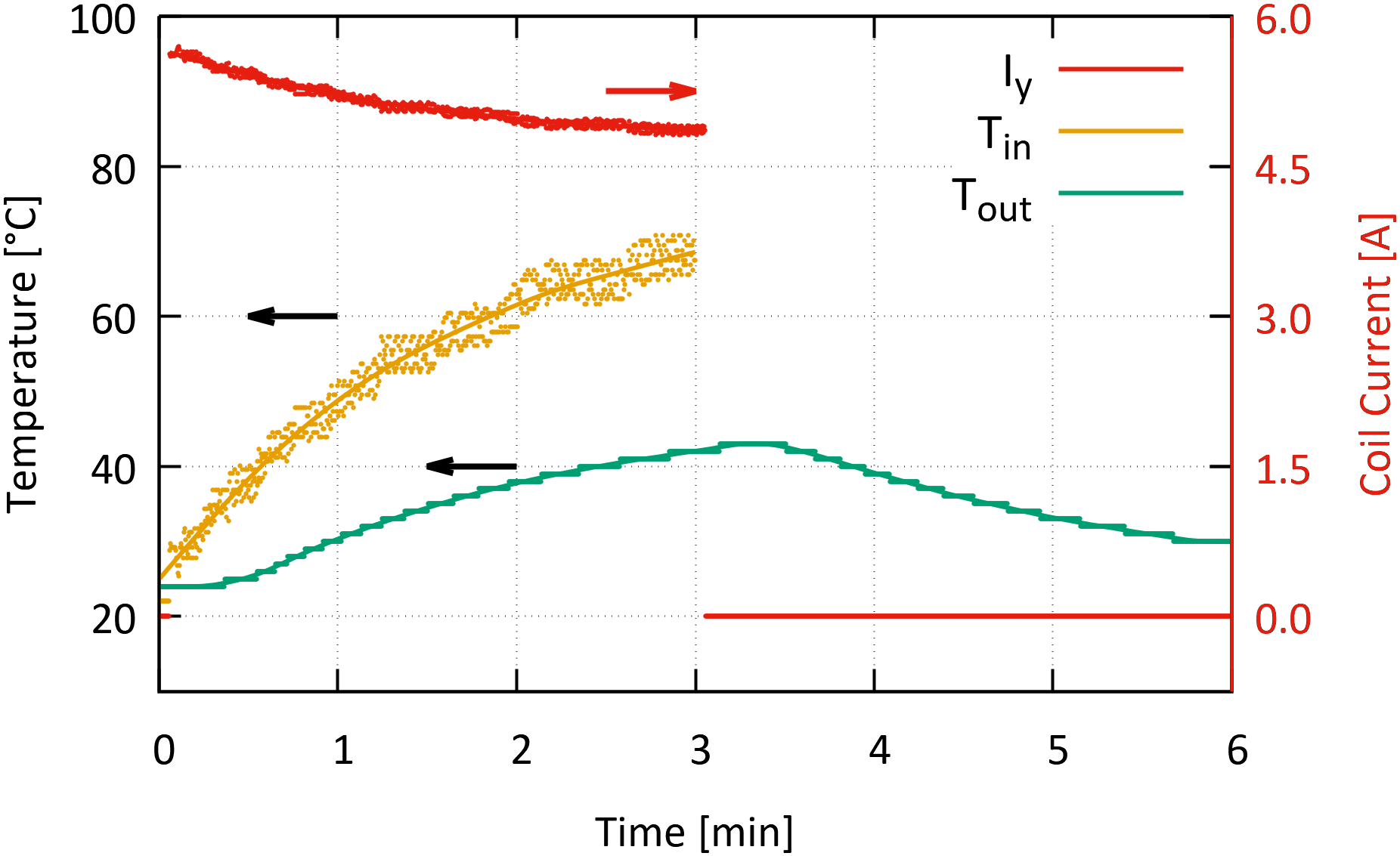}
    \caption{Thermal characterization of one of the coils as function of time. During the first 3 minutes the $y$-coil is driven with PWM=30\%, and then we let it cool over the remaining 3 minutes. $T_{in}$ is calculated by taking the thermally caused resistance variations into account while the current $I_y$ is `on', and $T_{out}$ is measured.}
    \label{fig:temp}
\end{figure}{}

One of the main physical limitations of EM-based systems is thermal effects due to Joule heating, to obtain large forces \cite{esmailie2017thermal}. The temperature is directly proportional to the actuation power ($P$) and the thermal dissipation obtained by the active and/or passive cooling. We evaluated the thermal behavior of our system for different power values. In this experiment, we set the current to on' for three minutes and then let the device cool down. Figure \ref{fig:temp} shows data from the middle coil actuated at PWM = 30\%. $T_{out}$ is the temperature measured at the coil boundary, measured with a \emph{Dallas DS18B20} sensor. $T_{in}$ is the average temperature of the copper wire obtained via the variation in resistance. We also plot the electrical current $I_y$ that drops as the coil heats up and the resistance increases. Note that no temperature compensation was used for building these thermal calibration curves. Each coil is able to accumulate some heat during the actuation and continuously dissipates it by the forced air circulation. Our system has a thermal time ($\tau_T$) on the order of minutes, in which it reaches the asymptotic temperature. The average power in the past $\tau_T$ seconds must be maintained within a safe value $P_{ave}$. Based on this plot, we choose $P_{ave} = 17 W$ per coil for our system. However, each coil can absorb peaks up to $15*P_{ave}$ for a few seconds.

\begin{figure}[!t]
    \centering
    \medskip
    \includegraphics[width=0.675\columnwidth]{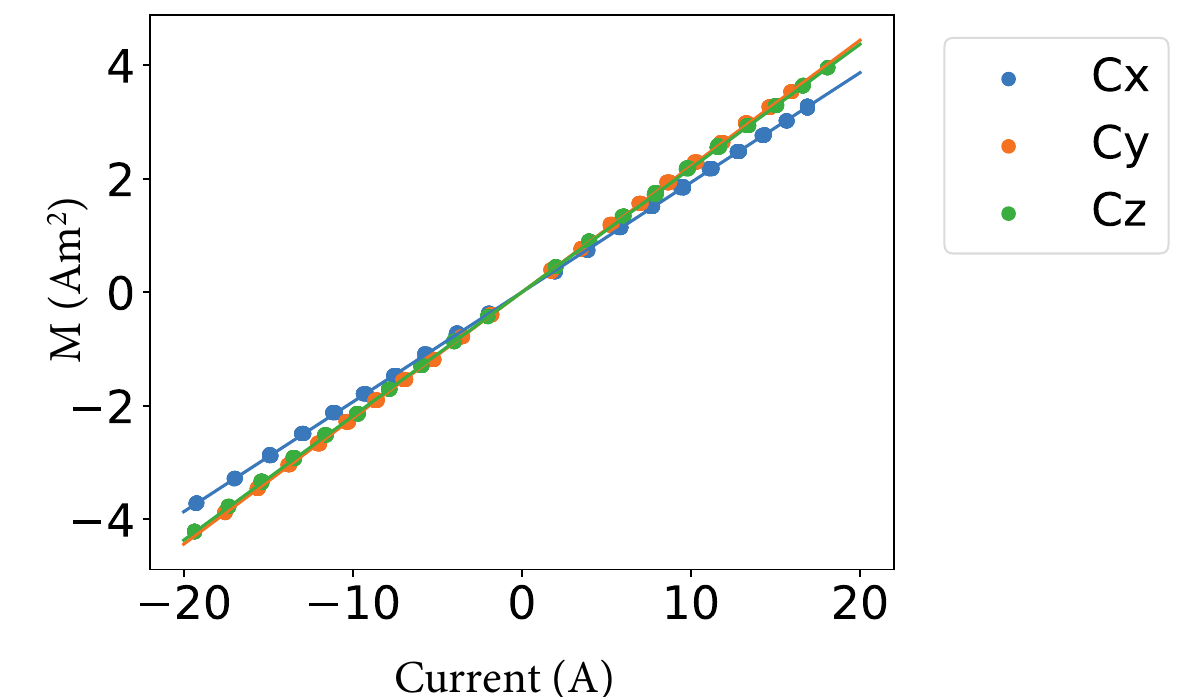}
    \caption{Electromagnet induced magnetization in each axis, $\meBold = (m_{e-x}, m_{e-y}, m_{e-z})$, as a function of the applied current settings ($I_x$, $I_y$, $I_z$). The magnetic field values are measured with hall sensors placed co-linear with each coil, and then transformed into $M$ values.}
    \label{fig:current_m}
\end{figure}

Within this safe range, we calibrate the values of $\me$ for each axis as a function of the current in each coil with the hall sensors around the sphere (see Figure \ref{fig:hardware_labels}) and with Eq. \ref{eq:I_vs_me}. Figure \ref{fig:current_m} shows the experimentally attained magnetization in the core $\meBold$ as a function of the current. For reference, applying a power $P_0 = 100$ W to each coil ($I_i = 12.9$ A), the equivalent dipole is $\meBold = [2.52; 2.7; 2.82]$ Am$^2$. We also obtain non-zero terms away from the diagonal since the coils are not perfectly orthogonal, and we use the calibration data to correct the PWM duty cycles.

Finally, values for the force acting on the permanent magnet can be attained via setting the magnetic dipole of the tool and Eq. \ref{eq:Fr} and \ref{eq:Ft}. In our experiments, we use a ring-shaped neodymium magnet (12 mm outside diameter, 5 mm inside diameter, 24 mm high). For any tool with this particular magnet, with a center-to-center distance between dipoles of 5 cm, we obtain a ratio of force per electrical current of 48 mN/A. This means the device can handle an averaged constant force of $F_r=258$ mN ($P = 17$ W) with a peak force of up to $F_r = 959$ mN ($P = 230$ W) at full strength (using PWM control). This force value can be increased by increasing the volume of the tool magnet, with the trade-off of losing angular resolution and adding weight to the tool.

\section{User Evaluation}
To assess the efficacy of our proposed approach, we validate the prototype in a perceptual study with 6 participants in order to 1) determine how well users can differentiate between different set-points, and 2) how accurate and precise users are with finding a set-point.

\begin{figure}[!t]
    \centering
    \begin{tabular}{cc}
     \includegraphics[width=0.4\columnwidth]{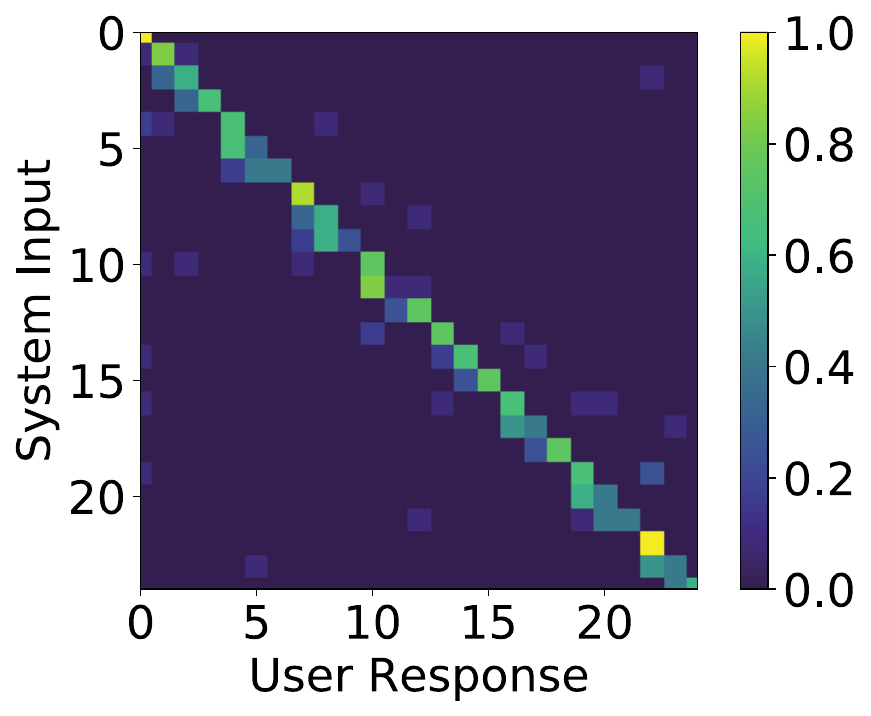} &
    \includegraphics[width=0.3\columnwidth]{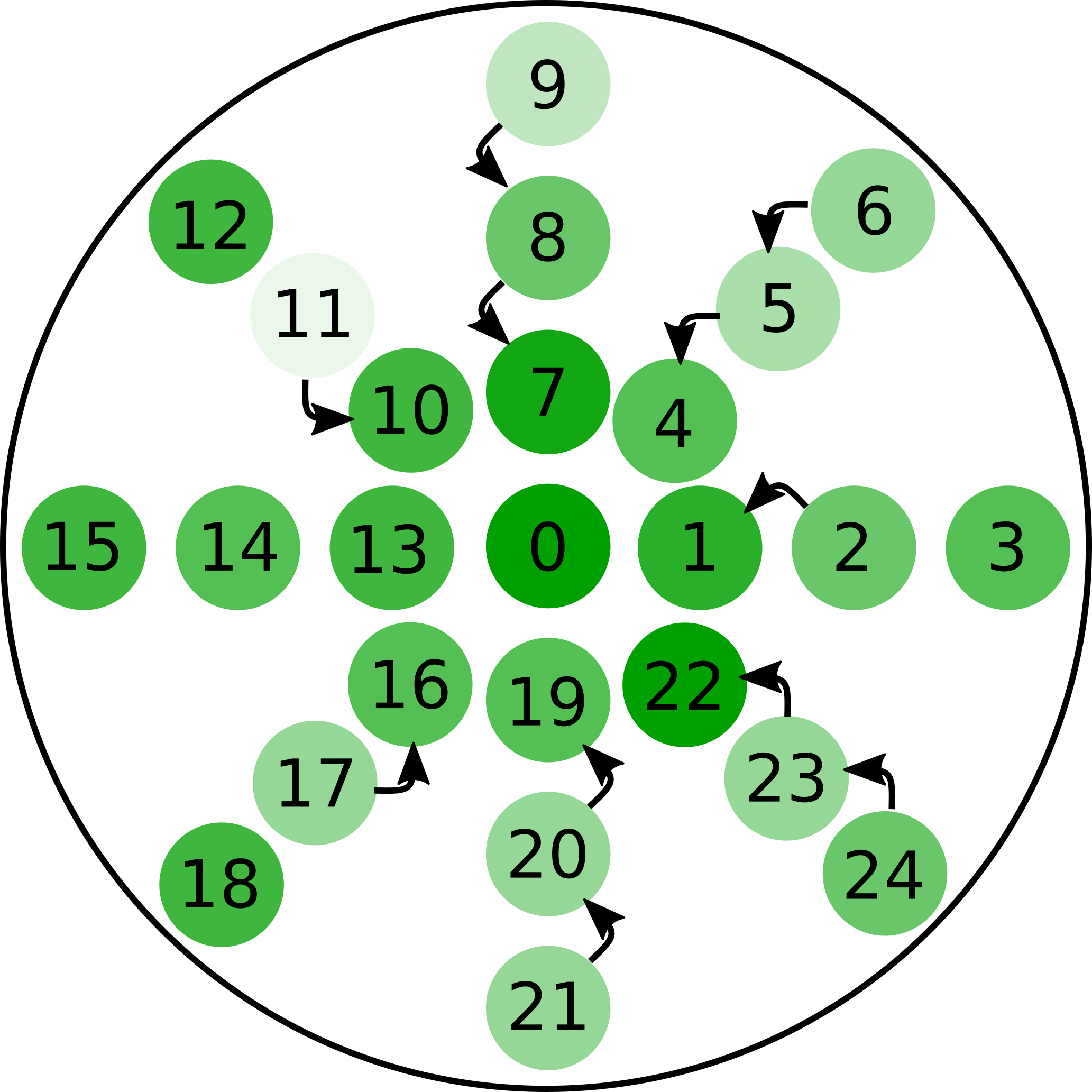}
    \end{tabular}
    \caption{\emph{Left:} confusion matrix of the 25 set-points, averaged over all users. High values on the diagonal indicate little confusion and the ability to differentiate between different set-points. \emph{Right:} Set-points used in the study. The opacity directly correlates with the percentage of correct identifications by the users. Arrows are drawn when 33\% or more of the \emph{wrong} answers were attributed to set-point that the arrow points to.}
    \label{fig:confusion_circle_error}
\end{figure}

\noindent\textbf{Procedure}: Based on a pilot study, we predetermined 25 evenly separated set-points (Figure \ref{fig:confusion_circle_error} right). We randomly selected a set-point, asked the user to find it, and report the corresponding number. Upon reporting, we also measured the Euclidean distance to the actual set-point. Every set-point was prompted exactly twice, resulting in 50 data points per user (300 in total). Only repulsive forces were tested. We used the same mapping parameters as in Figure \ref{fig:magnetic_model}.

\noindent\textbf{Location accuracy}: Figure \ref{fig:confusion_circle_error} depicts the resulting confusion matrix between set-points. It can be seen that users accurately perceive discrete actuation points. For those actuation points that do cause incorrect answers, users tend to pick the neighboring location (typically higher on the sphere). This effect is pronounced along the meridian arc facing away from the user, whereas the orthogonal meridian produces fewer erroneous detections. This could be due to the position of the hand and arm and differences in muscle groups that are involved in actuating the wrist versus the whole hand. The difference in coil diameters could be another contributing factor.
\begin{figure}[!t]
    \centering
    \medskip
    \includegraphics[width=0.51\columnwidth]{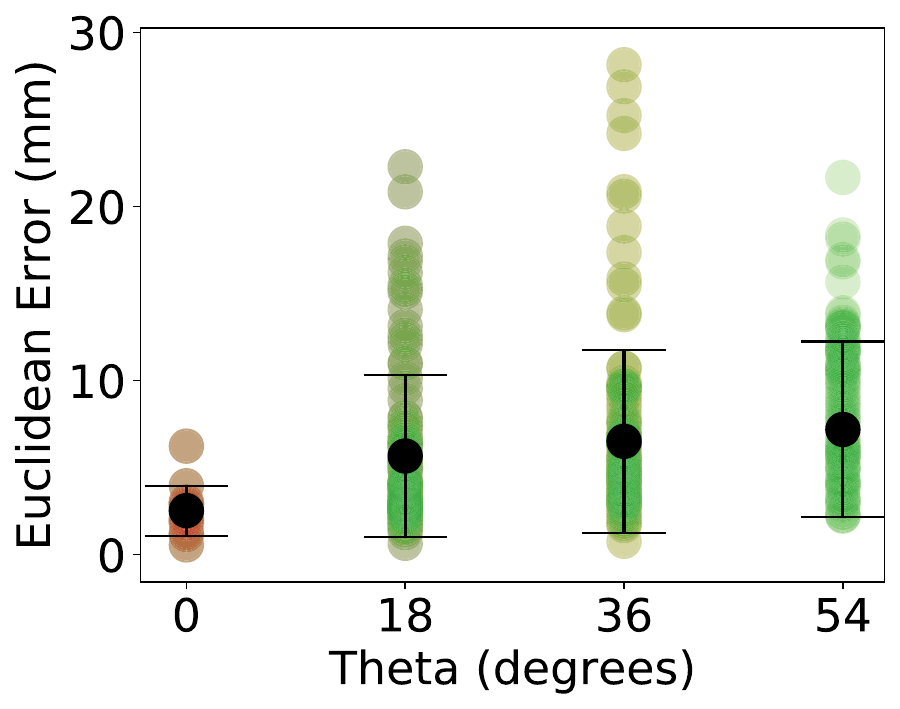}
    \caption{Euclidean distance between the true set-point position and the user reported position as a function of the azimuth ($\theta$), measured from the top of the sphere and averaged over all angles and users.}
    \label{fig:error_vs_angle}
\end{figure}

\noindent\textbf{Precision}: We report the precision with respect to the angle $\theta$. Figure \ref{fig:error_vs_angle} shows that the error increases as a function of the angle. A potential contributing factor here is that gravity has more impact on the pen the further down it moves on the hemisphere. This may make it more difficult for users to differentiate between the EM-actuation force and gravity. The mean errors of $2.5 mm \pm 1.4$, $5.7 mm \pm 4.6$, $6.5 mm \pm 5.2$, and $7.2 mm \pm 5.1$ are relatively small across the device.


\section{Discussion}
In this chapter, we presented a novel contact-free volumetric haptic feedback device. This device uses a symmetric electromagnet combined with a dipole magnet model and a simple control law to deliver dynamically adjustable forces onto a handheld tool, such as a stylus. The tool only requires an embedded permanent magnet, allowing it to be completely untethered. Despite being contact-free, the force is grounded via the electromagnet, enabling the user to feel relatively large forces.

While our proposed method offers many advantages, it also has some drawbacks. Heat generation limits the number of interactions possible within a certain time frame. Specifically, when operating at full power, continuous interaction is limited to 5 seconds.

It is also important to note that the interaction between magnets involves both forces and torques. In this work, we focused on controlling the three force components via the 3 degrees of freedom (DoFs) of the electromagnet, allowing the torque values to adapt accordingly. However, the same procedure can be applied to control a specific torque map, leaving the force values unconstrained, or to manage a combination of force and torque. In future work, we aim to explore the dynamic capabilities of our approach, including advanced control schemes to continuously shape the force map.

Finally, our current control method relies on knowing the location of the tool, achieved through external cameras for optical tracking. However, this setup is cumbersome, expensive, and requires line-of-sight. In the next chapter, we will address this limitation by tracking the permanent magnet embedded in the tool using Hall sensors and a gradient-based method.

\chapter{Volumetric sensing and actuation of passive magnetic tools for dynamic
haptic feedback}
\label{ch:shared:volumetric}
\chaptermark{Volumemetric Sensing and Actuation of Passive Magnetic Tools}

\contribution{
    In this chapter, we present \omniUIST, a self-contained 3D haptic feedback system capable of sensing and actuating an untethered, passive tool containing only a small embedded permanent magnet. \omniUIST overcomes the limitations of \omniHap in two significant ways. First, it eliminates the need for external tracking by utilizing a novel gradient-based method to reconstruct the 3D position of the permanent magnet in midair. This is achieved using measurements from eight off-the-shelf Hall sensors integrated into the base. Second, \omniUIST features an improved 3 DoF spherical electromagnet capable of delivering increased forces due to intertwined coils. The fully integrated \omniUIST system, with no moving parts and no need for external tracking, is easy and affordable to fabricate. We detail \omniUIST’s hardware implementation, our 3D reconstruction algorithm, and provide an in-depth evaluation of its tracking and actuation performance. Finally, we demonstrate its capabilities through a set of interactive usage scenarios.
}

\begin{figure}
\centering
\includegraphics[width=\textwidth]{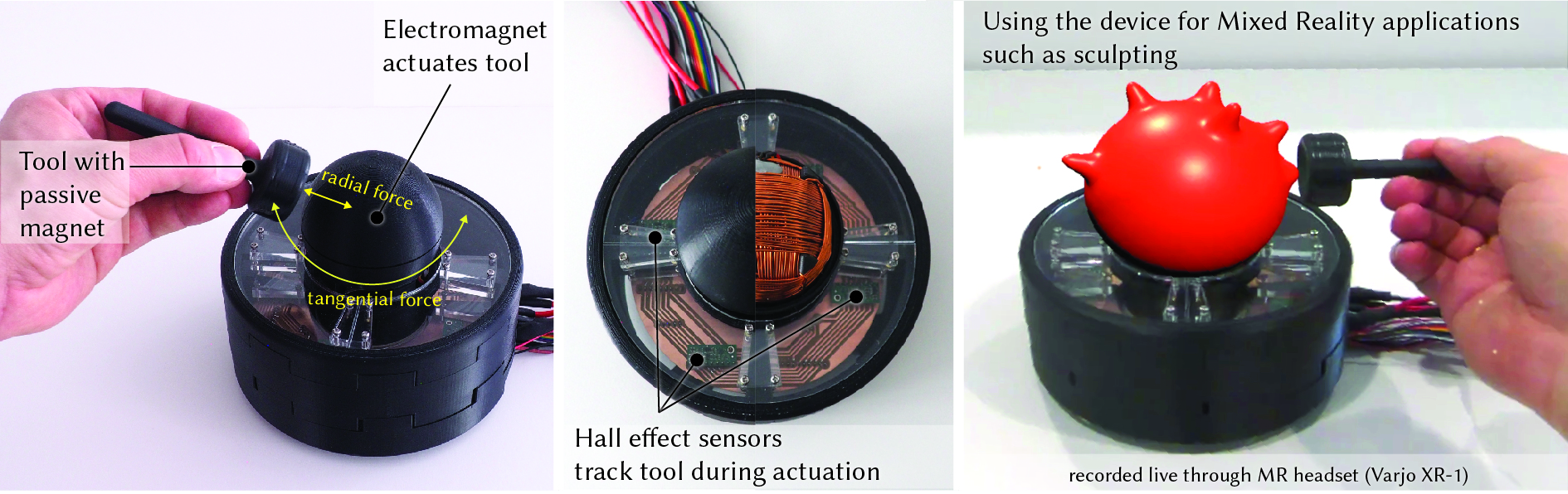}
    \caption{We present Omni, a device that can simultaneously actuate and sense the position of a passive handheld tool. This is enabled through integrated hall effect sensors and our novel gradient-based optimization scheme. Omni can for example be used in 3D applications such as MR sculpting.}
\label{fig:teaser_sensing}
\end{figure}

\section{Introduction}
In this dissertation, we have presented pen-based haptic control systems, including a novel spherical electromagnet (\chapref{ch:shared:contact}) and a sensing algorithm (\chapref{ch:shared:volumetric}). However, while these systems aim to control the pen, they inadvertently reduce user agency. A haptic system should guide the user while preserving their autonomy.

Pen-based interactive systems featuring \textit{haptic guidance} support users in various applications such as drawing, sketching, writing, and CAD design. The goal of such systems is to enable users to draw higher-complexity shapes with less effort and higher accuracy, or to support users through virtual haptic tools such as rulers or guides. Crucially, such systems aim to strike a balance between giving users a strong sense of control and agency, while providing feedback unobtrusively. This is enabled through embedding controllable electromagnets in the system, which can then guide input devices such as a pen, or provide feedback to users about the positions and boundaries of virtual objects.

Existing systems such as dePenD \cite{yamaoka2013depend} typically employ an \textit{open-loop control approach}. The magnet that drives the pen is set to a predefined trajectory, and users then must closely follow the movement of the system. In this case, it is not possible for users to adjust the trajectory, since it would lead to a loss of haptic guidance. This effectively leads to a decrease in control of users, and arguably a loss of user agency.

Alternatively, it is possible to extend haptic guidance systems with traditional closed-loop approaches, \eg by implementing a proportional–integral–derivative (PID) controller \cite{aastrom1995pid} (and as we did in \omniHap and \omniUIST) or based on heuristics \cite{Lopes16}. Such systems adjust to users' movement but are, \add{usually}, based on a timed reference, effectively dictating users' drawing speed. This can lead to unintended behavior such as snapping whenever the pen is too close to the actuator, a problem that is exacerbated for magnetic systems due to the non-linear nature of the magnetic force over distance.

We propose a real-time closed-loop control approach that allows users to retain agency and control while being assisted by an electromagnetic haptic guidance system. Our approach enables users to draw at their desired speed and adjust their target trajectory continuously.
It then adapts and complies with such modifications while giving corrective feedback. Our algorithm positions and regulates a variable-strength electromagnet such that it provides dynamically adjustable in-plane magnetic forces to the pen tip.

We contribute a novel optimization scheme for electromagnetic-based haptic guidance systems, \ie models and control algorithm, that enables formalizing this problem in the established Model Predictive Contour Control (MPCC) framework \cite{lam2010model}, which has previously only been employed in contexts such as RC-racing \cite{Liniger2014} or drone cinematography \cite{Naegeli:2017:MultiDroneCine}.

We provide an accurate system model, parameters, and an appropriate cost function alongside a method to optimize the model parameters given user inputs. Modeling the non-linear interaction of an electromagnetic force field typically makes use of the finite element method (FEM), which is not applicable for real-time scenarios. To overcome this challenge, we additionally contribute a novel approximate yet accurate model of the electromagnetic force field that can be evaluated analytically in real time. Compared to simpler control schemes such as Model Predictive Control (MPC) \cite{Faulwasser:2009} and \add{many implementations of} PID control, our approach does not require a timed reference and hence allows users to draw at their desired speed.

Furthermore, our optimization scheme allows for error-correcting force feedback, gently pulling the user back to the desired trajectory rather than pushing or pulling the pen to a continuously advancing setpoint on the trajectory. With our approach, the reference path can be updated at every timestep, thus allowing users to continuously change their desired trajectories. This enables applying the algorithm to fully dynamic references, for example virtual tools such as rulers or programmable French curves. 

To assess the proposed control algorithm, we developed a proof-of-concept hardware implementation (see Figure~\ref{fig:hardware}), leveraging an electromagnet that moves underneath the drawing surface or display on a bi-axial linear stage. The magnet provides variable strength guidance onto the tip of a minimally instrumented pen or stylus via an electromagnet positioned directly below a drawing surface, guided by our proposed approach.

We demonstrate the feasibility of our approach with a set of applications, specifically drawing guidance on conventional paper for sketching and writing, and a digital sketching application that features virtual haptic guides and rulers.

To evaluate our approach, we performed two experiments with twelve participants each.
We first compared free-hand drawing of shapes with varying complexity with and without our feedback system. Results showed that the haptic guidance using our approach improved the accuracy across shapes by up to 50\% to $\unit[1.87]{mm}$. We then compared our approach to our implementation of dePENd (open-loop) and a simple MPC-based closed-loop control scheme. Our approach showed significantly higher accuracy and was preferred by users.

In summary, we contribute
\begin{itemize}
\setlength\itemsep{0em}
\item A novel MPCC-based optimization scheme for electromagnetic haptic guidance systems including models, parameters, cost function and control algorithm.
\item A novel real-time approximate model for electromagnets that generalizes beyond our hardware implementation.
\item Evaluations showing the improved accuracy of our method.
\end{itemize}
\begin{figure}[!t]
\includegraphics[width=\columnwidth]{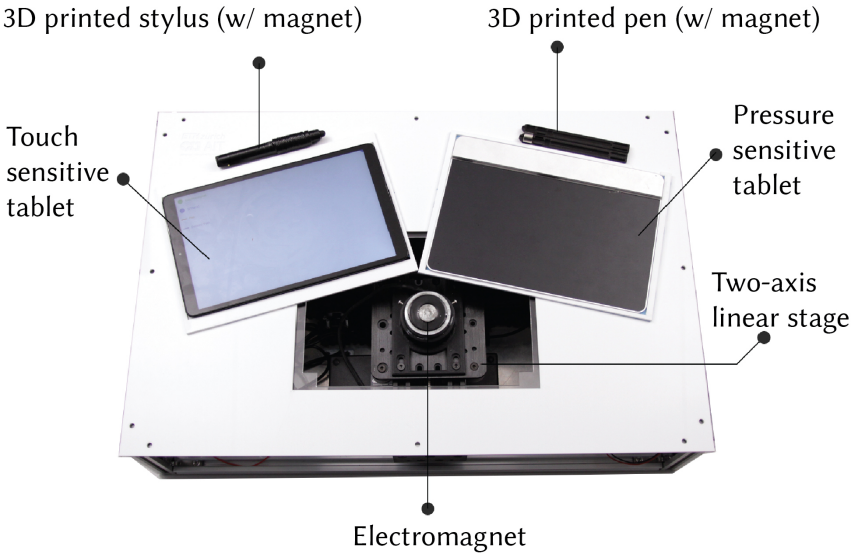}
\caption{We implement our proposed guidance system using a two-axis linear stage equipped with an electromagnet. All technical and user evaluations were completed using the Pressure Sensitive Tablet (\textit{right}). We additionally developed an all-digital implementation using a multi-touch tablet with display (\textit{left}).}
\label{fig:hardware}
\end{figure}

\begin{figure*}[!t]
    \centering
    \includegraphics[width=.75\textwidth]{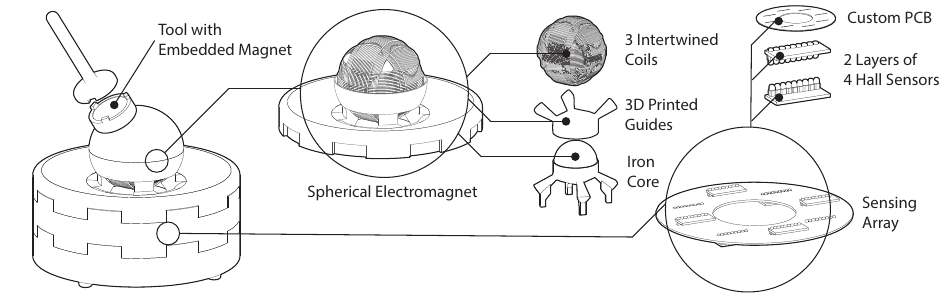}
   \caption{Overview of our system. A 3D printed base contains the 3 DoF intertwined coils and the circular PCB with an array of eight hall sensors. Arbitrarily shapes tools can be 3D printed and augmented with a permanent magnet, to interact with \omniUIST.}
    \label{fig:hardware}
\end{figure*}

\section{System Overview}

\omniUIST is a self-contained haptic feedback system that simultaneously integrates 3D tracking and actuation using the same modality (see Figure~\ref{fig:hardware}).
Through actuating an untethered, contact-free tool by means of a magnetic field, our device supports rendering precise haptic attractive and repulsive forces as well as accurate tracking without the need for any external infrastructure or markers. Our system allows for rich interactions with and haptic perception of dynamic virtual surfaces.

Our goal is to enrich AR/VR and other 3D applications via \omniUIST and a minimally instrumented tool. Simplicity of the haptic prop and a walk-up-and-use experience were important design goals of our work. Furthermore, to create rich immersive experiences, such a system must be able to deliver different types of high-fidelity haptic forces and precisely sense user input without requiring any external tracking. Moreover, we aim for a self-contained device that is affordable and easy to manufacture.

The actuation mechanism used in \omniUIST is based on the working principle proposed in \cite{zarate2020contact}. We build up a hemispherical shell base whose core houses a symmetric omnidirectional electromagnetic actuator. Three interwoven and mutually orthogonal coils generate the haptic forces.

By controlling the current in each coil, we can precisely configure the exerted force onto an external magnet, such as the one inside the 3D stylus tool. As the tool approaches the sphere, \omniUIST is able to provide independently controlled radial and tangential forces. Our design contributes several important improvements over \cite{zarate2020contact} that allow us to provide twice the amount of force ($2$ N in either direction) and for much longer periods without suffering from self (over-)heating. 

Although the tool is contact-free, the haptic force that the user perceives has its reaction force on the support base. In this sense, the user perceives grounded forces even if there is no mechanical link to the base.

In our current implementation, \omniUIST rests on a surface (\eg table), though it is compact and could be mounted on a robotic end effector to deliver large-scale 3D feedback. \add{This would allow for haptic feedback in a large volume, which would be beneficial for VR applications. In this case, however, geomagnetism should be taken into account more strongly.}

Beyond the improved actuator, our main contribution lies in the integration of the \omniUIST actuator with a fully self-contained real-time tracking method of the tool. To this end, eight Hall sensors are distributed below the interactive sphere. Each sensor reads a combination of the magnetic field generated by the tool, superimposed by the electromagnetic field generated by the actuator.We use a gradient-based optimization method to locate the magnet's 3D position based on the Hall sensors' readings, running at an interactive rate of 40 Hz.


\section{Method}
To support interactive experiences, our actuator needs to dynamically adjust its output according to the desired haptic feedback at a given time and tool position.
We now describe our real-time approach to reconstruct the tool's 3D position given the readings observed by the Hall sensors and the control strategy to govern the electromagnet-tool interaction.

\subsection{3D position estimation}
At the core of creating dynamic interactive experiences lies the ability to react in real-time to the movement of the user. Thus, a method to acquire the tool position with sufficient accuracy and precision with low latency is required. This is challenging due to the dynamic superimposition of the various magnetic fields (see Eq. \ref{eq:b_sum}). More precisely, directly computing the tool's position from the sensor readings would require inversion of Eq.~\ref{eq:basic_B}, which is non-linear and hence non-invertible, rendering an analytical solution for the tool position infeasible.

To overcome this difficulty, we introduce a reconstruction algorithm that optimizes an estimate of the tool's position given the Hall sensor readings in real time. We propose an iterative model fitting approach for 3D position estimation. We minimize the residual between the \emph{expected} sensor reading $\BiBold$, as predicted by our model of the magnetic field (Eq.~\ref{eq:b_sum}) given the current actuation, and the \emph{actual} measurements acquired by the Hall sensors $\BiTildeBold$. In this setting, the optimization variables are the tool's 3D position and its orientation. With a good initialization, which we attain by careful construction of the hardware, and exploiting the redundancy in the measurements, this algorithm provides accurate estimates of the tool's position in 3D, with a mean accuracy of 6.9$\pm 3.2$ mm, as shown in our technical evaluation.

For each Hall sensor ($\siBold \in \SBold$) defined by its 3D coordinate, $\siBold = [s_{x},s_{y},s_{z}]^T$, we seek to find the tool position $\RpenBold = [p_x,p_y, p_z]^T$ and orientation $\oBold = [\theta, \varphi]^T$ that provides the best model fit to the current reading. We use the global coordinate system for the sensors position $\RsiBold$ and tool position $\RpenBold$ with an origin in the center of the electromagnet, \ie $\ReBold = [0,0,0]$. The optimization problem is then given by:
\begin{equation}\label{eq:energy}
\arg\underset{\RpenBold,\ \oBold}{\min} \bigg[ \sum_{\siBold \in \SBold} \ \sum_{x,y,z}
\mathbf{w_i} \Big( \BpBold(\RsiBold - \RpenBold, \mpBold) + \BeBold(\RsiBold, \meBold) + \BnBold - \BiTildeBold \Big)^2 \bigg] \ ,
\end{equation}
where $\mathbf{w_i}$ selectively weighs the sensor axes depending on the value it reads (\ie a completely saturated sensor receives a weight close to 0). We pre-compute $\BeBold$ for different actuation strengths and $\BnBold$ denotes the background noise measured at startup.
We empirically found that including the tool orientation $\theta$ and $\varphi$ as free variables improves the position estimates by roughly 2 mm in Euclidean distance. However, the orientation estimates were too noisy to use in interactive settings.

We minimize Eq.~\ref{eq:energy} via iterative optimization. Specifically, we use PyTorch's second-order L-BFGS optimizer, which typically works well for non-smooth optimization instances such as ours and requires no parameter tuning. Gradients are computed automatically via auto-grad.

Our method relies on known sensor locations obtained via one-shot calibration. We empirically found that an initial estimate of the sensors' locations in the range of $1$ mm accuracy is required to support robust convergence of the algorithm.

\subsection{Actuation} \label{sc:actuation control}
Given the 3D pen-position, we can now deliver dynamically adjustable attractive and repulsive forces via the electromagnet to create desired haptic experiences. It remains in the hand of an application designer to decide with which intensity and in what direction the tool is pulled or pushed according to the desired user experience. \omniUIST is able to control 3 of the 6 DoF available, summing up forces and torques. We derive the case in which the goal is to control the three components of the haptic force $\mathbf{F_h}$, while controlling torques would follow an analogous derivation.

Under the magnetic dipole-dipole approximation, the force applied to the permanent magnet in the tool can be computed from the previous magnetic moments for the tool, $\mpBold$, and the electromagnet magnetization $\meBold$ we seek to control. Using the formulation of Yung \etal\cite{yung1998analytic} and rewriting it in matrix form allows us to derive a simple control law for the parameters of the electromagnet $\meBold^{set}$, given the location of the tool $\rpBold$, its dipole orientation $\mpBold$, and the desired haptic force $\mathbf{F_h}$:
\begin{equation}
\meBold^{set} = a_1 \ \left[\mathbb{D} \ + a_2 \ \mathbb{I}\right]^{-1} * \mathbf{F_h} \label{eq:force_me}
\end{equation}
\noindent where $a_1 = \frac{4\pi \rpen^{5}}{3\mu {0}}$, $a_2 = \langle\mpBold,\rBold\rangle$ can be computed from the tool position information (as described in the previous section).
The matrix $\mathbb{D}$ has elements
\begin{equation}\label{eq:d_ij}
d{i,j} = m_{p_i} \ r_{p_j} + m_{p_j} \ r_{p_i} - 5 \ a_2 \ (r_{p_i} / r_p^2) \ ,
\end{equation}
\noindent where $m_{p_{i}}$ and $r_{p_{i}}$ denote the $i$-component of the dipole $\mpBold$ (Eq. \ref{eq:mpBold}) and position $\rpBold$, respectively.
$\mathbb{I}$ is a diagonal identity matrix.

Finally, we use the calibration matrix from Eq.~\ref{eq:me_from_c_i} to find the electrical current to be applied to the individual coils:
\begin{equation}\label{eq:I_from_me}
\mathbf{I}^{set} = \mathbb{C}^{-1} * \meBold^{set} \ .
\end{equation}

By combining Eq.~\ref{eq:I_from_me} and \ref{eq:force_me}, the vector of desired haptic force $\mathbf{F_h}$ can be mapped into three actuation currents $\mathbf{I}^{set}$. This 3D-forces-onto-3D-currents mapping can always be decomposed into tangential and radial forces, using the tool's local coordinate system.

To only consider attractive and repelling forces and ignore the tangential component, Eq.~\ref{eq:I_from_me} can be further simplified to:
\begin{equation}
\mathbf{I}^{set} = \alpha \frac{2\pi \rpen^{3}}{3\mu _{0}} \ \rpBold \ ,
\end{equation}
\noindent where $\alpha$ is the intensity of the force and its sign denotes attraction or repulsion relative to the sphere. In this particular case, the electrical current vector, the direction of the tool, and the tool dipole can be assumed to be always collinear.

\subsection{Control implementation}
The tracking algorithm, electromagnet control, and user-facing components (AR applications) run on a commodity gaming PC (Intel Core i7-8086K with 6 cores at 4 GHz, 32 GB RAM, NVIDIA GeForce GTX 1080 Ti) on Windows 10. The system is implemented in Python 3.7 and uses PyTorch's L-BFGS solver. The optimization-based tracking algorithm runs at 40Hz at the highest precision (6.9 mm). The AR applications were implemented in Unity 2019, SteamVR, and the Varjo Unity Plugin v2.4.
\begin{figure}[!t]
\centering
\includegraphics[width=\columnwidth]{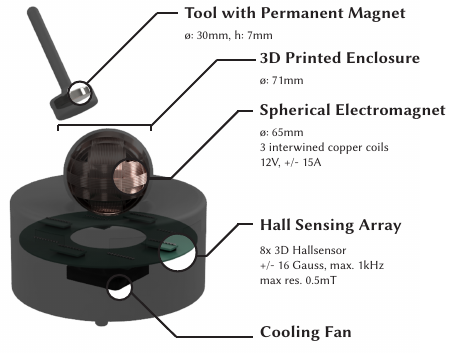}
\caption{\omniUIST hardware overview. Annotated view of the most important components of the system. All components can be acquired commercially or are easy to produce in a standard FabLab.
A top-down view of the physical device can be seen in Figure~\figref{fig:teaser_sensing}.}
\label{fig:control_Scheme}
\end{figure}

\subsection{Hardware integration}
Our hardware design is driven by two main factors. First, through a finite element analysis (FEA), we determined the physical characteristics of our hardware, such as coil diameter, core size, as well as the parameters of the permanent magnet embedded in the hand-held tool. This reference design strikes a balance between compact form-factor and force-generation capabilities. Second, we use off-the-shelf components for sensors and the voltage controller, in-house wound coils and in-house milled printed circuit boards (PCBs) to precisely mount the sensor's boards.

Reproducing \omniUIST only requires readily available components and few specialized tools, if any. The main hardware components are illustrated in Figure~\ref{fig:control_Scheme}.

\begin{figure}[!t]
\centering
\includegraphics[width=\columnwidth]{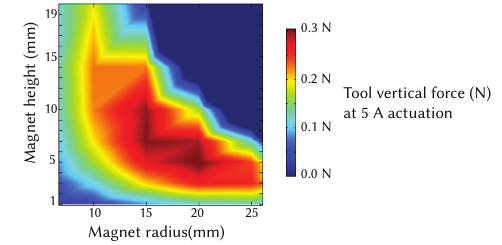}
\caption{Finite element analysis of actuation force as a function of the permanent magnet's dimensions. Our current design is based on a $15\times7$ mm magnet, which provides maximal force generation capabilities. However, the plot illustrates that there is a Pareto front of similarly well-performing shapes that could allow for different tool designs.}
\label{fig:comsol}
\end{figure}

Figure \ref{fig:comsol} plots the results from an FEA of the tool's magnet to identify the best configuration given our current spherical electromagnet design. We use a magnet with 15 mm diameter and 7 mm height (volume of 5 cm$^3$), which corresponds to a point in the dark red region of the plot, where the output vertical force at 5 A is maximal. Please note that these characteristics hold for magnets with similar volume: either wider and shorter cylinders or narrow and tall, providing ample room for the design of the handheld prop. Importantly, a bigger magnet would not necessarily perform better and may decrease performance (top right region of the plot), since the weight of the magnet counteracts the vertical actuation in repel mode.
\add{The omnidirectional electromagnet is based on a 30 mm diameter soft iron core encapsulated in 3D printed guides, aimed to assist during the manual winding (see Fig. \ref{fig:hardware}).}
\add{In contrast to~\cite{zarate2020contact}, we construct our coils layer by layer in an interwoven fashion.}

We iteratively add layers of x-winding, y-winding, and z-winding, respectively, until we reach an outer diameter of 65 mm. We use round copper wire with an external diameter of 0.9 mm (19 AWG) to obtain roughly 150 turns per coil. We employed a total cable length of $L_x = 21.4$ m, $L_y = 22.7$ m and $L_z = 24$ m for the x-, y- and z-coil respectively.
Since the x-coil is wound first, it has a smaller radius per layer, \ie a shorter perimeter per turn. The measured resistance of the coils are $R_x = 0.643 \ \Omega$, $R_y = 0.676 \ \Omega$ and $R_z = 0.708 \ \Omega$. The system supports up to 15 A of actuation current, which translates to a power of 157 W. To help remove the Joule heating generated within the coils, we place a brushless DC fan under the sphere (\emph{CUI Devices, 0.524 $m^3$/min}), and include air intakes on the side of \textit{Omni}.

To enable accurate and reliable tracking, it is paramount that the electromagnet and the Hall sensors are mounted rigidly with respect to each other. To ensure this, we fabricated custom PCBs using a desktop PCB milling machine (\emph{Bantam Tools}). The ring-shaped sensor PCB is located below the electromagnet, with two circular arrays that mount 4 Hall Sensors each (LIS3MDL, \emph{Pololu}).

All sensors are precisely aligned with the coil planes, such that each sensor's local coordinate system aligns with the global frame. The Hall sensors sample at up to 1kHz and are read out by a microcontroller (\emph{Teensy 4.0}), that communicates with the host PC.

We implemented an open-loop strategy to control the generated force. The approach is based on an analytical relation between the force, the coil actuation, and the tool location (Eq.~\ref{eq:force_me}), and relies on a few-point calibration. For actuation, we use three H-bridges (\emph{Pololu G2 18v17}) to control the current of each coil with a pulse width modulation (PWM) of the voltage. Given Ohm's law, we can directly control the current via the voltage if the resistance of each coil is known. Since electromagnets suffer from drift due to self-heating (and thus resistance changes), our system includes a coil-resistance drift compensation implemented directly on the actuation microcontroller. Two current sensors (\emph{INA260, Adafruit}) provide an independent measure of the voltage and current of each coil. A sliding window average of the measured current $I_i$ and voltage $V_i$ are used to stabilize force-generation.
\section{Evaluation}
\textit{Omni}'s capability of delivering convincing haptic sensations relies on the performance of two main components: tracking of the tool position and in-air actuation.
We performed technical evaluations on both aspects.

In summary, \textit{Omni} is able to reconstruct the position of the tool with an accuracy of $6.9 (\pm 3.2)$ mm and can deliver peak forces of up to $\pm$ 2 N, and 0.615 N continuously.
Besides this technical evaluation, we demonstrate \textit{Omni's} interactive capabilities in the application section.  We refer readers to Zarate \etal\cite{zarate2020contact} for a psychophysics evaluation of a comparable underlying actuation mechanism, showing that users can discriminate at least 25 discrete force locations.

\subsection{Tracking evaluation}
To evaluate \textit{Omni}'s tracking accuracy, we compared our position estimates to those of a 10-camera Optitrack setup, capturing a tracking space of 1.2 $\times$ 0.8 m with submillimeter accuracy at 100 Hz. We configured \textit{Omni} to run in precision mode at a frame rate of 40 Hz. For both tracking methods, we recorded the position and rotation angles.
We evaluated the accuracy of \textit{Omni} in two conditions: \emph{no actuation} and \emph{actuation}. In the \emph{no actuation} condition, no current was sent to the coils. In the \emph{actuation} condition, the coils were actuated using a sawtooth function sending current between -4 and 4 Amperes for each axis. 

For each condition, the pen was moved around the center of \textit{Omni} at a distance of up to 10 cm, covering the area around the device. We collected 1600 samples for the \emph{no actuation} condition and 2600 samples for the \emph{actuation} condition, both at roughly 5 Hz.

\subsubsection{Results}
\begin{figure}[!t]
\centering
\includegraphics[width=\columnwidth]{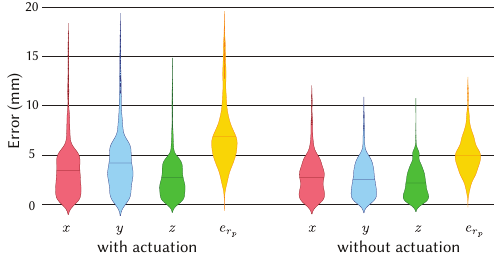}
\caption{Distribution of tracking error with and without current applied to the electromagnet.}
\label{fig:optitrack_eval}
\end{figure}

We found that the average difference between the two tracking systems is $e_{r_p} = 4.9 (\pm 1.8)$ mm in the \emph{no actuation} condition and $e_{r_p} = 6.9 (\pm 3.2)$ mm in the \emph{actuation} condition. Analyzing each axis separately, we found that $\mathbf{e_{r_p}} = [3.4;\ 3.1;\ 2.7]$ mm for the tracking errors and \emph{no actuation} and $\mathbf{e_{r_p}} = [4.4;\ 5.2;\ 3.3]$ mm in the \emph{actuation} condition. The results are summarized in Figure~\ref{fig:optitrack_eval}.

Finally, we tested our formulation with and without the orientation estimation of the magnet.
While we found that including these additional optimization variables improves the accuracy of the position estimates, these estimates are unstable and not yet useful for interactive applications.

Intuitively, it makes sense that including the orientation in the model fitting improves position estimation since the orientation of the magnet does influence the magnetic field. Furthermore, it is known that the model we leverage \cite{yung1998analytic} works best for spherical magnets (\eg point estimates of positions) and hence the models' approximation error may be a source of noise in the orientation estimates. We leave an extension of the reconstruction method to 5-DoF for future work.

\subsection{Actuator evaluation}
\begin{figure}[!t]
\centering
\includegraphics[width=1\columnwidth]{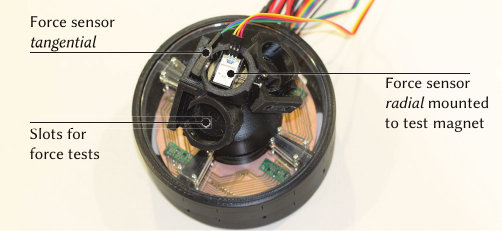}
\caption{Setup for actuator evaluation. An additional 3D printed hemisphere is placed on top of \textit{Omni} to hold the force sensors.}
\label{fig:eval_actuator}
\vspace{-1em}
\end{figure}

\textit{Omni}'s 3 DoF spherical electromagnet produces a force on the permanent magnet in the tool by dynamically adjusting the magnetic field through currents in the orthogonal coils.
To quantify this actuation, we measured the radial and tangential forces at different locations around the electromagnet in \textit{Omni}'s spherical base. We placed a 3D-printed hemisphere over the electromagnet (see Figure~\ref{fig:eval_actuator}). The hemisphere has three slots to place a test magnet (\emph{S-30-07-N, Supermagnete}, same as in tool) and two force sensors (\emph{FSAGPNXX1.5LCAC5, Honeywell}). The force sensors were attached between the electromagnet and the test magnet to measure radial force, and to the side of the test magnet to measure tangential forces.

\subsubsection{Results}
\begin{figure}[!t]
\centering
\includegraphics[width=1\columnwidth]{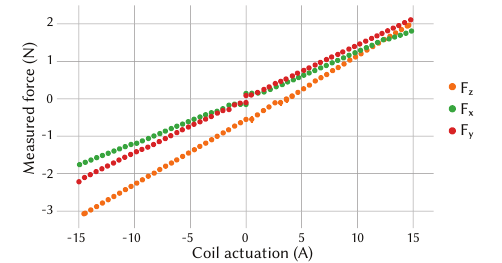}
\caption{Radial and tangential forces on the permanent magnet as a function of coil actuation $I_x$, $I_y$ and $I_z$, for the magnet located on top of the sphere (Fig~\ref{fig:eval_actuator}). Force was collected with a compression-like force sensor.}
\label{fig:Fz_vs_iz}
\end{figure}

We generally observed a linear response of actuation with respect to the applied current.
On top of the electromagnet, we measured a maximum vertical repulsive (radial) force of 1.95 N at $I_z = 14.6$ A and a maximum attractive force of -3.04 N at $I_z = -14.6$ A, shown in Figure~\ref{fig:Fz_vs_iz}.

When \textit{Omni} applies $I_z = +3.7 $ A, it compensates for the weight and snapping and the magnet starts to levitate\footnote{In this paper we use the term \emph{levitation} in the sense of \emph{compensate its weight completely}. A complete levitation would need to control the actuation in all three axes to keep the magnet floating in place.}. Note that at this position, the force is the sum of the electromagnetic actuation, the snapping to the core, and the gravitational attraction.

The weight of the tool produces a force of $F_r = -370$ mN (38 gr), while ferromagnetic snapping yielded additional 170 mN of force, combined these produce an attracting radial force of $F_r = -540$ mN without actuation ($I_z = 0$ A).All those components contribute to users' perception of force.

On top of the sphere, the weight and snapping are orthogonal to the $x$-axis and $y$-axis and do not influence the radial forces along those axes. Consistently, we measured a linear response on those axes of the form $Fr_{x-axis} = 0.122 ~[N/A] ~I_{x}$ and $Fr_{y-axis} = 0.142 ~[N/A] ~I_{y}$. For the other locations in our test setup (horizontal to vertical), we observed forces in the range of $\pm 2$ N at $\pm 15$ A with the corresponding corrections for weight and snapping. Note that the forces have been measured when the tool was in contact with \textit{Omni}'s hemisphere.

The force intensity decays with $1/(d_0 + g)^4$, where $g$ is the air-gap between the tool and the sphere. The parameter $d_0 = 41$ mm is the center-to-center distance between the electromagnet and the permanent magnet when the tool touches the sphere. For example, for $g = 10$ mm, the reachable range of forces drops to $\pm 835$ mN. \add{At 30mm from the hull this reduces to 0.2N}

\subsubsection{Evaluation of EM heating}
\begin{figure}[!t]
\centering
\includegraphics[width=\columnwidth]{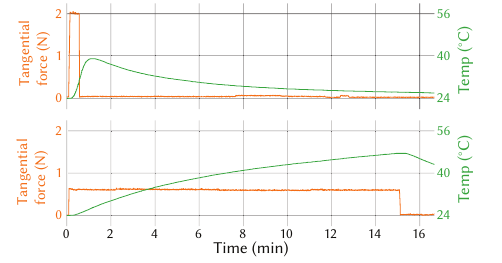}
\caption{Temporal evolution of the self-heating of the coils for two different types of actuation. Top: a \emph{peak-force} of 2 N ($I_y = 15$ A) during 25 seconds. Bottom: a \emph{constant-force} of 600 mN ($I_y = 5$ A) during 15 minutes.}
\label{fig:heating}
\end{figure}

To test the stability of the generated forces and the thermal capabilities of our system, we ran two experiments. First, we set the $y-$axis coils to maximum actuation current $I_y = 15$ A for 25 seconds and let it cool down afterwards to test the system under \emph{peak-force} conditions. Second, we set the same coil to 1/3 of the maximum actuation and we let it run for 15 minutes, to test under \emph{constant-force} conditions. Figure \ref{fig:heating} shows the evolution of the generated force and the temperature of the coil for both conditions.

During \emph{peak-force}, the system delivers a force of $2.04 \pm 0.04$ N. Starting from room temperature (24 °C), the actuator heats up to 39 °C but only 40 seconds after the actuation has been turned off, showing the system's thermal inertia. The $\Delta T = 15$ °C during this intense actuation peak shows that our system is capable of thermally buffering and dissipating the heat generated by intense forces even during tens of seconds.

In our \emph{constant-force} experiment (Figure \ref{fig:heating}, \textit{bottom}), the force remained constant within the limits $0.615 \pm 0.015$ N and for a duration of 15 min, even when the temperature of the coils (and their resistances) significantly changed.
In addition to compensating for the actuation drifts, we used the coils' resistance changes over time as the limiting factor to avoid overheating of the coils and the 3D printed parts, in case the system is required to apply maximum forces for minutes.

\section{Applications}
To further demonstrate the potential of our approach we illustrate possible usage-scenarios including calligraphy, outlining and inking. 
Finally, we combine the haptic feedback mechanism with a simple digital drawing application to initially explore the possibility of dynamic references.    

\subsubsection*{Calligraphy}
\begin{figure}[!t]
    \centering
        \includegraphics[width=\columnwidth]{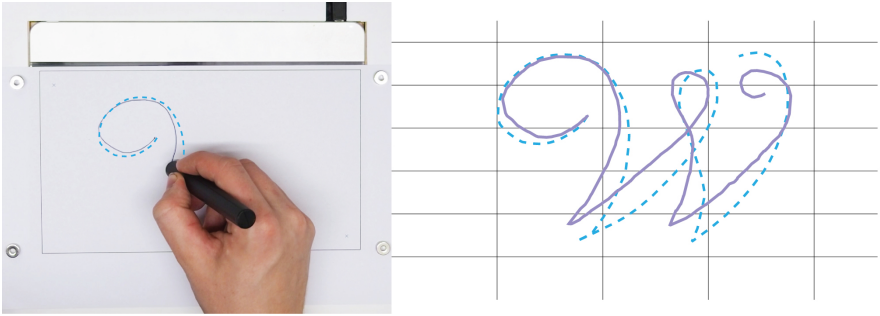}    
        \caption{Our approach can be used as a guidance system for calligraphy, where users either follow a target path very closely, or deviate if desired.}
    \label{fig:caligraphy}
\end{figure}

\figref{fig:caligraphy} illustrates writing of flourished characters, with only minimal visual guidance (single starting point). 
Our system takes the character as input, users can then draw at their desired speed. 
Although an offset from the reference path remains, the lines are smooth and the overall shape is close to the desired characters. 

%
%

\subsubsection*{Outlining \& inking}
\begin{figure}[!t]
    \centering
        \includegraphics[width=\columnwidth]{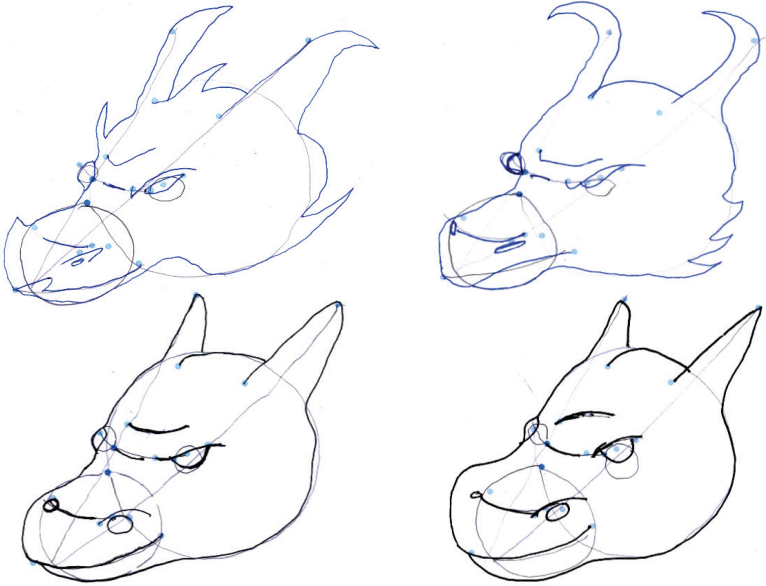}
    \caption{Different variants of the same dragon, drawn with identical system settings by a novice. Each pair of drawings used with different tools. First a pencil for proportions and a fine-liner (top) or pencil (bottom) to ink-in details. Multi-stroke lines are achieved by approaching each separate instance as a new drawing.}
    \label{fig:dragon}
\end{figure}

\figref{fig:dragon} illustrates the effect of two core capabilities of the proposed approach. 
Here we first outline the proportions of the dragon head (gray guidance lines) and then use different pens to ink-in the details. 
Note that the system provides haptic guidance but allows the user to draw the shape in different styles (\eg the ears of the two upper dragons) and with varying high-frequency detail, while maintaining similarity to the reference shape. 
This is a direct consequence of using time-free closed loop control approach.
In this case, all four variants were drawn without changes to the system or desired path. 

\subsubsection*{Virtual tools}
\begin{figure}[!t]
    \vspace{-.5em}
    \centering
    \includegraphics{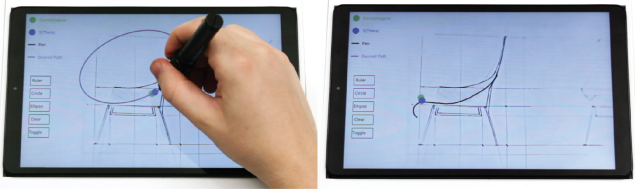}    \caption{Virtual tools can be used to dynamically construct a reference path combining haptic and visual feedback. Here  a simple drawing application combines freeform sketching with different virtual rules and guides that can be felt by the user.}
        \label{fig:tablet}
    \vspace{-.5em}
\end{figure}

Using a digital tablet with capacitive display (\figref{fig:tablet}) we explore integrating dynamically changing references. 
In a sketching application, artist select different virtual tools, and position and configure these anywhere. 
The canvas and the haptic feedback system then pull the stylus towards these virtual guides. 
In \figref{fig:tablet}, the user has selected a tool that helps them when drawing an ellipse that snaps to a previous part of the drawing, both visually and in terms of haptics.

\section{Discussion}
Our experiments indicate that the proposed approach indeed increases accuracy in drawing tasks and that users perceive the system favorably. 
While our system increased users' accuracy for complex shapes, it did not yield any improvements for the straight line. 
This limitation can be attributed to the maximum speed of the linear stage, as indicated by user feedback.
In our interviews, some users indicated that they had the feeling that their drawings without feedback were more accurate once they experienced the haptic guidance.
This suggests the potential for short-term muscle memory development when using our haptic guidance system.
Long-term learning is a very interesting area to explore for our approach and haptic guidance systems in general. We plan to conduct such experiments in the future.
 
Our experiments currently focused on drawing and sketching applications.
We believe our control strategy can be beneficial for a wide range of applications.
We started exploring the usage of our approach with a tablet and digital stylus.
Further experiments are necessary to determine the levels of complexity at which our approach is most beneficial.
Allowing users to adjust their input on-demand is crucial, particularly since systems typically lack complete knowledge of the users' target paths.
Our approach is a first step in the direction of balancing user input and system control for haptic guidance systems, and can be extended to other devices beyond electromagnetic systems if appropriate force models are provided.

In terms of hardware, the speed of the linear stage was a limiting factor. Future improvements may involve faster, more compact linear stages or a matrix of stationary electromagnets. The former requires no changes to our formulation, while the latter necessitates modifications to the EM model and dynamics model. A matrix of electromagnets could pave the way for a thinner form-factor design, which is an exciting research direction.

Addressing the hardware-induced speed limitation will open new avenues for efficient closed-loop control strategies, since faster pen motion would also tighten the latency and accuracy budget. 
Incorporating a mechanism to reconstruct the tilt of the pen would enhance sensing capabilities.
This could be achieved for example via an accelerometer built into the pen or via a grid of hall-sensors underneath the surface. 
Information on the pen tilt could then be combined with the angle dependent formulation of our EM model. 
Furthermore, we believe there are many research opportunities in combining our approach with ink beautification approaches (\eg~ \cite{simo2016learning,simo2018mastering,xing2015autocomplete}). 
Particularly interesting would be to leverage fully predictive models for non-drawing applications (\eg DeepWriting \cite{Aksan:2018:DeepWriting}). 

Future work could explore combining our approach with various types of haptic feedback, either environment mounted or body-worn, and different form factors such as spherical electromagnets \cite{Langerak:2020:Omni, zarate2020contact}. 
Electromagnetic feedback in combination with spatial actuation maybe interesting in other settings. 
For example, a magnet mounted to a robotic arm could deliver contact-less feedback in VR scenarios. 
It would also be interesting to investigate how to best exploit the system capabilities in the context of motor memory and learning.
All these scenarios make it necessary that a system interactively reacts to user input.
Our approach enables such applications, and can generalize to such systems that go beyond 2D haptic guidance systems.

Finally, our optimization is subject to system dynamics, including user influence on pen position. Explicitly incorporating user behavior into system dynamics is challenging due to its non-linear nature. In the next chapter, we will explore a model-free learning-based approach that can learn underlying task structures and human behavior without explicit system dynamics.

We have proposed a novel optimization scheme for electromagnetic haptic guidance systems based on the MPCC framework.
Our approach strikes a balance between user input and system control, allowing users to adjust their trajectory and speed on-demand.
It optimizes system states and inputs over a receding horizon by solving a stochastic optimal control problem at each timestep.
Our formulation provides dynamically adjustable forces and automatically regulates magnet position and strength.
It can be evaluated analytically and is hence suitable for iterative, real-time optimization approaches. 
We implemented our approach on a prototype hardware platform and experimentally demonstrated that the feedback is well-received by users and offers higher accuracy compared to open-loop and time-dependent closed-loop approaches.
We believe our approach offers a principled method for haptic guidance, enabling users to retain agency while receiving unobtrusive assistance. This approach has broad applications in areas such as drawing, sketching, writing, and guidance via virtual haptic tools.

In this chapter, we introduced \omniUIST, a novel electromagnetic platform that simultaneously tracks and actuates a permanent magnet in the space around it. Our self-contained base assembly integrates 3D magnetic sensing using Hall sensors and magnetic actuation through radial and tangential forces produced by three orthogonal electromagnetic coils within a single sphere.

Our core contribution lies in decomposing the natural interference caused by simultaneous magnetic tracking and actuation. This is enabled by our novel gradient-based optimization method, which minimizes the difference between estimated and observed magnetic fields. This approach affords 3D tracking capabilities with a mean error of 6.9 mm during actuation forces of up to 2 N.

\omniUIST's capabilities allow spatial interaction systems to integrate 3D tracking and actuation of untethered, free-ranging tools simply by embedding a small permanent magnet. We demonstrated a series of example applications leveraging \omniUIST's capabilities, showcasing its potential in various interactive scenarios.

However, an open question remains on how to control the system to enable user autonomy rather than guidance. Addressing this challenge will be a key focus for the next part of this dissertation, aiming to further enhance the usability and functionality of haptic devices in practical applications.

\chapter{Summary \& Insights}
\label{ch:control:conclusion}

\paragraph{Summary}
In summary, we introduced two optimal control strategies: \magpen and \marlui. In ''\magpenTitle'' (\chapref{ch:control:optimal}), we developed an optimization method for electromagnetic haptic guidance systems using the MPCC framework. This method balanced user input and system control, allowing for trajectory and speed adjustments as needed. It optimized system states and inputs over a receding horizon by solving a stochastic optimal control problem at each timestep. Our design dynamically adjusted forces and automatically modified magnet position and strength, suitable for real-time optimization. We implemented our method on a prototype hardware platform, \magpen, and demonstrated that users responded well to the feedback, achieving higher accuracy compared to open-loop and time-dependent closed-loop methods. This approach offered a principled method for haptic guidance, broadly applicable in areas such as drawing, sketching, writing, or virtual haptic tools. However, our MPCC-based framework relied on known system dynamics, which are not always available, especially when considering user behavior.

In our second project, ''\marluiTitle'' (\chapref{ch:control:multi}), we addressed this limitation by introducing a model-free multi-agent reinforcement learning strategy for adaptive user interfaces. \marlui incorporated a \useragent and an \interfaceagent. The \useragent aimed to achieve a task-specific goal quickly, while the \interfaceagent learned the task structure by observing the \useragent's interactions with the UI. This method did not require real user data, as the \useragent learned through trial and error. We evaluated our approach through simulations and real user tests across five different interfaces and various task structures. The results indicated that our framework supported adaptive UI development with minimal adjustments, effectively assisting real users in their tasks. \marlui, and the multi-agent approach in general, represent a promising direction for developing adaptive interfaces, reducing the need for interface- and task-specific strategies.

Overall, our methods enabled an autonomy-automation trade-off, resulting in better user performance in terms of speed, accuracy, and number of actions compared to alternatives. Our research is a starting point for considering and investigating the role of predictive user models in control loops.

\paragraph{Implications}
Previous systems commonly employ open-loop control \cite{yamaoka2013depend} to manage intelligent systems. This method does not take the user into account, thereby preventing the balancing of user agency with system automation. Alternatively, some systems rely on heuristics \cite{Lopes16, Browne1990, Smith2010, Stephanidis1997}, which involve the creation of rules by experts. Additionally, supervised learning \cite{Maes1995, Lashkari1997, McCreath2006, Faulring2010, Shen2009a, Shen2009b, Berry2011, Pejovic2014, Mehrotra2015} is widely used for controlling intelligent systems. However, supervised learning focuses on short-term decisions. By only optimizing for the immediate next step, they limit the system's ability to intelligently balance user agency with system automation.

In contrast, our approach minimizes a cost function over a receding horizon constrained by user and system dynamics. We have demonstrated how to integrate models of human behavior explicitly and implicitly into optimal control strategies for intelligent systems. This approach allows systems to consider future states and actions and optimize inputs accordingly. In \magpen, we embedded a user model explicitly in the control strategy, considering user behavior over a horizon, though mathematically formulating this behavior can be challenging. Conversely, in \marlui, we took a multi-agent RL approach where user behavior is implicitly learned, eliminating the need for explicit behavioral models or system dynamics descriptions, and allowing the method to generalize across interfaces and tasks.

Our method builds on existing cognitive models in the literature, leveraging previous research. This grounding in cognitive models enables advancements on two fronts: 1) developing cognitive models for user bounds, and 2) applying RL methods to more complex scenarios. Treating HCI, and specifically the interaction with intelligent systems, as a multi-agent reinforcement problem is intuitive and facilitates future extensions of our work.

\paragraph{Limitations}
Our research has limitations in the context of shared variable interfaces.

First, both \magpen and \marlui require an objective function. In task-driven scenarios, these objective functions are relatively intuitive, such as accuracy or task completion time. However, in open-ended scenarios, such as browsing the web or engaging in creative tasks, defining an appropriate objective is challenging. Further research, such as Inverse Reinforcement Learning, is required to extract objective functions from user data.

Second, our work does not explicitly account for how human behavior changes based on the system's actions. In \magpen, we assume a compliant user, and in \marlui, the interaction might be implicitly learned. However, explicit models focusing on the interaction and the influence of system actions on user actions are vital.

Finally, neither \magpen nor \marlui considers the user's learning phase. During interaction, users develop a mental model of the system. Ideally, the system would be aware of the user's mental model. Embedding theory-of-mind models into a control strategy could create methods that work for both novice and expert users and personalize interactions based on user behavior, resulting in more intuitive and efficient interactions.

\part{\control}
\cleardoublepage%

\def\dir{chapters/05_shared_control/mpc}
\chapter{Optimal control for electromagnetic haptic guidance systems}
\label{ch:control:optimal}
\contribution{
In \chapref{ch:shared:contact} and \chapref{ch:shared:volumetric}, we introduced novel actuator and sensing techniques for electromagnetic haptic devices. However, the challenge of effectively controlling the actuator given the sensor input remains unresolved. In this chapter, we address this by exploring an optimal control method for electromagnetic haptic guidance systems. Our approach assists users in pen-based tasks such as drawing, sketching, and designing, while ensuring that user agency is maintained.
Traditional methods force the stylus to follow a continuously advancing setpoint on a target trajectory, often resulting in loss of haptic guidance or unintended snapping. In contrast, our control approach gently pulls users towards the target trajectory, allowing for spontaneous adaptation and drawing at their own speed. To achieve this flexible guidance, we iteratively predict the motion of an input device (such as a pen) and dynamically adjust the position and strength of an underlying electromagnetic actuator.
To enable real-time computation, we introduce a novel, fast approximate model of an electromagnet. We validate our approach on a prototype hardware platform featuring an electromagnet on a bi-axial linear stage and demonstrate its effectiveness through various applications. Experimental results indicate that our method is more accurate and preferred by users compared to open-loop and time-dependent closed-loop approaches.
}

\begin{figure}
    \centering
  \includegraphics[width=\textwidth]{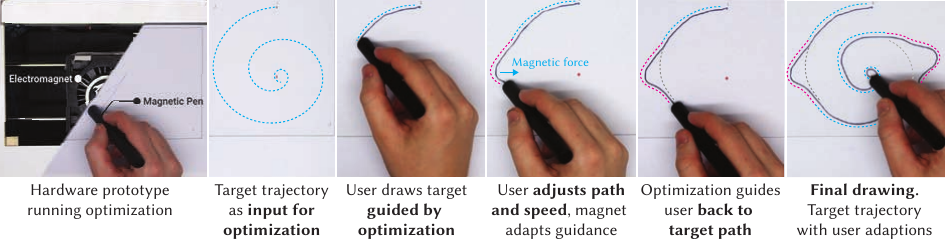}
  \caption{We propose an optimal control scheme for electromagnetic guidance systems (\textit{left}). A target trajectory is provided, on which users are guided. They can always adapt the trajectory, our optimization then guides users back to the target (\textit{offset between target and drawing for illustration purposes}).}
  \label{fig:teaser_mpc}
\end{figure}

\section{Introduction}
In the previous chapter, we demonstrated the use of model predictive control to determine the actuation of a haptic feedback device, relying on known system dynamics that included predictive models of user behavior. However, these models have limitations. To overcome this, we investigate a model-free reinforcement learning approach for learning system, task, and user dynamics. In this chapter, we shift our focus from haptics to adaptive point-and-click interfaces, allowing us to emphasize control aspects, as actuation and sensing in graphical user interfaces are deterministic.

Point-and-click interfaces form the core interaction paradigm in modern Human-Computer Interaction (HCI) \cite{reilly2005just, kemp2008point, lafuente2023comparing}. These interactions include clicking toolbar items, navigating hierarchical menus, and selecting objects—from UIs on traditional 2D desktops to emerging spatial UIs in immersive Mixed Reality. While individual point-and-click interactions are simple, complexity increases with the number of available interactive items. More items require users to process more information and evaluate more options, increasing cognitive load and diminishing the user experience, potentially extending task completion times.

Adaptive User Interfaces (AUIs) aim to mitigate this complexity by dynamically adjusting the interface to display the most relevant items for the user's current task while omitting less relevant information. This reduces cognitive load and simplifies navigation through complex interfaces, such as hierarchical menus.

Developing effective AUIs poses numerous challenges: AUIs must (1) infer users' intentions by observing their behavior and input, and (2) adapt the user interface based on these inferred intentions~\cite{oulasvirta2018computational}. Previous research has predominantly relied on machine learning (ML) techniques~\cite{Shen2009a, Shen2009b, todi2021adapting, gebhardt2019learning}, or heuristics~\cite{Browne1990,Stephanidis1997,Smith2010}. However, these methods depend on manually collected data or carefully hand-crafted rules, which is a significant limitation. Each new user interface or task requires new data collection and updated or redesigned adaptations, even when the interaction paradigm remains unchanged.

For example, customizing a game character through a toolbar interface illustrates these challenges \figref{fig:teaser_rl}. The toolbar has limited slots, but the set of clothing items is much larger. An AUI must assign the most relevant items to the slots based on the user's previous selections and likely final outfits. Similarly, in a VR game where a user builds a castle from different blocks, the AUI recommends the most likely block to be selected for each operation, considering the user's past choices and probable final designs. Despite the similar point-and-click paradigm, ML or heuristic approaches would require separate data collection and rule design for each task. This requirement makes developing more complex AUIs time-consuming and costly.

In this paper, we propose \marlui, a proof-of-concept framework for easy development of AUIs across various point-and-click UIs. We frame the point-and-click interaction paradigm in a Multi-Agent Reinforcement Learning (MARL) framework. An \interfaceagent learns to adapt a UI by selecting the most relevant subset of items at each interaction step. The \interfaceagent observes the \useragent and minimizes its task completion time, while the \useragent learns human-like point-and-click behavior to interact with the UI and train the \interfaceagent. Unlike previous approaches requiring real user data to train an \interfaceagent, our method relies on a human-like \useragent. The \useragent and the \interfaceagent jointly learn by exploring the interface through trial-and-error. Our \interfaceagent then transfers the learned adaptation strategy to real scenarios, selecting the most relevant items after a real user's click or selection. Switching tasks within the \marlui framework, such as from dressing a game character to constructing a block castle, requires minimal modification and only training on the respective interface. This approach eliminates the need for developers to gather or create task- and interface-specific data or heuristics, streamlining AUI development. Future work can focus on user personalization, extending to more complex interfaces, adapting to changing goals, and addressing different tasks.

To model the point-and-click paradigm as a MARL problem, we propose a simulated \useragent to learn interaction with a UI for task completion. We model the \useragent hierarchically, decomposing decision-making, such as selecting a toolbar item, from motor control actions, such as moving the cursor to a desired menu slot. Targeting human-like behavior in the \useragent, we incorporate cognitive and motor control bounds, following advancements in Computational Rationality~\cite{oulasvirta2022computational}. Concurrently, we train an \interfaceagent. The \useragent aims to reach a goal state, while the \interfaceagent adapts the UI to help the \useragent achieve its goal more efficiently, minimizing the number of clicks. The \useragent and \interfaceagent operate in a turn-based manner: the \useragent acts, followed by the \interfaceagent's adaptation, continuing until the task is completed. This novel approach enables our data- and heuristic-free method.

To demonstrate our framework's effectiveness across various tasks and interfaces, we explore five use cases: (1) customizing a game character using a toolbar, (2) adaptive numeric keypad design, (3) building a block tower, (4) selecting initially out-of-reach objects, and (5) photo editing with anticipatory menu opening. These use cases cover a broad range of interface types, from heads-up displays to world-anchored interfaces, showcasing the versatility of our approach.

We evaluate our framework in two stages: in-silico studies on the character creation task, showing our \interfaceagent can generalize to unseen goals, and evaluations with real participants, comparing \marlui against data-driven baselines. Our findings indicate that training the \interfaceagent with our simulated \useragent enables real users to reduce the number of actions needed compared to previous frameworks.

In summary, we make three key contributions in this paper:
\begin{enumerate}
\item The first attempt at a multi-agent Reinforcement Learning approach that does not rely on real user data or hand-crafted heuristics to adapt point-and-click user interfaces in real-time.
\item A \useragent that learns to operate a user interface and enables an \interfaceagent to learn the task's underlying structure purely by observing the \useragent's actions.
\item An \interfaceagent that learns the task's underlying structure by observing actions in the interface by the \useragent.
\item Results from evaluations demonstrating the effectiveness of our approach and five use cases showcasing its general applicability for point-and-click tasks.
\end{enumerate}

\section{Method Overview}
The goal of our online optimal control scheme is to allow users to maintain control and agency over the input device (e.g., pen, stylus), while experiencing dynamic guidance from the system. Importantly, it leverages \textit{time-free references}, and thus the dynamics are entirely driven by the pen position over time, which is different from approaches such as MPC.

The proposed optimization scheme allows us to adjust the magnet position and strength such that it gently pulls the pen tip towards a desired stroke, while allowing users to draw at their desired speed and without fully taking over control. The algorithm is generally hardware agnostic and works for devices with electromagnetic actuators underneath an interaction surface. This can be implemented via bi-axial linear stage as in our prototype (see \figref{fig:hardware}) or via a matrix of electromagnets which would lend itself better to miniaturization. Furthermore, the algorithm requires a reference trajectory over the optimization horizon. This can be defined a priori, such as a known shape to be traced, or may be provided dynamically, e.g., the output of a predictive model (e.g., Aksan et al. \cite{Aksan:2018:DeepWriting}).

At each time step, we minimize a cost functional over a receding time horizon to find optimized values for system states $\mathbf{x}$ and inputs $\mathbf{u}$.
As a high-level abstraction, the cost function
\begin{equation}
    \underset{\mathbf{x},\mathbf{u}}{\text{minimize}} \sum 
    \underbrace{\mathcal{C}{\text{force}}(\mathbf{x},\mathbf{u})}{\text{Eq. \ref{eq:err_F}, \ref{eq:err_d} \& \ref{eq:erralpha} }} + 
    \underbrace{\mathcal{C}{ \text{path}}(\mathbf{x},\mathbf{u})}_{\text{Eq.  \ref{eq:errLC}}} +
    \underbrace{\mathcal{C}{\text{progress}}(\mathbf{x},\mathbf{u})}_{\text{Eq. \ref{eq:errtheta}}}
      \label{eq:min1},
\end{equation}
serves three main purposes: 1) ensuring that the user perceives haptic feedback of dynamically adjustable force ($\mathcal{C}{\text{force}}$), 2) stays close to the desired path but does not rigidly prescribe it ($\mathcal{C}{\text{path}}$), and 3) makes progress along it ($\mathcal{C}_{\text{progress}}$) but allows the user to vary drawing speed freely.

 \begin{table}[!t]  \label{tab:control_params}
  \caption{Overview control parameters and values}
  \begin{tabular}{cp{0.25\columnwidth}p{0.5\columnwidth}}
    \toprule
    Name & Range / Value & Description\\
    \midrule
     $\posp$   			& $\mathbb{R}^2$ 						& Position of pen
     \\     
     $\posm$  			& $\mathbb{R}^2$ 						& Position of electromagnet
     \\
     $\mathbf{F_a}$  			& $\mathbb{R}^3$ 						& Electromagnetic force vector
     \\
     $\alpha$ 				&$\left[0,1\right]$	 & Electromagnetic intensity
     \\
     $\mathbf{s}$  		& $\theta \in[0,L]$ 						& Target trajectory of length $L$
     \\
    $\mathbf{x}$ 		& $[\mathbf{p}_{m},\dot{\mathbf{p}}_{m}, \alpha, \theta]$ & System states 
    \\
    $\mathbf{u}$ 		& $[\ddot{\mathbf{p}}_{m}, \dot{\alpha}, \dot{\theta}] $	& System inputs 
    \\
     \bottomrule
\end{tabular}
\end{table}

\section{Method}
Our main contributions are models and a control strategy that enables using the MPCC framework \cite{lam2013model} for electromagnetic haptic guidance.
MPCC is a closed-loop \emph{time-independent} control strategy that minimizes a cost function over a fixed receding horizon. 
There are several advantages in using our formulation over open-loop (as used in dePENd~\cite{yamaoka2013depend}) or time-dependent strategies (\eg MPC). 
First, closed-loop control allows to react to user-input, whereas open-loop control removes all user agency. 
Both MPC and MPCC are closed-loop control strategies. 
However, MPC tracks a timed reference, requiring a fixed velocity by users. 
MPCC follows a time-free trajectory, which allows the user to progress at their own speed. 
\figref{fig:control} illustrates the expected behavior for the different strategies, given that the user slows down or stops moving the pen. 
The desired behavior here would be that the algorithm essentially ``waits'', \ie provides guidances towards a slowly or no-longer advancing setpoint.
In this situation, open-loop approaches would lead to lost haptic guidance. 
Closed-loop time-dependent approaches would guide the pen towards a constantly advancing setpoint (although users do no longer move), which can lead to problems such as the user being guided backwards (\eg timestep $t=3$ is in front of $t=2$).

\begin{figure}[!t]
    \centering
    \vspace{-.5em}
    \includegraphics[width=\columnwidth]{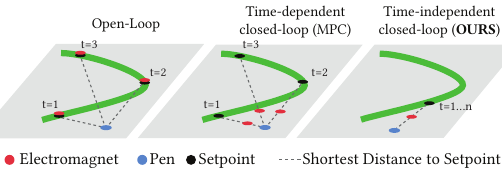}
    \caption{
    Overview of different control strategies on a target trajectory (\textit{green}), with constant pen position.
    For open-loop, the position of the electromagnet is identical to the constantly advancing setpoint, leading to loss of haptic guidance.
    For MPC, although the pen is static, the guidance changes at every timestep since the setpoint advances.
    In our approach, the setpoint is also based on the pen position, therefore remains stationary in this case and guides the user towards the target trajectory.}
    \label{fig:control}
    \vspace{-1em}
\end{figure}

Our method is designed to exert a force $\mathbf{F}_\theta$ of desired strength onto the pen to guide the user towards the target trajectory $\mathbf{s}$. 
The path $\mathbf{s}$ of length $L$ is parametrized by $\theta \in[0,L]$. 
Note that we do not prescribe how fast users draw and hence for each given pen position $\posp$ we first need to establish the closest position on the path parameterized by $\mathbf{s}(\theta)$.
The vector between the pen position and $\mathbf{s}(\theta)$ is defined as $\Rtheta$.
We leverage a receding horizon optimization strategy and the global reference can hence be adjusted or replaced entirely at every iteration. 
The path $\mathbf{s}$ is then a local fit to the global reference.
Furthermore, we seek to find optimized values for the electromagnet intensity $\alpha$ and the in-plane electromagnet position $\mathbf{p}_{m}$. 
Solving the error functional given in Eq. (\ref{eq:J_k}) at each timestep yields optimized values for system states $\mathbf{x}$ and inputs $\mathbf{u}$.   
 
As common in MPC(C), the system is initialized from measurements at $t=0$. 
The system state is then propagated over the horizon with the dynamics model $f(\mathbf{x},\mathbf{u})$. 
The system state vector $\mathbf{x}$ contains variables that are controlled by the algorithm (magnet intensity and position, current path progress). 
The first of the optimized inputs ($u_0$) is then applied to the physical system, transitioning the system state to $x_1$, before iteratively repeating the process to allow for correcting modeling errors. 


\begin{figure}[!t]
    \centering
    \includegraphics[width=.8\columnwidth]{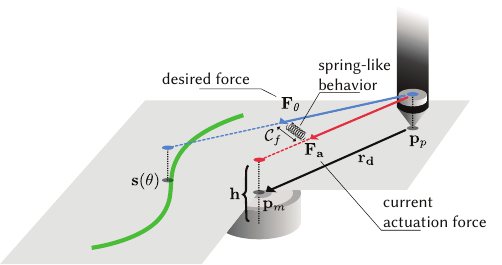} 
    \caption{Illustration of actuation force $\mathbf{F_a}$, desired force $\mathbf{F_{\theta}}$, and the force cost-term $\Cost_f$ associate with the difference between those two forces. 
    }
    \label{fig:em_model}
\end{figure}

\subsection{Haptics model: controlling the force of the electromagnet} \label{sc:em_costs}
The main goals of our approach is that users can move freely in terms of position and speed, and that the actuator continuously pulls them towards an advancing setpoint $\mathbf{s}(\theta)$ on the target trajectory $\mathbf{s}$.
At any time, the magnet exhibits an actuation force $\mathbf{F_a}$ on the pen,  given by our electromagnetic force model (see \nameref{sec:Implementation} section).
Therein lies the challenge, illustrated in \figref{fig:em_model}. 
The setpoint is continuously advancing based on the movement of the pen to ensure progress.
The actuator needs to pull the pen towards the setpoint by exhibiting force $\mathbf{F_\theta}$, but currently exhibits $\mathbf{F_a}$.
The two forces only align if the pen is exactly at the setpoint, which is rarely the case.
To overcome this challenge, we propose modeling this interaction by a spring-like behavior that ``pulls'' $\mathbf{F_a}$ towards  $\mathbf{F_\theta}$.
In this way, the magnet continuously guides the pen towards the setpoint, and the force linearly increases with distance between the pen and the target setpoint denoted as:
\begin{equation}
     \mathbf{F}_{\theta} (\Rtheta) = c \ F_0 \ \mathbf{r_{\theta}} \ \mathbf{e_{r_{\theta}}} \ . \label{eq:Fd}
\end{equation}
Here $\mathbf{e_{r_{\theta}}}$ is a unit vector in the direction of $\mathbf{r_{\theta}}$, $c$ is a scalar that regulates the stiffness of the spring (in our case $c=5/h$), $F_0$ a scaling of the EM force (\ie the force felt by users) and $h$ the distance between dipoles in $z$ (see Fig. \ref{fig:em_model}). 
Although simple, this formulation ensures that the haptic guidance is strong under large deviation from the path while vanishing as the user approaches the target path ($r_{\theta} \to 0$). 
Note that Eq. \ref{eq:Fd} is a design choice. 
Different formulations can be used to achieve different user experiences. 
Furthermore, replacing our hardware prototype and force-model would allow for adaptation of the remainder of the method to different actuation principles.

The above haptics model serves as basis for our problem formulation of electromagnetic guidance in the MPCC framework.
Using the vectors of the current actuation force $\mathbf{F_a}$ and desired force $\mathbf{F_\theta}$, we formulate a quadratic cost term to penalize the difference between desired force and actual force as:
\begin{equation}\label{eq:err_F}
    \Cost_f(\posm, \posp, \alpha) = \norm{ \ \mathbf{F}_{\theta}(\Rtheta) \ - \ \mathbf{F_a}(\mathbf{d}) \ }^2. 
    \end{equation}
where $\mathbf{d}$ is the in-plane vector between the magnet and the pen.
Since the actuation force $\mathbf{F_a}$ declines rapidly with distance $\mathbf{d}$, the gradient of $\Cost_f$ goes to 0 for large values of $\mathbf{d}$ causing the optimization to become unstable. 
To counterbalance this issue we encourage the electromagnet to stay close to the pen:
\begin{equation}
    \Cost_d(\posm,\posp) = d^2. \label{eq:err_d}
\end{equation}

Finally, we prioritize proximity between the magnet and the pen rather than increasing its force by penalizing excessive use of magnetic intensity $\alpha$:
\begin{equation}
    \Cost_{\alpha}(\alpha) = \alpha^2. \label{eq:err_alpha}
\end{equation}
\subsection{Controlling the position of the electromagnet} 
\begin{figure}[!t]
    \centering
    \includegraphics[width=.9\columnwidth]{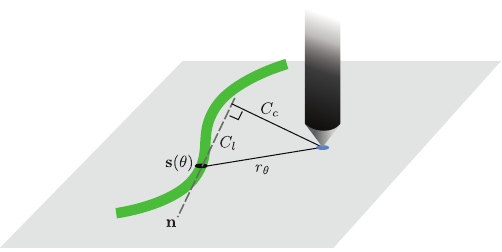}
    \caption{Illustration of lag- and contouring error decomposition.}
    \label{fig:elc}
    \vspace{-1em}
\end{figure}

We continuously optimize the position of the electromagnet with the goal of keeping the distance between the desired path and the pen minimal. 
To give the user freedom in deciding their drawing speed we first need to find the reference point $\mathbf{s}(\theta)$ on the target trajectory $\mathbf{s}$. 
Finding the closest point on the path is an optimization problem itself and hence can not be used within our optimization. 
Similar to recent work in robot trajectory generation \cite{Naegeli:2017:MultiDroneCine, Gebhardt:2018}, we decompose the distance to the closest point into a contouring and lag error, as shown in Figure~\ref{fig:elc}. 
$\Rtheta$ is the vector between the pen $\posp$ and a point $\mathbf{s}(\theta)$ on the spline, and $\mathbf{n}$ as the normalized tangent vector to the spline at that point, which is defined as $\mathbf{n} = \frac{\partial \mathbf{s} (\theta)}{\partial\theta}$.
The vector $\Rtheta$ can now be decomposed into a lag error and a contour error (\figref{fig:elc}). 
The lag-error $\Cost_l$ is computed as the projection of $\Rtheta$.
The contour-error $\Cost_c$ is the component of $\Rtheta$ orthogonal to the normal:
\begin{equation}
    \begin{aligned}\label{eq:errL_C} 
\Cost_l (\posp, \theta) &= \norm{\langle\Rtheta,\mathbf{n}\rangle}^2 , \\
\Cost_c (\posp, \theta) &= \norm{ \Rtheta - \left( \langle\Rtheta,\mathbf{n}\rangle \right) \mathbf{n} }^2.
\end{aligned}
\end{equation}
Separating lag from contouring error allows us, for example, to differentiate how we penalize a deviation from the path ($\Cost_c$), versus encouraging the user to progress ($\Cost_l$). 
We furthermore include cost terms to ensure that the magnet stays ahead of the pen ($\Cost_{\theta}$) and to encourage smooth progress ($\Cost_{\dot{\theta}}$) computed as
\begin{equation} \label{eq:err_theta}
  \begin{aligned}
\Cost_{\theta}(\theta) &= - \theta ,\\
\Cost_{\dot{\theta}}(\dot{\theta}) &= (\dot{\theta}_t-\dot{\theta}_{t-1})^2 .
\end{aligned}  
\end{equation}

\subsection{Dynamics model}
To phrase electromagnetic haptic guidance in the MPCC framework, we contribute a a dynamics model $f(\mathbf{x},\mathbf{u})$ describing the system dynamics given its states $\mathbf{x}$ and inputs $\mathbf{u}$.  
\begin{equation}
\begin{gathered}
\label{eq:model}
    \dot{\mathbf{x}} = f(\mathbf{x}, \mathbf{u}) ~\text{with}\\
    \mathbf{x} = [\mathbf{p}_{m},\dot{\mathbf{p}}_{m}, \alpha, \theta] \in \mathbb{R}^6
    ~\text{and} ~
    \mathbf{u} = [\ddot{\mathbf{p}}_{m}, \dot{\alpha}, \dot{\theta}] \in \mathbb{R}^4.
\end{gathered}
\end{equation}

The system state $\mathbf{x}$ consists of the position of the electromagnet $\mathbf{p}_{m} \in  \mathbb{R}^2$ and its velocity $\dot{\mathbf{p}}_{m}$, the magnet intensity $\alpha$ and the current path progress $\theta$.
The inputs to the system $\mathbf{u}$ consist of the in-plane electromagnet accelerations $\ddot{\mathbf{p}}_{m}$, and velocities $\dot{\alpha}$ and $\dot{\theta}$ for magnet intensity and the spline progress respectively.
Note that we empirically found that magnet accelerations yield smoother motion than using velocities. 
The system model is given by the non-linear ordinary differential equations using first and second derivatives as inputs:
\begin{equation}
  \ddot{\mathbf{p}}_{m} = v_{m}, \quad \dot{\alpha} = v_{\alpha} \quad \text{and} \quad \dot{\theta} = v_{\theta} ,  
\end{equation}
where $v_{\left(\cdot\right)}$ are the external inputs. 
The continuous dynamics model $\dot{\mathbf{x}} = f(\mathbf{x}, \mathbf{u})$ is discretized using a standard forward Euler approach: $\mathbf{x}_{t+1} = f(\mathbf{x}_t, \mathbf{u}_t)$ \cite{gibbs2011advanced}.

In our hardware implementation, we derive the sets of admissible states $\boldsymbol{\chi}$ and inputs $\boldsymbol{\zeta}$ empirically to conform to the physical hardware constraints of the linear stage (\eg max x,y-position) and EM specifications (\eg max voltage). 
These are used in the constrained optimization problem solved in Eq. \ref{eq:mpcc-formulation}.
The pen position is propagated via a standard linear Kalman filter \cite{gibbs2011advanced}. 
While not an accurate user model, it works well in practice since the states are recalculated at every timestep.


\begin{table}[!t]
	\begin{tabular}{clc}
    \toprule
    Term & Description of cost & Eq.\\
    \midrule
     $\Cost_f$  		& Decreases difference in magnetic force		& \ref{eq:err_F}
     \\
    $\Cost_d$ 		& Decreases distance between magnet and pen		& \ref{eq:err_d}
    \\
    $\Cost_{\alpha}$ 		& Encourages close distance over large force		& \ref{eq:err_alpha}
    \\
     $\Cost_l$   				& Decreases lag to path contour		& \ref{eq:errL_C} 
     \\     
     $\Cost_c$  			& Decreases distance to path contour		& \ref{eq:errL_C} 
     \\
     $\Cost_{\theta}$  				& Magnet stays ahead of pen		& \ref{eq:err_theta}
     \\
     $\Cost_{\dot{\theta}}$ 		& Ensures smooth progress		& \ref{eq:err_theta}
    \\
     \bottomrule
\end{tabular}
\caption{Summary of costs terms used in optimization.}
\label{tab:costs}
\end{table}

\subsection{Optimization}
We combine the cost terms (Table \ref{tab:costs}) to control the force and position of the actuator to form the final stage cost:
\begin{align}\label{eq:J_k}
J_k= \quad 
     & w_f \Cost_f(\mathbf{p}_{m,k}, \mathbf{p}_{p,k}, \alpha_k, \theta_k) + \nonumber \\
     & w_d \Cost_d(\mathbf{p}_{m,k}, \mathbf{p}_{p,k}) + w_\alpha \Cost_{\alpha}(\alpha_k)+ \nonumber \\
	 &	 w_l \Cost_l(\mathbf{p}_{p,k}, \theta_k) +  w_c \Cost_c(\mathbf{p}_{p,k}, \theta_k) + \nonumber \\
     & w_{\theta} \Cost_{\theta}(\theta_k) +  w_{\dot{\theta}} \Cost_{\dot{\theta}}(\dot{\theta}_k),
\end{align}
where the scalar weights $w_l,w_c,w_{\theta},w_{\dot{\theta}},w_f,w_d, w_{\alpha}>0$ control the influence of the different cost terms. 
The values used in our experiments and applications can be found in the \nameref{sec:Implementation} section.
The system states and inputs are computed by solving the $N$-step finite horizon constrained non-linear optimization problem at time instance $t$. 

The final objective therefore is:
\begin{align}
\label{eq:mpcc-formulation}
\underset{\mathbf{x}, \mathbf{u}, \theta}{\text{minimize}}\quad & \sum_{k=0}^{N} w_k\left ( J_k + \mathbf{u}_k^T \mathbf{R} \mathbf{u_k} \right ) && \\
\text{Subject to:}\quad & \mathbf{x} _{k+1} = f(\mathbf{x_k}, \mathbf{u_k}) & \text{(System Model)} \nonumber\\
                        & \mathbf{x}_0 = \hat{\mathbf{x}}(t) & \text{(Initial State)} \nonumber \\
                        & \theta_0 = \hat{\theta}(t) & \text{(Initial Progress)} \nonumber \\
                        & \theta_{k+1} = \theta_k + \dot{\theta}_k dt & \text{(Progress along path)} \nonumber \\
                        & 0 \leq \theta_k \leq L& \text{(Path Length)} \nonumber \\
                        & \mathbf{x}_k \in \boldsymbol{\chi} & \text{(State Constraints)} \nonumber \\
                        & \mathbf{u}_k \in \boldsymbol{\zeta} & \text{(Input Constraints)} \nonumber
\end{align}

Here $k$ indicates the horizon stage and the additional weight $w_k$ reduces over the horizon, so that the current timestep has more importance than later timesteps. 
$\mathbf{R}\in\mathbb{S}_+^{n_u}$ is a positive definite penalty matrix avoiding excessive use of the control inputs. 
In our implementation we use a horizon length of $N=10$. 
Experimentally we found that this is sufficient to yield robust solutions to problem instances and longer horizons did not improve results, yet linearly increases computation time. 

\section{Implementation} \label{sec:Implementation}
In this section, we detail our electromagnetic force model used in our optimal control scheme as well as the implementation of our hardware prototype.

\subsection{Electromagnetic force model}\label{sc:em_model}
Our approach requires a model of the interaction between a variable-strength electromagnet (EM) and the permanent magnet in the stylus that is sufficiently accurate and can be evaluated in real time.
Accurately modeling the non-linear EM field of the electromagnet's core is typically done through finite element analysis (FEA), which cannot be performed in real time.
Similarly, precomputing the volumetric map of the EM field $\mathbf{B_m}$ via FEA for all levels of electrical current is not computationally feasible.
We therefore contribute a novel fast approximate yet accurate electromagnetic model that provides a good balance between speed and accuracy to enable haptic guidance in applications such as writing or sketching.

In general, we aim at finding the actuation force on the pen $\mathbf{F_p}$, which is given by integrating over the volume of the permanent magnet in the pen:
\begin{equation}
    \mathbf{F_p} = \iiint \nabla \left( \mathbf{M_p} \cdot  \mathbf{B_m}(\cdot)\right) dxdydz , \label{eq:gradB2}
\end{equation}

\noindent where $\mathbf{M_p}$ is the magnetization of pen magnet and $\mathbf{B_m(\cdot)}$ is the EM field evaluated at the pen position, which is too costly to evaluate in real time.
Our model approximates this underlying physical phenomena, can be efficiently evaluated at every iteration of our optimization procedure and provides a very good fit to empirical data.
In this section, we briefly discuss the most important aspects of our model, for a full derivation and analysis we refer readers to the Appendix \ref{ap:em_model}.

We make the following two assumptions in our derivation:
\begin{inparaenum}[1)]
    \item the electromagnet and the permanent magnet can be approximated as dipoles (\ie oriented point magnets), and
    \item for the smaller dipole (the permanent magnet in the pen) the out-of-plane vector component is much larger than the in-plane counterpart. This allows us to use only the vertical component in the calculation of the force.
\end{inparaenum}

\begin{figure}[!t]
    \centering
    \includegraphics[width=\columnwidth]{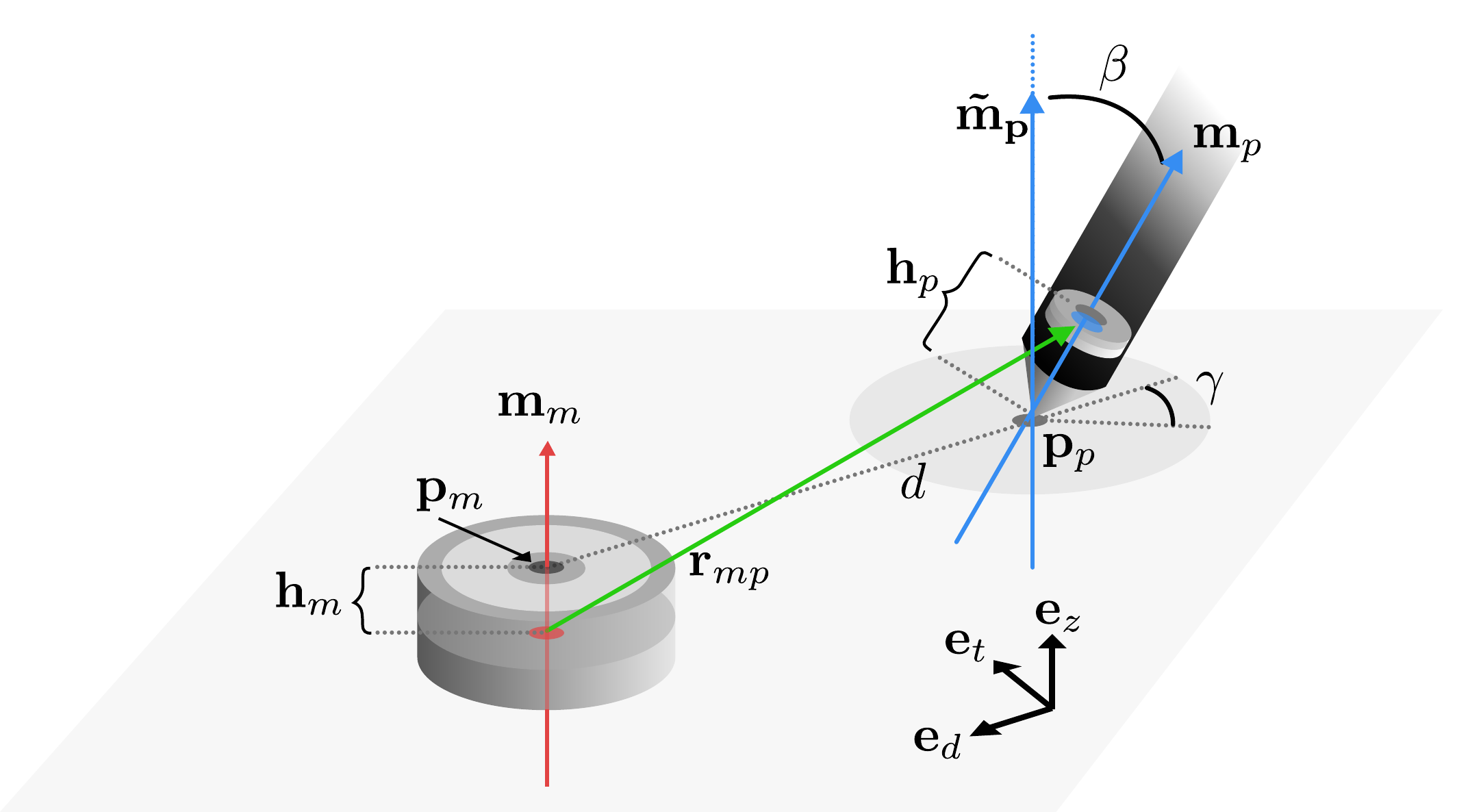} \\
    \caption{Illustration of the model to compute the force $\mathbf{F_p}$ on dipole $\mpBold$ due to dipole $\mmBold$. }
    \label{fig:dipole_dipole}
    \vspace{-1em}
\end{figure}

The first assumptions allows us to use the standard model by Yung~\etal~\cite{yung1998analytic} to compute the force exerted by the electromagnet $\mmBold$ onto the pen $\mpBold$ (see \figref{fig:dipole_dipole}) as:
\begin{multline}
   \mathbf{F_p} = {\dfrac  {3\mu _{0}}{4\pi \Rmagtopen^{5}}}
   \left [ \left(\langle\mpBold,\RmagtopenBold\rangle \right) \mmBold + 
   \left(\langle\mmBold,\RmagtopenBold\rangle\right) \mpBold \right . +
   \\
   \left(\langle\mpBold,\mmBold\rangle\right) \RmagtopenBold - 
    \left . {\dfrac{5\left(\langle\mpBold,\RmagtopenBold\rangle\right)
    \left(\langle\mmBold,\RmagtopenBold\rangle\right)}{\Rmagtopen^{2}}} \RmagtopenBold \right ] \ , \label{eq:F21-dip}
\end{multline}
where $\mu_0$ is a constant (vacuum permeability $4\pi \ 10^{-7}$ [H/m]) and  $\RmagtopenBold$ is the 3D vector between the centers of the electromagnet and pen dipoles.
In contrast to FEA, this expression is analytic and differentiable, thus suited for iterative optimization.
\figref{fig:dipole_dipole} shows all quantities needed to compute the total magnetic force exerted on the pen.
The expression does, however, lead to an actuation force $\mathbf{F_p}$ that depends on the tilt of the pen.
In pre-tests, we found that users can not perceive a difference in strength when tilting the pen in-place.
We therefore leverage our second assumption, which reduces the EM model from 6 DOF to 3 DOF, to avoid this computation. 

Based on the second assumption, we can retrieve the two vertical force vectors of the electromagnet $\mathbf{m_m}$ and the pen $\mpBoldt$.
The vector between the two centers can now be computed as $\RmagtopenBold$.
We then project this vector onto the plane, yielding the final vector $d$ between the pen tip and the in-plane projection of the actuator dipole.
The total force acting on the pen (Eq. \ref{eq:F21-dip}) can now be decomposed as:
\begin{equation}
    \mathbf{F_p} = F_a \ \mathbf{e_d} + F_z  \ \mathbf{e_z} \ . \label{eq:Fp_decomp}
\end{equation}

Here $\mathbf{F_a} = F_a \ \mathbf{e_d}$ represents the in-plane quantity we seek to control, as it is the magnitude $F_a$ of the force vector $\mathbf{F_a}$ in the direction of a unit vector $\mathbf{e_d}$ along $d$.
$F_z$ is the vertical force components which pulls the pen downwards. 
During our experiments there was no significant change in perceived friction when comparing the drawings with and without electromagnet (\ie with or without $F_z$). 
For this reason we do not actively account for $F_z$ in our optimization and only maintain the in-plane force contribution ($\mathbf{e_d}$ direction).
The actuation force as function of pen-magnet separation is obtained as:
\begin{equation}
    \mathbf{F_a} = \alpha \ F_0 \ \left( \frac{d \left(4 - \frac{d^2}{h^2}\right)}{h \left(1 + \frac{d^2}{h^2}\right)^\frac{7}{2}} \right)  \ \mathbf{e_d} , \label{eq:Fa}
\end{equation}
where  $\alpha \in \left[0,1\right]$ is a dimensionless scalar to control the desired strength of the force that should be felt by users, \add{$h$ is the center-to-center distance between both magnets projected on to the z-axis} (\figref{fig:em_model}) and $F_0$ is a constant force parameter given by the expression,
\begin{equation}
 F_0 = \frac{3 \ \mu_0 \ m_p \ m_m}{4 \ \pi \ h^4} \ . \label{eq:F0}
\end{equation}
The actuation force $F_a$ is zero if the two magnets are aligned with one another ($d=0$), it has a maximum $F_a^{max} = 0.9 \ F_0$ at $d=0.39h$, and we can assume there is no more attraction for distances $d>2h$. 
Note that the second assumption (only use in-plane component) lead only to a small approximation error
Compared to an angle dependent formulation (see Appendix \ref{sc:ap.angle-dipole}), a tilt of up to $\angt = 30^{\circ}$ leads to a max error in our model (Equation \ref{eq:Fa}) equivalent to shifting the distance $d$ by $\pm$ \unit[3]{mm}. 
This uncertainty in $d$ is comparable with the in-plane positioning error (dispersion) of our hardware prototype.
An angle dependent formulation of our model (\ie 6 DOF) can be found in Appendix \ref{sc:ap.angle-dipole} for future use in cases where the pen angle is tracked. 
This model remains valid for other hardware implementations involving a single moving electromagnet or can be easily extended onto a grid of fix electromagnets.


\subsection{Hardware prototype}
\label{sc:hardware}
We have developed one possible hardware instance that utilizes our optimization scheme for an in-plane haptic guidance system (see \figref{fig:hardware}). 
Our system consists of 3 main components: 1) a moving platform that controls the 2D location of the electromagnetic actuator, 2) an input device such as a stylus, and 3) an output device such as a digital tablet or digitizer used in combination with a non-digital drawing surface. 

\subsubsection{Motion platform}
The motion platform consists of a controllable electromagnet on a bi-axial linear stage directly underneath the drawing plane. 
The linear stage has two orthogonal \unit[12]{mm} lead screws, which are driven by two \unit[24]{V}, \unit[4.0]{A} NEMA23 high-torque stepper motors. 
We control the motors with a DQ542MA stepper driver and an Arduino UNO. 
As electromagnet, we use an Intertec ITS-MS-5030-12VDC magnet (\unit[5]{cm} diameter, \unit[3]{cm} height, \unit[12]{V}), controlled via pulse-width modulation.
It can deliver up to \unit[488]{mN} of lateral force at 11 W. 
\add{
We used FEM analysis to select this magnet from a range of commercially available magnets \cite{comsol}.
It provides a balance between power consumption, size, and force, \ie it delivers a strong perceivable force while having a small footprint relative to our hardware.
}

To measure the positional dispersion of the motion platform, we moved the electromagnet at full strength ($\alpha=1$) to 300 random locations and then always back to the center of the drawing surface. 
During theses trials, a user held the pen upright and followed the magnet passively.
By measuring the difference in target and actual position, we found that our system yields \unit[2.8]{mm} ($\pm$ \unit[0.8]{mm}) of point dispersion.
We believe this is sufficient for most applications and our experiments.
One of the factors that contribute to this dispersion is the vanishing of the actuation force $F_a$ as $d\rightarrow0$. 
This can lead to the pen motion stopping slightly before it reaches the target.

\subsubsection{Software}
Our software runs on a standard PC (Intel Core i7-4770 CPU 4 cores at \unit[3.40]{GHz}) in all our experiments. 
The solver is implemented in FORCES Pro \cite{forcespro}, which produces efficient C-code. 
The following weight values are used for our control scheme \add{(Equation \ref{eq:J_k})}:

  \begin{tabular}{ccccccccc}
    \toprule
    $w_l$&$w_c$&$w_\theta$&$w_{\dot{\theta}}$&$w_f$&$w_d$&$w_\alpha$&$w_v$&$w_m$\\
    \midrule
    1.5&1.5&10.&0.1&10.&0.05&7.&1.&1.\\
   \bottomrule
   \end{tabular}
  \vspace{5pt}
  

Due to the steepness of the electromagnetic force $\mathbf{F_p}$ and the potentially fast pen motion, runtime and latency are crucial performance metrics. 
The optimization algorithm contributes to both, whereas latency is dominated by the hardware and I/O. 
The mean solve time for a problem instance is \unit[7.4]{ms} ($\pm$ \unit[3.0]{ms}). 
This can be expected to be mostly constant since we do not manipulate the system state space and the only measured input comes from the pen. 
The hardware and overall system latency adds to the solve time. 
We use a high-speed camera (\unit[1000]{fps}) to establish the motion (pen) to motion (magnet) latency. 
This yields an approximate latency of \textasciitilde{}\unit[10]{ms}. 
Given the combined latency envelope of \textasciitilde{}\unit[20]{ms}, we did not experience any abrupt pen snapping in our experiments.

\subsubsection{Input and Output Devices}
Our primary input and output devices for our user experiments consist of a 3D printed ballpoint pen with a permanent ring magnet mounted in the shaft (see \figref{fig:hardware}) and a piece of paper. 
The strokes are recorded by a Sensel Morph pressure sensitive touch pad \cite{morph}. 
\add{If the system cannot locate the pen (\eg when it is lifted) the last known position is used.}
We chose the Sensel for its high spatial resolution (\unit[6502]{DPI}), high speed (\unit[500]{Hz}) and low latency (\unit[2]{ms}), according to specification, and since the sketching surface does not interfere with the input recognition.
Additionally, we developed an all digital input/output system with a multi-touch tablet + stylus. 
We use a Galaxy Tab A tablet with capacitive input and an off-the-shelf stylus with a permanent magnet placed near the tip. 
This magnet is slightly bigger (\unit[12]{mm} radius, \unit[12]{mm} height) to compensate for the increase in tablet thickness. 
\section{Evaluation}
We first evaluated if our optimization scheme is beneficial for users in providing haptic guidance compared to a no-feedback baseline.
In a second experiment, we compared our method with an open-loop and a closed-loop approach.

\subsection{Experiment 1 - Haptic feedback}
\begin{figure}[!t]
    \centering
    \includegraphics[width=\columnwidth]{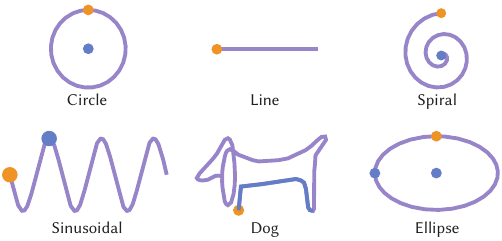}   
     \caption{Shapes of our user tests. 
    The drawing surface only contained sparse visual references (shown in blue) and starting positions (orange).}
    \label{fig:usertest_examples}
    \vspace{-1em}
\end{figure}

We compared our MPCC formulation with a no-feedback baseline to gather insights on task performance and user perception.
Users were asked to draw several shapes (see \figref{fig:usertest_examples}) and we evaluated accuracy and subjective feedback. 

\subsubsection{Procedure and tasks}
We recruited 12 participants from the local university, all without professional drawing experience. 
Users were given an introduction to the system functionality and got to experience the system in a self-timed training phase.
During the experiment we asked each participant to draw six basic shapes, each with and without haptic feedback.
The presentation order of shapes and interface condition was counterbalanced. 
The drawing surface (white piece of paper) only contained a starting point and, in the case of more complex shapes, limited additional visual guidance (see \figref{fig:usertest_examples}).
Furthermore, the participants were shown a scaled version during task execution (scaled to prevent 1:1 copying). 
After the full experiment, users completed a questionnaire on their preference.

\subsubsection*{Quantitative Results}
\label{sc:quantitative_results}
We compute the Hausdorff-like distance \cite{rockafellar2009variational} between the drawn path and the reference path as error metric. 
To make the metric robust to drawing speed, we re-sample both paths equidistantly.
To ensure fairness we also compute the inverse distance (reference to drawn path). 
A Kolmogorov-Smirnov test \cite{kolmogorov1933sulla} indicated that the set of uni-directional distances is not significantly different from the set of inverse distances.
We therefore only report uni-directional distances. 
The quantitative results for each target averaged over all participants are summarized in Table \ref{tab:accuracy}.
Our method on average resulted in a 66\% ($\pm$ 24.5\%) lower error, \ie it improve the average point-to-path difference by \unit[1.54]{mm}.
A two-way ANOVA on the mean error revealed a main effect for the feedback type (F=46.187, p<.001) and for the shapes (F=11.771, p < .001). 
Post-hoc analysis revealed that the line was statistically significantly different from all other shapes.
This indicates that our method is beneficial for non-trivial shapes.

\begin{table}[!t]
\caption{Mean accuracy ($mm$). * indicates statistical significance ($\alpha .05$).}
\vspace{.5em}
    \begin{tabular}{l|cc|cc|c}
    \multicolumn{1}{c}{} &\multicolumn{2}{c}{With}&\multicolumn{2}{c}{Without}&\multicolumn{1}{c}{}\\
    \midrule
Scenario& Mean & SD & Mean & SD & Err $\%$ \\
 \midrule
 Circle* & \textbf{2.19} & 0.90 & 4.26 & 2.39 & 0.51 \\
 Line  & 1.18 & 0.80 & \textbf{1.03} & 0.84 & 1.15  \\
 Spiral* & \textbf{2.55} & 0.75 & 4.38 & 1.64 & 0.58  \\
 Sinus*  & \textbf{2.53} & 0.70 & 5.08 & 2.19 & 0.50  \\
 Dog*  & \textbf{2.31} & 0.54 & 3.81 & 1.32 & 0.60 \\
 Ellipse*  & \textbf{2.40} & 0.56 & 3.84 & 1.22 & 0.62 \\
\end{tabular}
\label{tab:accuracy}
\end{table}

\subsubsection*{Qualitative Results}
A brief exit interview shows that users subjectively rate the system favorably, on a 5-point Likert scale (5 is best), in terms of accuracy ($4.33 \pm0.62$), speed ($4.00\pm0.91$), force ($3.50\pm0.86$) and overall performance ($4.50\pm0.9$).
While we acknowledge that this might be in part due to novelty effects, we believe that the quantitative results indicate that our system is beneficial for users in general.
The ratings indicate that participants generally see benefit in our approach and are not disturbed or hindered when using our approach.


\subsection{Experiment 2: Comparison of control strategies}
\label{Sc:preliminary_user_evaluation}
In this second experiment, we compared our time-free closed-loop optimization strategy to a simpler MPC variant and our implementation of dePENd \cite{yamaoka2013depend}, denoted as dePENd$^{*}$. 

\subsubsection{Procedure and tasks}
We asked twelve new participants (students and staff from a local university) to draw one complex shape (dog in \figref{fig:usertest_examples}) in three different conditions: \emph{dePENd$^{*}$}, time-dependent closed loop (\emph{MPC}), and time-free closed loop (\emph{Ours}), counterbalanced using a latin square. 
\add{The speed of the system in the time-dependent cases was decided empirically based on pre-tests to work well at regular drawing speeds.}
After receiving instructions and a brief training, participants completed the three drawings.
Participants were also encouraged to provide comments during the individual conditions.

\subsubsection{Data collection}
We analyze three measures: 
1) the mean distance from the pen to the path, 2) the mean distance from the pen position projected onto the path and the setpoint along the path, denoted as $d(pen, \mathbf{s(\theta)})$, and 3) the mean distance from the pen to the electromagnet. 
By taking the mean of the error terms over subjects we ensured equal numbers of datapoints, accounting for differences in speed. 
Participants were instructed to draw at roughly constant speed.
This was done to ensure fairness in comparing our approach with the open-loop approach, which would deteriorate if the variability of the drawing speed were to high.
Note that our approach does generally not require this assumption.
%


\subsubsection{Quantitative results}

\begin{table}[!t]
    \centering
    \caption{Mean distances in $mm$ for Experiment 2). }
    \begin{tabular}{l|ccc}
         &|pen-path|& d(pen, $s(\theta)$) & |pen-em| \\
         \midrule
         \emph{dePENd$^{*}$} & $4.1(\pm 0.7)$ & $38.0(\pm 56.9)$  &$38.2(\pm 25.1)$ \\ 
         MPC & $3.9(\pm 1.3)$& $45.0(\pm 50.8)$ & $8.6(\pm 1.6)$ \\ 
         \textit{Ours} & \textbf{2.0}$(\pm 0.6 )$& \textbf{6.2}$(\pm 0.8)$ & \textbf{4.6}$(\pm 0.9)$\ 
    \end{tabular}
    \label{tab:strategy_results}
\end{table}

Table \ref{tab:strategy_results} summarizes our quantitative findings. 
Not surprisingly, the distance from the electromagnet to the pen and  $d(pen, \mathbf{s(\theta)})$ for \emph{dePENd$^{*}$} is quite large. 
Since the force exerted on the pen falls-off quadratically with distance, participants often lost any haptic guidance early on, confirmed via user comments such as ``I don't feel anything'' (P3) and ``Is the system on?'' (P6). 

\begin{figure}[!t]%
	\centering
    \includegraphics[width=1\columnwidth]{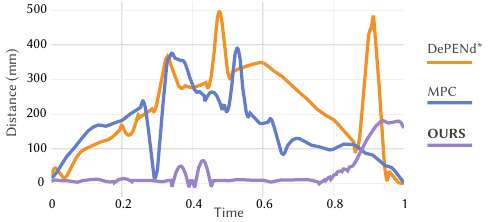}%
    \caption{Comparison of error (path distance pen-$\mathbf{s}(\theta)$) over time for a single participant (P1). The inverse u-shape illustrates that the setpoint $\mathbf{s}(\theta$) moves at a different speed than the user for \emph{dePENd$^{*}$} and MPC. The data is smoothed to increase readability.} 
    \label{fig:single_user_control}
    \vspace{-1em}
\end{figure}

A Kruskal-Wallis test revealed that our approach has the highest accuracy compared to \emph{dePENd$^{*}$} and MPC (H(2)=20.76, p<.001).
Furthermore, the setpoint $\mathbf{s}(\theta)$ (H(2)=7.362, p<.05) and the electromagnet (H(2)=27.12, p <.001) are closest to the pen using our approach. 
Thus our time-free formulation overcomes both problems of wrong setpoints (\emph{MPC}) and a run-away electromagnet (\emph{dePENd$^{*}$}).
 Figure \ref{fig:single_user_control} shows one typical example of a user. 
 Both the distance along the path and the pen-magnet distance fluctuate strongly for \emph{dePENd$^{*}$} and  \emph{MPC} control strategies.
 Our approach yielded a continuously low error.

While \emph{MPC} reduces the distance from the pen to the magnet, it does not optimize for the progress along the path and hence may pull the pen into undesired directions. 
Furthermore, we saw that \emph{MPC} produced extreme corner cutting behavior to catch up to the advancing setpoint.
Both \emph{dePENd$^{*}$} and \emph{MPC} also produce results with high standard deviations.
This is likely due to the absence of direct coupling between user feedback and path progress, which makes it possible for the user to lag behind the setpoint significantly (albeit at the cost of reduced force feedback). 
In our approach, the path progress is adjusted to the user's drawing speed, drastically reducing the standard deviation and in consequence ensuring delivery of force feedback throughout the drawn path. 
 
\subsubsection{Qualitative results} 
From our observations we saw that $\mathbf{s}(\theta)$ was either in front or behind the user for MPC. 
This was also confirmed in our interview, where people especially commented on the MPC strategy: ``The system tries to push me in the wrong direction'' (P2) and ``It is counteracting me'' (P11). 
In contrast with our formulation the magnet remains always slightly ahead of the pen, resulting in users commenting on our approach as the most preferred condition. 
In the words of one subject this is: ``since I still had control'' (P9). 

In summary, theses initial results indicate that our approach outperforms existing open-loop and time-dependent closed-loop approaches. 
\emph{dePENd$^{*}$} causes numerous problems, including users not perceiving any haptic feedback. 
This is especially troublesome in settings where autonomy is desired. 
In MPC the haptic feedback is perceived, but can be erroneous. 
This is especially evident when users do not conform to the expected behavior. 
We plan to perform more in-depth experiments to investigate, for which applications our approach can be especially beneficial, and for which levels of autonomy.


\def\dir{chapters/05_shared_control/rl}
\chapter{Multi-Agent Reinforcement Learning for Goal-Agnostic Adaptive User
Interfaces}
\chaptermark{Multi-Agent Reinforcement Learning for AUIs}
\label{ch:control:multi}

\contribution{
In the previous chapter, we demonstrated the use of model predictive control to determine the actuation of a haptic feedback device. However, it relied on known system dynamics, including predictive models of user behavior. One way to overcome this limitation is through learned system, task, and user dynamics. In this chapter we investigate model-free reinforcement learning for such purpose. Furthermore, we switch away from haptics as use case and focus on adaptive point-and-click interfaces. This switch allows us to focus on the control, as the actuation and sensing in graphical user interfaces are deterministic.
A core challenge in developing adaptive interfaces is inferring user intent and choosing adaptations accordingly. Current methods often depend on tediously hand-crafted rules or extensively gathered user data. Furthermore, heuristics need to be recrafted and data regathered for every new task and interface.
To address this issue, we formulate interface adaptation as a multi-agent reinforcement learning problem. Our approach learns adaptation policies without relying on heuristics or real user data, enabling the development of adaptive interfaces across various tasks with minimal adjustments.
In our formulation, a \useragent mimics a real user and learns to interact with an interface via point-and-click actions. \del{Simultaneously, an \interfaceagent learns interface adaptations, by observing the \useragent's behavior, to maximize the \useragent's efficiency.} Simultaneously, an \interfaceagent learns interface adaptations, to maximize the \useragent's efficiency, by observing the \useragent's behavior. 
For evaluation, we replaced the simulated \useragent with actual users. Our study involved twelve participants and focused on automatic toolbar item assignment. Results demonstrate that the policies developed in simulation effectively apply to real users. These users were able to complete tasks with fewer actions and in similar times compared to methods trained with real data. "Additionally, we showcased the method's efficiency and generalizability across four different interfaces and tasks.
}

\begin{figure}
    \centering
    \includegraphics[width=\textwidth]{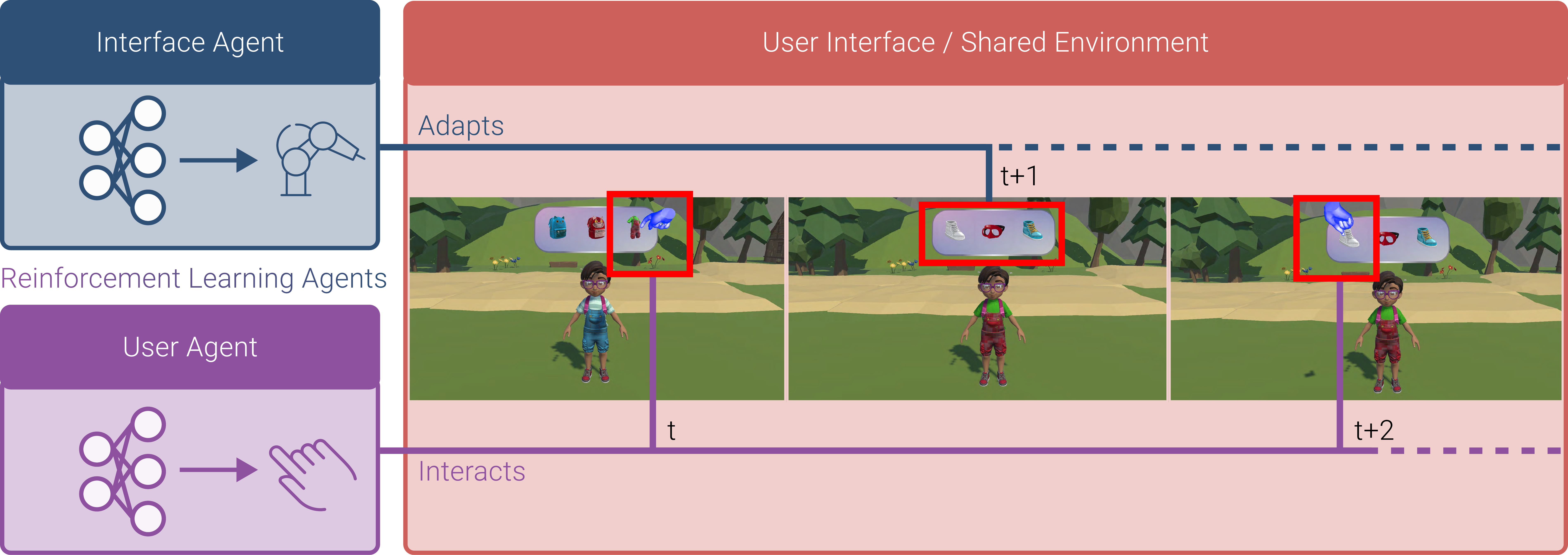}
    \caption{ 
     We formulate online user interface adaptation as a multi-agent reinforcement learning problem. Our approach comprises a user- and interface agent. The \useragent interacts with an application in order to reach a goal and the \interfaceagent learns to assist it. In the depicted example the \useragent interacts with a Virtual Reality toolbar, while the \interfaceagent assigns relevant items for the \useragent. The \interfaceagent does not know the goal of the \useragent. Crucially, our approach does not rely on labeled offline data or application-specific handcrafted heuristics.}
    \label{fig:teaser_rl}
\end{figure}

\section{Framework Overview}

\begin{figure*}[!t]
    \centering
    \includegraphics[width=0.9\textwidth]{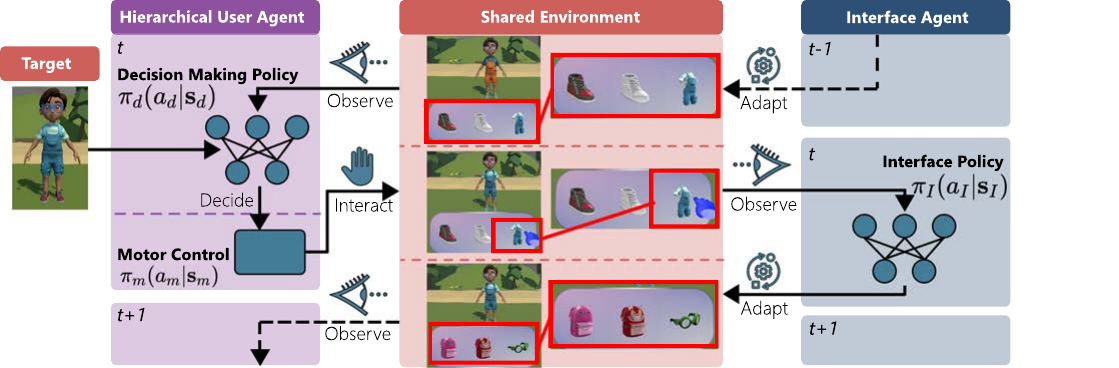}
    \caption{Our \interfaceagent and \useragent act in the same environment. The \useragent is modeled as a two-level hierarchy with a high-level decision-making policy $\policy_d$ and a low-level motor control policy $\policy_m$. The \useragent interacts with the UI. The high-level policy of the \useragent observes that state of the environment (\Eq{sd}) and chooses a specific menu slot as target accordingly (\Eq{ad}). The lower level receives this action and computes a movement (\Sec{ll}). The \interfaceagent policy $\policy_I$ adapts the interface to assist the \useragent in achieving its task more efficiently. It observes user actions in the UI (\Eq{si}) and decides on adaptations. Note that the \interfaceagent cannot access the goal of the user, making the problem partially observable.}
    \label{fig:overview}
\end{figure*}

\add{We define ''adaptation'' as the online alteration of an interface given current and past user input to support users in completing their task more efficiently. As such AUIs dynamically assign relevant interactive elements to an interface, thereby decreasing number of decisions and actions users need to take.}
MARLUI is a framework to create AUIs \del{that dynamically assign the most relevant interactive elements to a point-and-click interface.
MARLUI} and can learn adaptive policies, \add{that is neural networks that given an input predict suitable adaptations}. \add{MARLUI learns} without needing data and only necessitates minimal adjustments to learn these policies for different AUIs. \add{We specifically, assume the users to be Computational Rational \cite{oulasvirta2022computational} to make the problem tractable.}
In this section, we provide a high-level overview of the workings of our proposed framework.
As shown in \Fig{overview}, MARLUI consists of three main components: a user interface as a shared environment \add{(which gets adapted)},  a \useragent \add{(which learns to interact with the AUI)}, and an \interfaceagent that learns an adaptation policy. 
Referencing the game character customization example illustrated in \Fig{teaser_rl}, we first describe the function of each component individually, followed by an explanation of how MARLUI can support real users.

\paragraph*{User Interface / Shared Environment}
Our approach models UI adaptation as a MARL problem, where an \interfaceagent learns from a simulated \useragent's interactions with a user interface, i.e., the shared environment. In this shared environment, user- and \interfaceagent take actions in a turn-based manner. The \useragent performs a point-and-click maneuver, which changes the state of the UI, e.g., altering the clothing of the game character. The \interfaceagent observes the action and the change to the interface and based on that selects a new subset of items to display, e.g., adjusting the clothing items displayed in the toolbar.
In turn, the \useragent observes this adaptation and takes a new action. This cycle continues till the task is complete, e.g., the desired clothing configuration is reached. 

\paragraph*{User Agent}
The \useragent aims to achieve its goal as fast as possible. In the context of game character creation, goals might involve selecting specific attributes for a character, such as the color of a shirt or backpack. The \useragent can observe the visible parts of the interface, and knows its internal state. Based on these information, the \useragent acts on the interface with point-and-click maneuvers, aiming to align the current state of the UI with its goal, e.g., aligning the current- with the desired clothing configuration. 
By constraining its motor behavior in a physiologically plausible manner, the \useragent is bound to exhibit human-like behavior for pointing and selecting items.
Through trial and error, the \useragent will eventually learn a policy that allows it to realize this alignment.

\paragraph*{Interface Agent}
The \interfaceagent adapts the UI in a turn-based manner to minimize the number of actions the \useragent must perform to complete a task. Despite not knowing the \useragent's specific goal, it learns the task structure \add{(i.e., sequence of states)} through observing the \useragent's interactions with the UI. 
It can then learn to select the subset of items that are most relevant to the \useragent at its current state, e.g., dynamically populating the toolbar with the cloting items that are most likely to be picked.
Through trial and error, it can learn UI adaptation policies without relying on pre-collected user data or predefined heuristics.



\paragraph*{Interaction with Real Users}
By mimicking human-like point-and-click behavior through the \useragent, the \interfaceagent can learn to adapt UIs such that it also assist real users in accomplishing the same task.
To apply the learned adaptive interface to real users, the setting is changed such that \interfaceagent interacts with the actual users instead of the \useragent. 
In a turn-based fashion, it selects the most relevant next items after each click or selection of the actual user according to what it has learned through interactions with the \useragent.

    

\section{Method}
We first present an outline the model of our \useragent, consisting of a high- and low-level policy. Then we present the \interfaceagent (\Fig{overview}).

\subsection{General Task Description}
We model tasks to be completed if the \useragent achieves their desired goal. For game character creation, a goal can be the desired configuration of a character with a certain shirt (red, green, blue), and backpack (pink, red, blue). We represent the goal as a one-hot vector encoding $\gattr$ of these attributes. A one-hot vector can be denoted as  $\mathbb{Z}_{2}^{j}$, where $j$ is the number of items. For the previous example, $\gattr$ would be in $\mathbb{Z}_2^6$ as it possesses six distinct items. 

Furthermore, the \useragent can access an input observation denoted by $\tools$, e.g., this can correspond to the current character configuration. The current input observation, $\tools$, and the goal state $\gattr$ are identical in dimension and type. 

The \useragent interacts with the interface and attempts to match the input observation and goal state as fast as possible, such that $\tools = \gattr$. Each interaction updates $\tools$ accordingly, and a trial terminates once they are identical. In the character creation example, this would be the case if the shirt and backpack of the edited character are the same as the desired configuration. The \interfaceagent makes online adaptations to the interface. It does \emph{not} know the specific goal of a user. Instead, it needs to observe user interactions with the interface to learn the underlying task structure that will yield the optimal adaptations, e.g., the user likely wants to configure the shoes after configuring the backpack.

\add{The \interfaceagent learns to adapt the UI to the \useragent by maximizing the same expected discounted reward. Specifically, the \interfaceagent learns to infer optimal next adaptations over an infinite horizon by updating implicit probabilities of likely next actions of the \useragent, given its current sequence of past actions. As such it does not learn an explicit or implicit goal probability distribution, but a distribution of the most likely next actions of the \useragent. An example is to suggest a white and blue shoe to the \useragent in the tool, as the \useragent has not interacted with the category of shoes thus far (\Fig{fig:teaser_rl}). }

\subsection{User Agent}
\label{sec:user_agent}

The \useragent interacts with an environment to achieve a certain goal (e.g., select the intended attributes of a character). The agent tries to accomplish this as fast and accurately as possible, hence it minimizes task completion time. Thus, the \useragent first has to compare the goal state and input observation and then plan movements to reach the target. We model the \useragent as a hierarchical agent with separate policies for a two-level hierarchy \cite{Langerak:2021:Generalizing}. First, we introduce a high-level decision-making policy $\policy_d$ that computes a target for the agent (high-level decision-making), we approximate visual cost with the help of existing literature \cite{10.1145/1240624.1240723}. Second, a WHo Model Fitts'-Law-based low-level motor policy $\policy_m$ that decides on a strategy to reach this target. We now explain both policies in more detail.

\subsubsection{High-level Decision-Making Policy}
The high-level decision-making policy of the hierarchy is responsible to select the next target item in the interface. The overall goal of the policy is to complete a given task while being as fast as possible. Its actions are based on the current observation of the interface, the goal state, and the agent's current state. More specifically, the high-level state space $\StatePerPolicy_d$ is defined as:

\begin{equation}
    \StatePerPolicy_d = \left (\pos, \menu, \tools, \gattr \right ),
    \label{eq:sd}
\end{equation}
which comprises: i) the current position of the \useragent's end-effector normalized by the size of the UI, $\pos \in I^n$ (where $n$ denotes the dimensions, e.g., 2D vs 3D), ii) an encoding of the assignment of each item $\menu \in \mathbb{Z}_2^{\nitems \times \nslots}$, with $\nitems$ and $\nslots$ being the number of menu items and environment locations, respectively, iii) the current input state $\tools \in \mathbb{Z}_2^{\nitems}$, and iv) the goal state $\gattr \in \mathbb{Z}_2^{\nitems}$. Here, $I$ denotes the unit interval $[0,1]$, and 
$\mathbb{Z}_{2}^{n}$ is the previously described set of integers.
The item-location encoding $m$ represents the current state of a UI. It can be used, for instance, to model item-to-slot assignments. The action space $\ActionPerPolicy_D$ is defined as:
\begin{equation}
    \ActionPerPolicy_d = \target,
    \label{eq:ad}
\end{equation}
which indicates the next target slot $\target \in \mathbb{N}_{\nslots}$. The reward for the high-level decision-making policy consists of two weighted terms to trade-off between task completion accuracy and task completion time: i) how different the current input observation $\tools$ is from the goal state $\gattr$, and ii) the time it takes to execute an action. Therefore, the high-level policy needs to learn how items correlate with the task goal as well as how to interact with any given interface. With this, we define the reward as follows: 

\begin{equation}
    \RewardPerPolicy_d =  \alpha \underbrace{\error_{gd}}_{i)} - (1-\alpha)\underbrace{\left(\dect + \mt\right)}_{ii)} + \mathbbm{1}_{\text{success}},
    \label{eq:rd}
\end{equation}
where $\error_{gd}$ is the difference between the input observation and the goal state, $\alpha$ a weight term, $\mt$  the movement time as an output of the low-level policy, $\dect$ the decision time, and $\mathbbm{1}_{\text{success}}$ an indicator function that is 1 if the task has been successfully completed and 0 otherwise. 

In addition to movement time, we also need to determine the decision time $\dect$. To this end, we are \addiui{inspired by} the SDP model \cite{10.1145/1240624.1240723}. This model interpolates between an \addiui{approximated} linear visual search-time component \del{($T_s$)} and the Hick-Hyman decision time \cite{hick1952rate} \del{($T_{hh}$)}, both functions take into account the number of menu items and user parameters. 

We define the difference $\error_{gd}$ between the input observation $\tools$ and the goal state $\gattr$ as the number of mismatched attributes:
\begin{equation}
    \error_{gd} = - \sum_{x \in \gattr, y \in \tools }\frac{\mathbbm{1}_{x \neq y}}{n_{attr}},
\end{equation}
where $\mathbbm{1}$ is an indicator function that is $1$ if $x\neq y$ and else $0$,  $x$ and $y$ are individual entries in the vectors $\gattr$ and $\tools$ respectively, and $n_{attr}$ is the number of attributes (e.g., shirt, backpack, and glasses).

\subsubsection{Low-Level Motor Control Policy}
\label{sec:ll}
The low-level motor control policy is a non-learned controller for the end-effector movement. In particular, given a target, it selects the parameters of an endpoint distribution (mean $\mu_\pos$ and standard deviation $\sigma_{\pos}$) . We set $\mu_\pos$ to the center of the target. The target $\target$ is the action of the higher-level decision-making policy ($\ActionPerPolicy_D$). 
Following empirical results \cite{fitts1954information}, we set $\sigma_{\pos}$ to 1/6th of a menu slot width to reach a hitrate of 96\%.

Given the current position and the endpoint parameters (mean and standard deviation), we compute the predicted movement time using the Fitts' Law derived WHo Model \cite{guiard2015mathematical}.
\begin{equation}
        \mt = \left ( \frac{k}{(\sigma_{\pos}/d_{\pos}-y_0)^{1-\beta}} \right ) ^{1/\beta} + \mt^{(0)},
\end{equation}
where $k$ and $\beta$ are parameters that describe a group of users, $\mt^{(0)}$ is the minimal movement time, and $y_0$ is equal to the minimum standard deviation. The term $d_{\pos}$ indicates the traveled distance from the current position to the new target position $\mu_\pos$. We follow literature for the values of other parameters \cite{guiard2015mathematical, jokinen2021touchscreen}. We sample a new position from a normal distribution: $\pos \sim \mathcal{N}\left(\mu_{\pos}, \sigma_{\pos}\right)$.

\subsection{Interface Agent}
The \interfaceagent makes discrete changes to the UI to maximize the performance of the \useragent. 
In our running example of character customization, it assigns items to a toolbar to simplify their selection for the \useragent.
Unlike the \useragent, we model the \interfaceagent as a flat RL policy. The state space $\StatePerPolicy_I$ of the interface agent is defined as:

\begin{equation}
    \StatePerPolicy_I = \left (\pos, \tools, \menu, \stack \right ),
    \label{eq:si}
\end{equation}

which includes: i) the position of the user $\pos \in I^2$, ii) the input observation $\tools \in \mathbb{Z}_2^{\nitems}$, iii) the current state of the UI $\menu \in \mathbb{Z}_2^{\nitems \times \nslots}$, and iv) a vector including the history of interface elements the \useragent interacted with (commonly referred to as stacking). The action space $\ActionPerPolicy_I \in \mathbb{Z}$ and its dimensionality is application-specific. 
The goal of the \interfaceagent is to support the \useragent. Therefore, the reward of the \interfaceagent is directly coupled to the performance of the \useragent. We define the reward of the \interfaceagent to be the reward of the \useragent's high-level policy:

\begin{equation}
    \RewardPerPolicy_{I} = \RewardPerPolicy_{D}.
    \label{eqn:ri}
\end{equation}

Note that the \interfaceagent does \emph{not} have access to the \useragent's goal $\gattr$ or target $\target$. To accomplish its task, the \interfaceagent needs to learn to help the \useragent based on an implicit understanding of i) the objective of the \useragent, and ii) the underlying task structure. Our setting allows the \interfaceagent to gain this understanding solely by observing the changes in the interface as the result of the \useragent's actions. This makes the problem more challenging but also more realistic. 

The \interfaceagent learns an implicit distribution of possible goals, and by observing the \useragent it narrows down the distribution over goals. At every time step the \interfaceagent suggests the items that are most likely needed, given the goal distribution. 
\section{Implementation}
We train the user and \interfaceagent's policies simultaneously in a shared environment (the AUI). All policies receive an independent reward, and the actions of the policies influence a shared environment. We execute actions in the following order: (1) the \interfaceagent's action, (2) the \useragent's high-level action, followed by (3) the \useragent's low-level motor action. The reward for the two learned policies is computed after the low-level motor action has been executed. The episode is terminated when the \useragent has either completed the task or exceeded a time limit.

We implement our framework in Python 3.8 using RLLIB \cite{liang2018rllib} and Gym \cite{brockman2016openai}. We use PPO \add{\cite{schulman2017proximal, yu2022surprising}} to train our policies. We use 3 cores on an Intel(R) Xeon(R) CPU @ 2.60GHz during training \add{and an NVIDIA TITAN Xp GPU}. Training takes $\sim$36 hours. \del{We utilize an NVIDIA TITAN Xp GPU for training.} The \useragent's high-level decision-making policy $\policy_d$ is a 3-layer MLP with 512 neurons per layer and ReLU activation functions. The  \interfaceagent's policy  $\policy_I$ is a two-layer network with 256 neurons per layer and ReLU activation functions. \addiui{We sample the full state initialization (including goal) from a uniform distribution. We use stochastic sampling for our exploration-exploitation trade-off.} 

We use curriculum learning to increase the task difficulty and improve learnability. Specifically, we adjust the difficulty level every time a criteria has been met by increasing the mean number of initial attribute differences. More initial attribute differences result in longer action sequences and are therefore more complex to learn. We increase the mean by 0.01 every time the successful completion rate is above 90\% and the last level up was at least 10 epochs away.

We randomly sample the number of attribute differences from a normal distribution with standard deviation $1$, normalize the sampled number into the range $[1, n_a]$ and round it to the nearest integer, where $n_a$ is the number of attributes of a setting (in the case of game character $n_a=5$).

\del{The difference between agents of different applications is their respective state- and action spaces.}
\section{Evaluation}
\begin{figure}[!t]
    \centering
    \includegraphics[width=0.6\columnwidth]{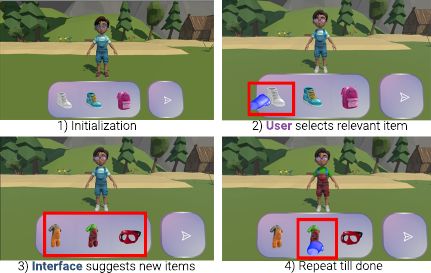}
    \caption{In our proposed task the \interfaceagent and the \useragent interact in a turn-based manner, in which the user (agent) matches a game character selection to a target state  (1). First, the user operates a toolbar with three slots (2). Second, The interface agent assigns the most relevant items to the available slots (3). Finally, This cycle continues till the two characters match (4).}
    \label{fig:task}
\end{figure}

MARLUI aims to learn UI adaptations from simulated users that can support real users in the same task. Specifically, we want our framework to produce AUIs that are competitive with baselines that require carefully collected real user data. In this section, we evaluate if our approach achieves this goal. Thus, we first conduct an \emph{in silico study} to analyze how the \interfaceagent and the \useragent solve the UI adaptation problem in simulation. Then, we conduct a \emph{user study} to investigate if policies of the \interfaceagent that were learned in simulation benefit real users in terms of number of actions needed to complete a task.

\subsection{Task \& Environment}
\label{sec:task}
To conduct the evaluation, we introduce the \emph{character-creation task} (see \Fig{task}).
In this task, a user creates a virtual reality game character by changing its attributes. A character has five distinct attributes with three items per attribute: i) shoes (red, blue, white), ii) shirt (orange, red, blue), iii) glasses (reading, goggles, diving), iv) backpack (pink, blue, red), and v) dance (hip hop, break, silly). The characters' attribute states are limited to one per attribute, i.e., the character cannot be dancing hip hop and break simultaneously. This leads to a total of 15 attribute items and 243 character configurations. We sample the goal uniformly from the different configurations. 

The game character's attributes can be changed by selecting the corresponding items in a toolbar-like menu with three slots. The user can cycle through the items by selecting "Next." The static version of the interface has all items of an attribute assigned to the three slots, and every attribute has its own page (e.g., all shoes, if the user presses next, all backpacks). The character's attribute states correspond to the current input state $\tools$ and the target state $\gattr$, where $\gattr$ is only known to the \useragent. The goal of the \interfaceagent is to reduce the number of clicks necessary to change an attribute, by assigning the relevant items to the available menu slots. For the \emph{user agent}, the higher level selects a target slot, and the lower level moves to the corresponding location. 

\subsection{In Silico}
\paragraph{Training}

\begin{figure}[!t]
    \centering
        \centering
        \includegraphics[width=\textwidth]{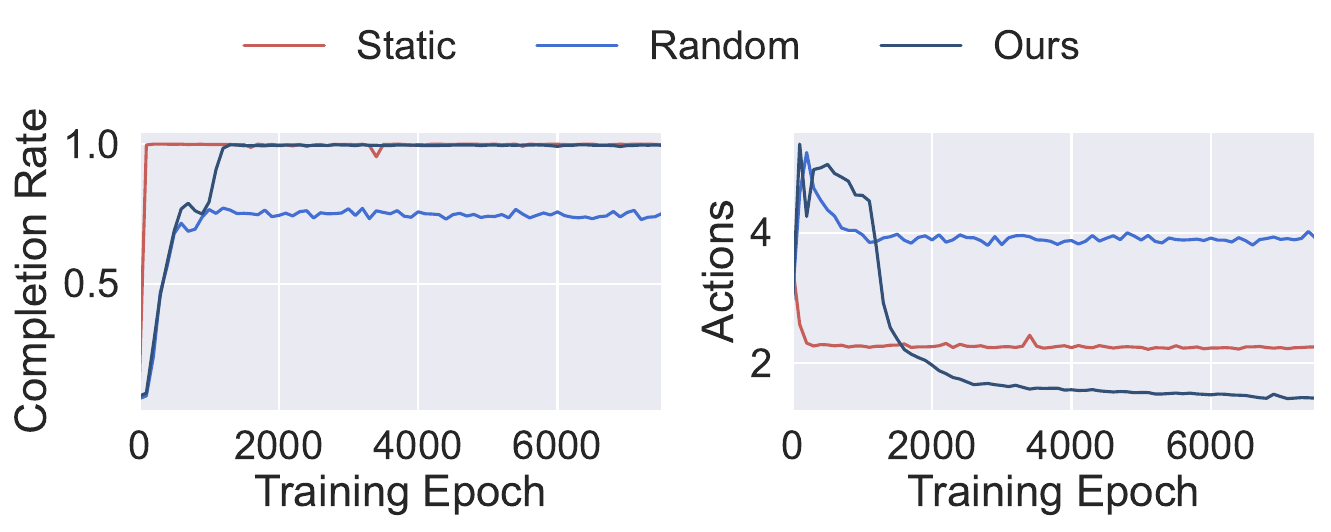} 
        \caption{We train our agent till convergence. Left: the fraction of successfully completed episodes per epoch. Ours and Static reach a 100\% successful completion rate. Random does not converge. Right: The number of actions needed on average during a successful episode. Our framework needs less actions compared to Static and Random.}
        \label{fig:training}
\end{figure}

\begin{figure}[t]{!t}
        \centering
        \includegraphics[width=0.5\textwidth]{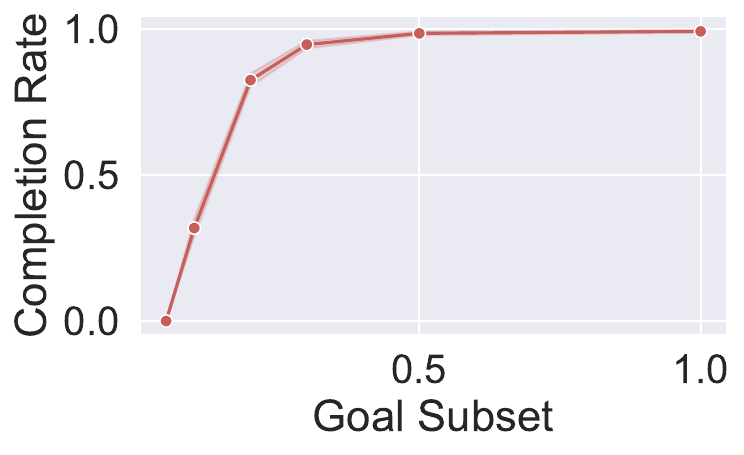}
        \caption{The fraction of successfully completed episodes as function of the fraction of the total number of goals. The graph shows that it is sufficient to see half of the goals to learn policies that generalize to all goals.}
        \label{fig:goal}
\end{figure}

We evaluate the training of our framework against a static and a random interface. In the random interface, items are randomly assigned to the slots. \add{Our method would perform on par, or worse, with random if it was incapable of learning. Hence, the random baseline provides a lower bound for performance. The static baseline has no adaptations, allowing us to evaluate the general effectiveness of our adaptive interface. For most tasks, a task-specific heuristic could be found that presents an upper baseline. However, this has to be designed specifically for each task. On the other hand, our goal is to provide a general framework that works across tasks with minimal changes to the reward function and observation spaces.}

\Figure{training} shows the \useragent's task completion rate and number of actions per task of all three interfaces during training. Ours and the static baseline converge, whereas the random baseline does not. Furthermore, the mean number of actions of ours is lower than the mean of the static interface.

\paragraph{Generalization to unseen goals} 
To understand how well our approach can generalize to unseen goals, we ablate the fraction of goals the agents have access to during training. We then evaluate the learned policies against the full set of goals, which is defined as all possible combinations of character attributes. 
The results are presented in \Figure{goal}. We find that having access to roughly half of the goals is sufficient to not impact the results. This indicates that our approach generalizes to unseen goals of the same set for this task. 

\begin{figure}[!t]
    \centering
    \includegraphics[width=\columnwidth]{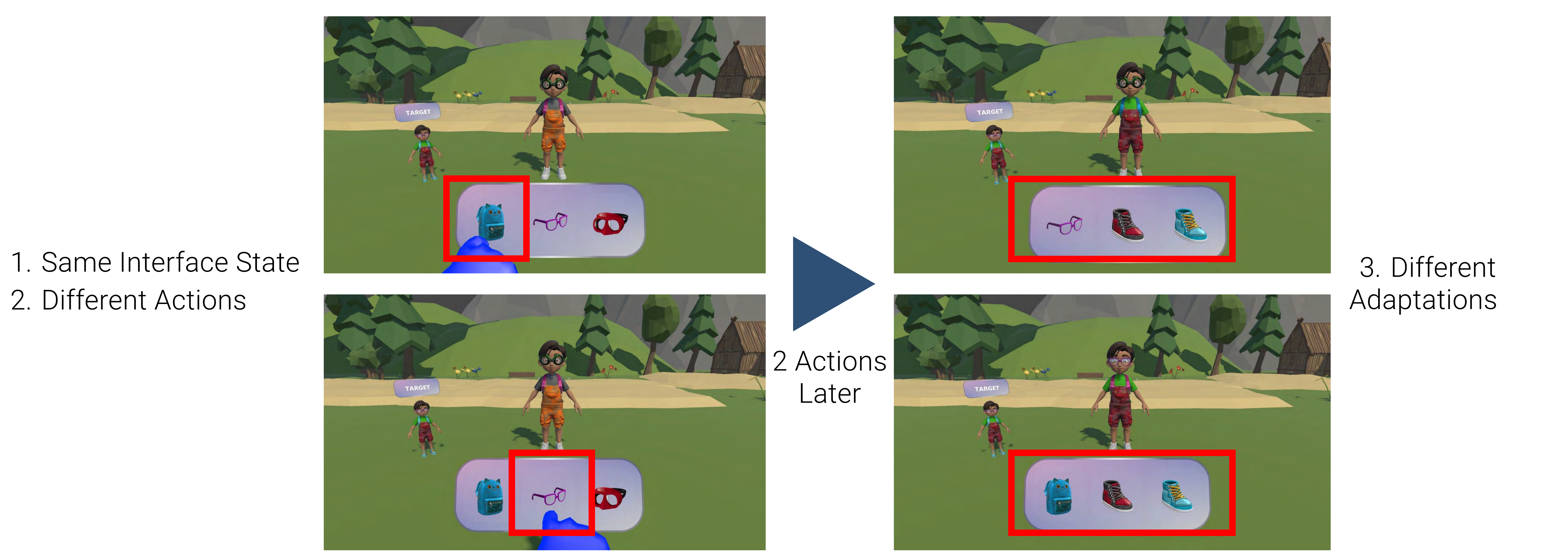}
    \caption{With our framework, multiple relevant items can be assigned simultaneously; yet the user can only select one (left). Depending on the user's action (top: select backpack, bottom: select glasses), other item gets assigned later (top: shoes, bottom: backpack). Note that the user has not selected any backpack in those two actions. This shows that our framework actively adapts to user input.}
    \label{fig:sequence}
\end{figure}

\paragraph{Understanding policy behavior}
We qualitatively analyzed the learned policies of our \interfaceagent to understand how it supports the \useragent in its task.
In \Figure{sequence}, we show a snapshot of two sequences with identical initialization. 
To reach the target character configuration, the user agent can either select the blue bag or the purple glasses (both are needed). 
Depending on which item the user selects at this time step, the \interfaceagent proposes different suggestions in subsequent steps. For instance, it might happen that the \interfaceagent suggest two relevant items; while the user can only select a single on. As example, the blue backpack the user did not select initially (\Figure{sequence}, bottom) gets suggested again later; as the user has not selected a single backpack during the interaction cycle.
This behavior shows that the \interfaceagent implicitly reasons about the attribute that the user intends to select based on previous interactions. \addiui{In short, the \interfaceagent learns to suggest items that the user is not wearing, or that the user has not interacted with.}

\subsection{User Study}
\label{sec:userstudy}
Our goal was to create a \useragent whose behavior resembles that of real users, so the \interfaceagent can support them in the same task. To this end, we evaluated the sim-to-real transfer capabilities of our framework by conducting a user study where the \interfaceagent interacted with participants instead of the \useragent.

\subsubsection{Baselines}
We compared our framework to two supervised learning approaches and the static interface (see \Sec{task}).
In line with previous work \cite{gebhardt2019learning}, we used a Support Vector Machine (SVM) with a Radial basis function (RBF) kernel as a baseline. \add{It represents a direct competitor to our approach, as it generalizes on a method level. However, it needs data recollection for every interface and task.}

We used the implementations of scikit-learn \cite{scikit-learn} and optimized the hyperparameters for performance. The feature vector of the baseline was identical to that of our framework. The baseline learned the probability with which a user will select a certain character attribute next. We assigned the three attributes with the highest probability to the menu slots. Note that we did not consider "Next" to be an item. 

\paragraph{Dataset} We collected data from 6 participants to train the supervised baselines. These participants did not take part in the user study. They interacted with the static interface, which resulted in a dataset with over 3000 logged interactions. \del{We found that more data points did not improve the performance of the SVM classifier through k-fold cross-validation}\add{We saw that the SVM performance saturates when increasing to more than 2700 data points} and reached around 91\% \addiui{top-3 classification accuracy (i.e., the percentage of how often the users' selected item was in the top three of the SVM output)} on a test set. Furthermore, we found that the baseline generalize well to unseen participants (again through cross-validation). \add{We used all data points for the final model.} 

\paragraph{Metrics}
We used two metrics (dependent variables) to evaluate our approach. 
\begin{enumerate}
    \item \emph{Number of Actions:} the number of clicks a user needed to complete a task, which is a direct measure of user efficiency \cite{card1980klm}. 
    \item \emph{Task Completion Time:} the total time a user needed to complete a task.
\end{enumerate}

\subsubsection{Procedure}
Participants interacted with the \interfaceagent and the two baselines. The three settings were counterbalanced with a Latin square design, and the participants completed 30 trials per setting. In each condition, we discarded the first six trials for training. The participants were instructed to solve the task as fast as possible while reducing the number of redundant actions. They were allowed to rest in-between trials. We ensured that the number of initial attribute differences (IAD), which refers to the number of clothing attributes that differ between the current selection and the target at the start of a trial, was uniformly distributed within the participant's trials. Participants used an Oculus Quest 2 with its controller. 

We recruited 12 participants from staff and students of an institution of higher education (10 male, 2 female, aged between 23 and 33). All participants were right-handed and had a normal or correct-to-normal vision. On average, they needed between 35 to 40 minutes to complete the study.

\subsubsection{Results}
\begin{figure}[!t]
     \centering
     \begin{subfigure}{0.3\columnwidth}
        \centering
        \includegraphics[width=\textwidth]{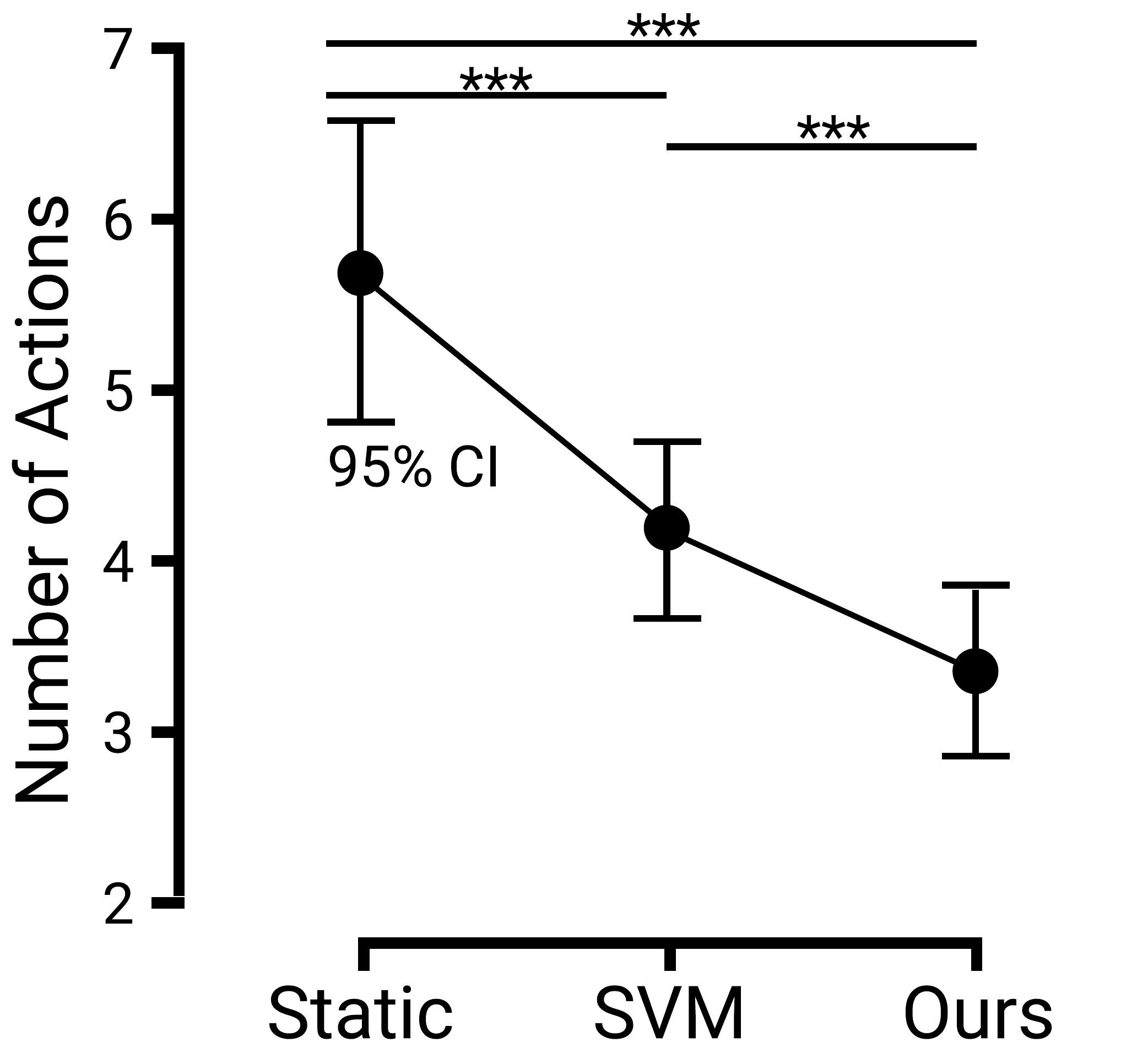}
        \label{fig:userstudy_actions}
    \end{subfigure}
    \hspace{0.5cm}
    \begin{subfigure}{0.3\columnwidth}
        \centering
        \includegraphics[width=\textwidth]{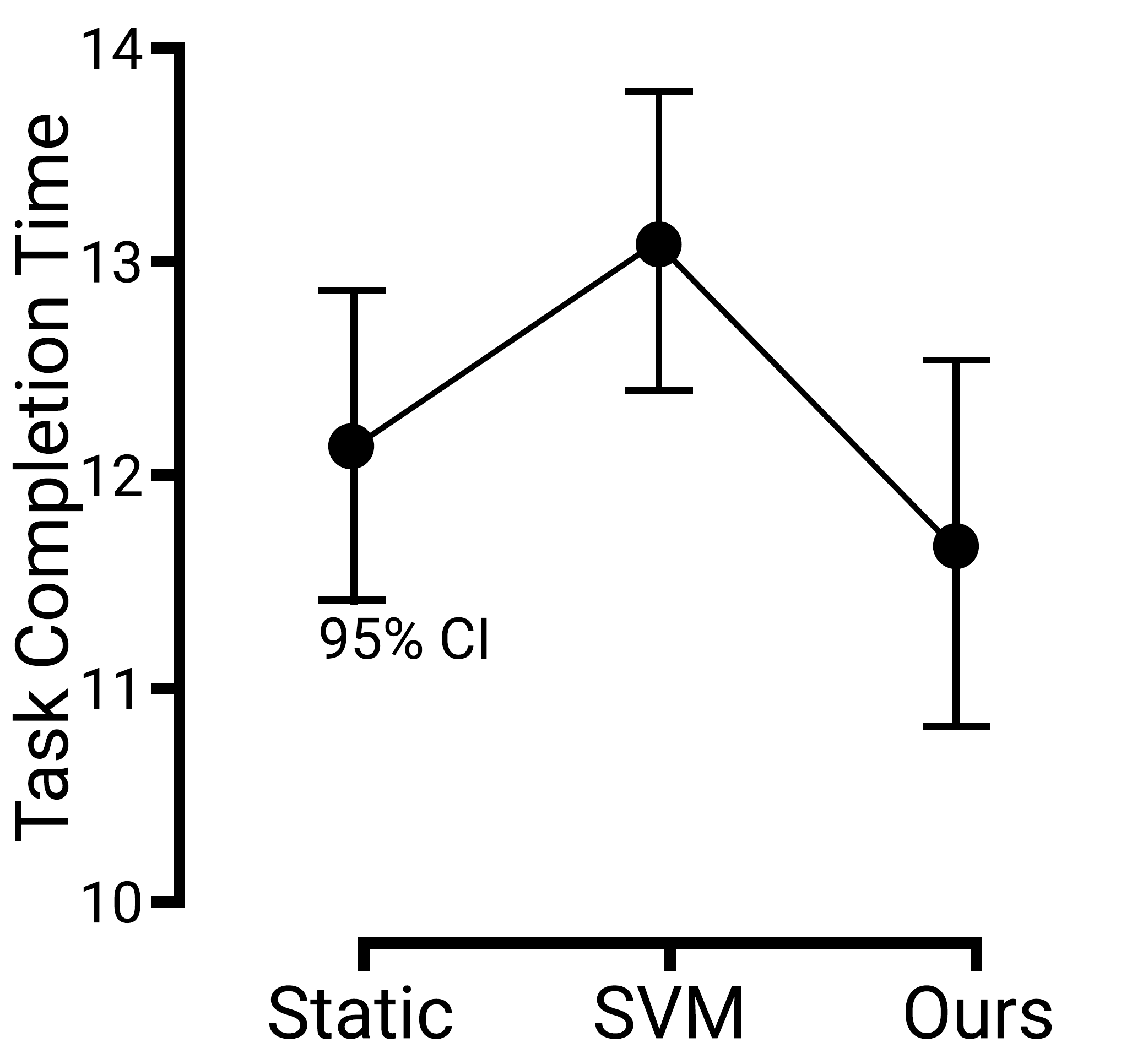}
        \label{fig:userstudy_time}
    \end{subfigure}    
    \caption{The average number of actions (left) and the task completion time (right) participants needed to finish the tasks of our user study, plotted with the 95\% confidence interval. We compare Ours against Static, and an SVM. Our approach outperforms both baselines on the number of actions. Averaged over all participants and all trials.}
    \label{fig:userstudy}
\end{figure}

We present a summary of our results in \Figure{userstudy}. We analyzed the effect of conditions on the performance of participants with respect to the number of actions and task completion time. 

Participants needed on average $3.34$ actions to complete a task with our framework, compared to $5.73$, and $3.87$ for the static, and SVM baselines respectively. \add{The normality and sphericity were not violated.} We performed a repeated-measures ANOVA We found a significant effect on method ($F(2, 22)=209.68, p<.001$). With a Holm-corrected post-hoc we find that both the SVM and Ours significantly outperform the static interface (both $p<.001$) in terms of number of actions. We also find that ours significantly outperforms the SVM ($p=0.006$).  

It is important to decompose the results per initial attribute difference. This is crucial because a task with a single IAD is expected to be shorter in duration than tasks with multiple IADs. We report the values in \Table{desc}. Using a repeated-measures ANOVA, we found a significant effect ($F(8, 0.25)=35.537, p<.001$) on the interaction between the approach and the IAD. We limited our analysis to scenarios where the interaction involves "Ours" and only considered scenarios where the IADs between the methods were identical (i.e., we do not discuss, for example, the SVM with five initial attribute differences versus the static method with a single difference). Our findings from a Holm-corrected post-hoc analysis showed no difference between the static baseline and the other approach for a single initial attribute difference (all $p=1.00$) and a significant difference for more than one IAD (all $p<.001$). When comparing the SVM versus our method, we found no difference for one IAD (p=$1.00$) or for five (p=$0.26$); for all other cases, a difference was found (all $p<.001$). We observed that the \interfaceagent's strategy, which groups similar category items (e.g., showcasing pink and blue backpacks if the character has a red one cf. \Fig{sequence}), increases the chance of selecting a useful item. However, this strategy doesn't apply when all attributes need changing.

\begin{table}[!t]
\centering
\caption{User study result reported by IAD and method (Mean $\pm$ SD).}
\label{tab:desc}
\begin{adjustbox}{width=\textwidth}
{
\begin{tabular}{lccccc}
\toprule
& \textbf{1} & \textbf{2} & \textbf{3} & \textbf{4} & \textbf{5} \\
\cmidrule[0.4pt]{2-6}
\textbf{Heuristic} & $1.708 \pm 0.396$ & $3.889 \pm 0.451$ & $6.139 \pm 0.797$ & $7.903 \pm 0.845$ & $9.000 \pm 0.000$ \\
\textbf{SVM} & $1.456 \pm 0.219$ & $3.747 \pm 0.515$ & $4.607 \pm 0.174$ & $5.439 \pm 0.259$ & $5.827 \pm 0.389$ \\
\textbf{Our} & $1.625 \pm 0.569$ & $\textbf{2.036} \pm \textbf{0.297}$ & $\textbf{3.281} \pm \textbf{0.657}$ & $\textbf{4.719} \pm \textbf{0.613}$ & $5.311 \pm 0.186$ \\
\bottomrule
\end{tabular}
}
\end{adjustbox}
\end{table}

When looking at the task completion time, participants using our framework needed $12.14$ seconds to complete the task. The completion time was $12.36$ for the static interface and $12.71$ \add{for the SVM}. The task completion time was normally distributed (Shapiro-Wilk $p>0.05$). We found no significant difference in overall task completion time with a Greenhouse-Geisser (for sphericity) corrected repeated-measures ANOVA ($F(1.568, 17.243)=0.86, p=0.42$). This could be, because despite requiring more action for a static baseline, the participants formed a strategy.

\subsection{Discussion}
To analyze if our multi-agent framework is competitive with a baseline that requires carefully collected real user data, we compared it against a supervised SVM. We did not find significant differences between the two approaches in the task performance metrics of the number of actions and task completion time. This suggests that our approach is a competitive alternative to data-driven approaches for creating adaptive user interfaces. 

The adaptive approaches significantly reduce the number of actions necessary to complete the task compared to the static interface. However, no significant differences in task completion times were found. We argue that this could be due to real users being more familiar with the ordering of items in the static interface that is kept constant across trials. This familiarity is not captured by our current cognitive model or incentivized in the reward function. In future work, we will model familiarity and investigate its effect on task completion time. 

We have shown qualitatively that our \interfaceagent learns to take previous user actions into account. This characteristic is core to meaningful adaptations. At the moment, our agent's capabilities are limited by the size of the stack $\stack$. In the future, recurrent approaches such as LSTM could be investigated to overcome this limitation. 

Furthermore, we presented evidence that our framework can generalize to goals that were not seen during training. 
It is important to mention that the results of this experiment are subject to its task and that seen and unseen goals are from the same distribution. Nevertheless, the study provides first indications that our approach generalizes to applications where not all users' goals might not always be encountered during training.
\section{Applications}
To demonstrate the versatility of our framework, we introduce four additional point-and-click interfaces to demonstrate how our approach generalizes to different scenarios. Every scenario offers a distinct adaption, task, and interface. Our method requires minimal to no adaptations for the different scenarios (i.e., mostly a change in the dimensions of the observation and action spaces). We showcase both 2D interfaces as well as Mixed Reality interfaces. We show how the \interfaceagent selects from a set of pre-designed UI widgets, hand user's the correct blocks when building a tower, move out-of-reach items closer, and make hierarchical menus more efficient during photo editing. 

Please refer to our supplementary video for visual demonstrations of the tasks. Due to the diverse nature of use cases, we will report either number of clicks or task completion time as success metric. All scenarios are showcased through the interaction of a real user with a trained \interfaceagent. The numerical results are obtain in simulation.

\subsection{Number Entry}
\label{sec:number}
\begin{figure}[!t]
    \centering
    \includegraphics[width=0.6\columnwidth]{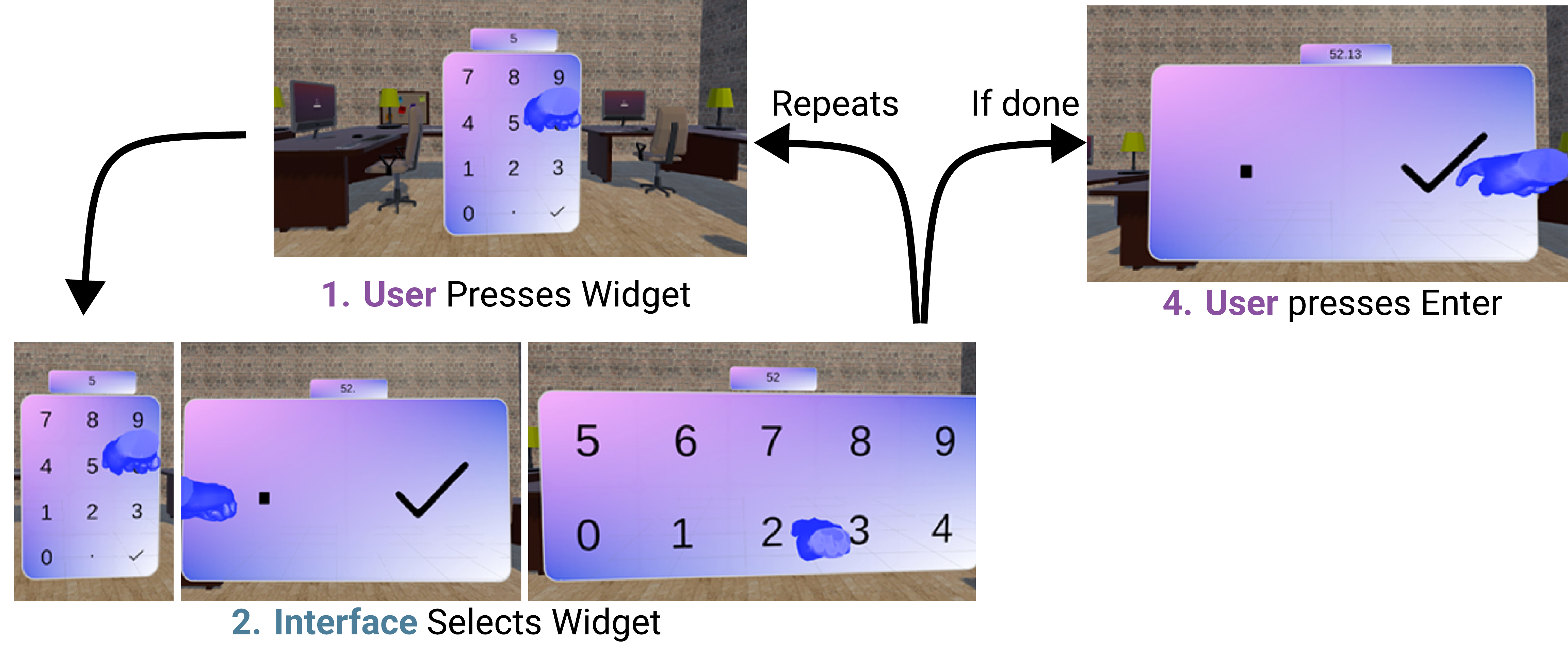}
    \caption{Adaptive keypad: the \useragent is asked to enter a randomly initialized price by using a keypad (1). The \interfaceagent selects from three pre-designed different widgets either a normal keypad,  a digits-only keypad or a non-digits-only keypad. The \useragent selects a button of the chosen widget. The task ends when the \useragent presses enter (4).}
    \label{fig:price}
\end{figure}

We introduce a price entry task on a keypad. The \interfaceagent selects a widget from a pre-designed set. We show that our approach can support applications requiring users to issue command sequences and provide meaningful help given a user's progress in the task (see \Fig{price}). 
The task assumes a setting where the simulated user must enter a product price between $10.00$ and $99.99$.
To complete the task, the \useragent has to enter the first two digits, the decimal point, the second two digits, and then press enter. 
The \interfaceagent can select one of three different interface layouts: i) a standard keypad, ii) a keypad with only digits and iii) a widget with only the decimal point and the enter key. 

The goal difference penalty (\Eq{rd}) in this case is based on whether the current price $\tools$ matches the target price $\gattr$: 
\begin{equation}
    \error_{gd} = -\sum_t\mathbbm{1}_{\tools_t \neq \gattr_t},
\end{equation}
where $\mathbbm{1}$ is an indicator that is 1 if $\tools_t \neq \gattr_t$ and $0$ otherwise, and $t$ is the current timestep. Every time a button is hit, $t$ increases by 1. This is similar to the penalty in all other tasks. However, it considers that the order of the entries matters. On average, the \useragent needs $4.0$ seconds to complete the task in cooperation with the \interfaceagent, compared to $4.9$ seconds when using a static keypad. The number of clicks is identical, since the full task can be solved on the standard keypad.

\subsubsection*{Qualitative Policy Inspection} We observe that the \interfaceagent learns to select the UI that has the biggest buttons for an expected number entry (e.g., only digits or only non-digits). From this we can conclude that the \interfaceagent implicitly learns the concept of Fitt's law and prioritizes larger buttons where appropriate. 

\subsection{Block Building}
\label{sec:building}
\begin{figure}[!t]

        \centering
        \includegraphics[width=0.5\columnwidth]{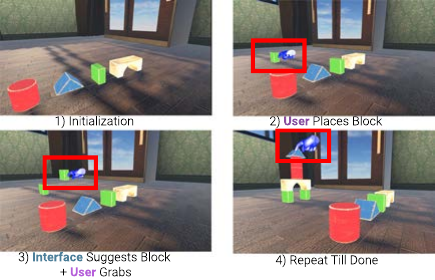} 
        \caption{Block Building: The user is building a castle from blocks (1). The user places the first block (2). The \interfaceagent suggests a next block to place (3). This is repeated till the castle is built (4). }
        \label{fig:building}
\end{figure}

The second scenario is a block-building task (\Fig{building}) where the user constructs various castle-like structures from blocks. Compared to the game character task, only the dimensionality of the observation and action space needs to be changed. It can choose between 4 blocks (wall, gate, tower, roof) and a delete button. The agent needs to move the hand to a staging place for the blocks (see \Figure{building}) and then place the block in the corresponding location. The block cannot be placed in the air, i.e., it always needs another block on the floor below. The \interfaceagent suggests a next block every time the user places a block. However, the user can put the block down, in case it is unsuitable. An action is picking or placing a block. 

This task represents a subset of tasks that do not have a Heads-Up-Display-like UI to interact with, but are situated directly in the virtual world. This is a common interactive experience of AR/VR systems. The user needs on average $1.1$ actions with our framework, compared to $2.0$ actions without the \interfaceagent. Thus 1.1 indicates that the \interfaceagent suggests the correct next block, most of the time. 

\subsubsection*{Qualitative Policy Inspection} We observe that the policy learns to always suggest a block that is usable given the current state of the tower. This indicates that the policy has an implicit understanding of the order of blocks and can distinguish between those belonging to the foundation versus the upper parts of a tower.


\begin{figure}[!t]
        \centering
        \includegraphics[width=0.5\columnwidth]{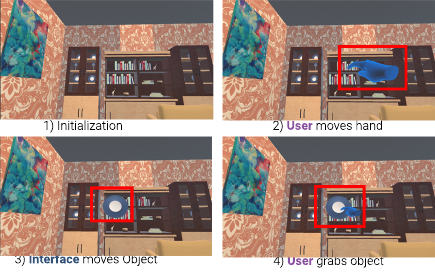} 
        \caption{Out-of-reach object grabbing: the \useragent attempts to grab a specific object, that is initially out of reach, in a space containing multiple objects (1). The \useragent learned to move towards an object to indicate its intention to grab it (2). Based on that, the \interfaceagent learned to move the intended object within the \useragent's reach (3). The \useragent then grabs the object to finish the task (4).}
        \label{fig:objectgrab}
\end{figure}

\subsection{Out-of-reach Item Grabbing}
\label{sec:out-of-reach}

In the third usage scenario, the user needs to use their hand to grab an object that is initially out of reach. Thus, the \interfaceagent needs to move an object within reach of the user, which can then grab it. The \interfaceagent observes the location of the user's hand. The task environment includes several objects. Uniquely, in this scenario the user and the \interfaceagent are forced to collaborate to select the correct target object and complete the task (see \Figure{objectgrab}); as it is impossible for the \useragent to complete the task on its own. Compared to the game character task, only the dimensionality of the observation and action space needs to be changed. 

In this use case, we changed the lower level of our user to learn motor control with RL instead of using the Fitts-Law-based motor controller. This highlights the modularity of our approach and can be useful in scenarios where existing models, such as Fitts' Law, are not sufficient.
The low-level motor control policy controls the hand movement. In particular, given a target slot, the policy selects the parameters of an endpoint distribution. Given the current position and the endpoint parameters (mean and standard deviation), we compute the predicted movement time using the WHo Model \cite{guiard2015mathematical}. The low-level policy needs to learn i) the coordinates and dimensions of menu slots, ii) an optimal speed-accuracy trade-off given a target slot, and its current position. Refer to \Appendix{learned} for more details.

\subsubsection*{Qualitative Policy Inspection} \del{We qualitatively evaluate the learned policy.} We find that the the policy selects objects positioned in the direction of the user's arm movement rather than the closest ones. This indicates that the policy implicitly learns the correlation between directionality of movement and intent. 


\subsection{2D Hierarchical Menu}
\label{sec:photo}
In this task, a user edits a photo by changing its attributes. A photo has five distinct attributes with three states per attribute: i) filter (color, sepia, gray), ii) text (none, Lorem, Ipsum), iii) sticker (none, unicorn, cactus), iv) size (small, medium large), and v) orientation (original, flipped horizontal, and vertical). The photo's attribute states are limited to one per attribute, i.e., the photo cannot be in grayscale and color simultaneously. This leads to a total of 15 attribute states and 243 photo configurations. 

The graphical interface is a hierarchical menu, where each attribute is a top-level menu entry, and each attribute state is in the corresponding submenu. By clicking a top-level menu, the submenu expands and thus becomes visible and selectable. Only one menu can be expanded at any given time. 

The photo attribute states correspond to the current input state and the target state. The \interfaceagent selects an attribute menu to open. Its goal is to reduce the number of clicks necessary to change an attribute, e.g., from two user interactions (filter->color) to one (color). For the \emph{\useragent}, the higher level selects a target slot, and the lower level moves to the corresponding location. 

\subsubsection*{Qualitative Policy Inspection} 
We observe that the \interfaceagent intelligent\add{ly} decides which submenu to open next. We notice that this is never a menu the user recently interacted with as the probability of having to change, e.g., the color twice in a row is minimal and only a result of errors.

\begin{figure}[!t]
    \centering
    \includegraphics[width=\textwidth]{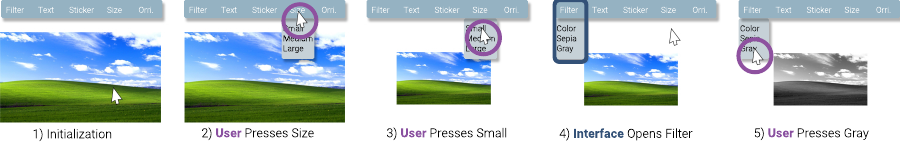}
    \caption{We introduce a photo editing task where (1)~a user matches a photo to a target by operating a hierarchical menu. (2)~The user selects the submenu `size`. (3)~The user then selects the attribute `small`, which alters the image. (4)~After the user has changed an attribute, the interface observes the new state of the photo and finds the most likely submenu for the next user action. (5)~The user clicks on an item in the submenu to complete the task.}
    \label{fig:hierarchical_good}
\end{figure}
\section{Discussion}
\label{sec:limitations}
MARLUI models the interaction with point-and-click adaptive interfaces as a multi-agent cooperative game by teaching a simulated \useragent and an \interfaceagent to cooperate. Learned policies of the interface agent have shown their capability to effectively assist real users. Demonstrating our approach in a wide variety of use cases is a first step towards general methods that are not tied to specific applications nor dependent on manually crafted rules or offline user data collection. However, there are limitations that require further research.

In this work, we have focused on point-and-click interfaces. However, it would be interesting to extend the user agent to model other interaction paradigms. By enhancing our user agent to replicate behavior for other interaction types than clicking, we could extend the possible use cases that MARLUI can support. For example, research has shown that gaze-based selection, similar to cursor movement, follows Fitts' Law \cite{schuetz2019explanation}. \add{Furthermore, similar concepts can also be applied to human-robot interactions, such as using simulated humans to train human-robot handshakes or human-to-robot handovers \cite{christen2023learning, christen2019guided, christen2023synh2r}.} 

Moreover, user goals can change during human-computer interaction, particularly in creative tasks where users constantly adjust their objective based on intermediate results. This presents challenges for standard RL approaches, which assume goals to remain stationary. Future research on MARL for AUIs needs to focus on finding strategies to easily adapt trained interfaces to changing or new user goals. This is required to establish more robust and flexible adaptive interfaces that can support real-world use cases.

We have demonstrated that our formulation solves problems with up to 5 billion possible states, as in the character creation application (\Sec{task}). However, the complexity of the problem grows exponentially with the number of states. This makes it challenging for MARLUI to scale to interfaces with even larger state spaces. To overcome this, we could explore different input modalities, such as representing the state of the UI as an image instead of using one-hot encoding. This approach is similar to work on RL agents playing video games \cite{mnih2013playing}, which showed that image representations can effectively cope with large state spaces.

We have shown that the simulated \useragent's behavior was sufficiently human-like to enable the \interfaceagent to learn helpful policies that transfer to real users. The \interfaceagent's performance is inherently limited by the \useragent. Therefore, increasing realism in the model of the simulated user is an interesting future research direction, for instance, modeling human-like search \cite{chen2015emergence} or motor control with a biomechanical model \cite{fischer2021reinforcement}. \add{Along similar lines, our work helps with the creation of policies that adapt interfaces given user interactions. However, it does not adapt to the user themselves (e.g., different levels of expertise). Such personalization is an interesting direction for future research.}

Our framework has theoretical appeal because it provides a plausible model of the bilateral nature of AUIs: the adaptation depends on the user, whereas also the user action depends on the adaptation. Modeling this unilaterally as in supervised learning does not reflect reality well. Related to ongoing research \cite{murray2022simulation}, we believe that future work can leverage our framework to gain a better theoretical understanding of how users interact with a UI.  Our setup has the potential to scale to multiple users with different skills and intentions. This could lead to bespoke assistive UIs for users with specific needs or UIs for users with specific expertise levels. 

Finally, our proposed framework enables adaptive policies for different point-and-click tasks and interfaces. We have shown how our framework produces policies that support users in a variety of these interfaces and tasks. Building on this, future work can investigate the transition from framework to developer tool. Tools that enable developers to use our framework easily and consistently will streamline the development process of AUIs.


The question addressed in this paper is whether it is possible to develop a \emph{general framework} for point-and-click AUIs that does not depend on task-specific heuristics or data to generate policies offline. To this end, we have introduced \marlui, a multi-agent reinforcement learning approach. Our method features a \useragent and an \interfaceagent. The \useragent aims to achieve a task-dependent goal as quickly as possible, while the \interfaceagent learns the underlying task structure by observing the interactions between the \useragent and the UI. Since the \useragent is RL-based and thus learns through trial-and-error interactions with the interface, it does not require real user data. We have evaluated our approach in simulation and with participants, by replacing the \useragent with real users, across five different interfaces and various underlying task structures. The tasks ranged from assigning items to a toolbar, handing out-of-reach objects to the user, selecting the best-performing interface, providing the correct object to the user, and enabling more efficient interaction with a hierarchical menu. 
Results show that our framework enables the development of AUIs with minimal adjustments while being able to assist real users in their task.
We believe that \marlui, and a multi-agent perspective in general, is a promising step towards tools for developing adaptive interfaces, thereby reducing the overhead of developing adaptive strategies on an interface- and task-specific basis.

\cleardoublepage%
\part{Conclusion}
\cleardoublepage%

\def\dir{chapters/06_conclusion}

\chapter{Conclusion}
\label{ch:summary}
In this dissertation, we investigate the trade-off between user agency and system automation in interactive intelligent systems, focusing specifically on interactions involving shared variables. These interactions allow both a user and an artificial agent to manipulate the same variable. Despite rapid advancements in contextual intelligent systems, research on their interaction remains insufficient. We approach this problem from two angles: first, by investigating a haptic interface as the interaction medium, and second, by leveraging optimal control strategies to balance agency and automation more effectively.

\section*{The Design of Shared Variable Interfaces}
In a shared variable interface, both the user and the system act on a shared variable, which can take forms such as kinesthetic haptic devices or graphical menus. Kinesthetic haptic feedback serves as a special instance of this shared variable, where both action and perception occur in the same modality from the user's perspective, such as using hands to change or perceive a joystick's position. This contrasts with interfaces where perception and interaction modalities differ, like touch interfaces. Previous kinesthetic haptic feedback systems often involved complex devices requiring user instrumentation, which limited full user agency. To overcome these limitations, we introduce novel haptic interfaces that elicit grounded forces for system automation and are untethered for full user agency.

\subsection*{1. \omniHapTitle}
In \chapref{ch:shared:contact}, we introduce \omniHap, a contact-free, non-planar haptic feedback device with a spherical electromagnetic actuator. The actuator consists of three orthogonal coils integrated into a single sphere, capable of eliciting radial and tangential forces on a permanent magnet embedded in a tool. Despite the tool being untethered, the force remains grounded via the electromagnet. Our device can produce forces up to 1N, which can be increased by enhancing the volume of the embedded permanent magnet, albeit at a weight trade-off. Users were able to discriminate at least 25 points spread evenly on the device's sphere (\figref{fig:confusion_circle_error}), with mean errors of $2.5mm \pm 1.4$, $5.7mm \pm 4.6$, $6.5mm \pm 5.2$, and $7.2mm \pm 5.1$ across different azimuth angles (\figref{fig:error_vs_angle}). \omniHap showed that enabling feedback in the same modality as the user's exertion and with a shared variable results in natural interactions. However, the reliance on external tracking for tool position remained a limitation due to its expense and complexity.

\subsection*{2. \omniUISTTitle}
\chapref{ch:shared:volumetric} introduces on-device sensing with \omniUIST, integrating Hall sensors into the \omniHap base. This core contribution allows us to decompose the natural interference caused by simultaneous magnetic tracking and actuation. Our novel gradient-based optimization method minimizes the difference between estimated and observed magnetic fields, affording 3D tracking capabilities with a mean error of $6.9mm$ during actuation (\figref{fig:optitrack_eval}). Furthermore, we improved the actuator design compared to \omniHap, doubling the peak force to $\pm 2N$ at 15A (\figref{fig:Fz_vs_iz}). We demonstrated \omniUIST's capabilities through three Virtual Reality applications: sculpting, gaming, and soft object exploration (\figref{fig:usecases}). \omniUIST showed how it is possible to sense the tool state and take it into account in the control loop without user instrumentation or cumbersome setups.

\subsection*{Implications}
Our work lays the foundation for investigating haptic devices from a new perspective, that of \interfacesLower. We have taken the first step towards understanding haptic \interfacesLower and demonstrate new applications in gaming, object exploration, and design. This integration of sensing and actuating the tool through electromagnetism is both cost-effective and efficient. The primary advantages of \omniHap and \omniUIST are their high accuracy and substantial force output, achieved without mechanically moving parts, thereby avoiding wear and increasing user agency. By eliciting forces on the tool manipulated by the user, we enable haptic feedback in the same modality as the user's exertion, resulting in natural interactions. However, more intelligent system control is needed for complex dynamics.

\section*{The Control of Shared Variable Interfaces}
The intelligent control of interfaces with shared variables is crucial to trade off user agency with system automation. Previous systems typically used methods that either do not take the user into account at all or do not predict user behavior. Therefore, previous methods have no or limited ability to intelligently trade off user agency and system automation. In contrast, we embed, explicitly and implicitly, a predictive model of user behavior into our control strategy. A predictive model allows the system to take future human decisions into account in the control loop. Using future predictions enables a better balance between user agency and system automation.

\subsection*{3. \magpenTitle}
In \chapref{ch:control:optimal}, we explore Model Predictive Contour Control to enable haptic guidance for electromagnetic systems. Our real-time approach assists users in tasks like calligraphy (\figref{fig:caligraphy}), sketching (\figref{fig:dragon}), or designing (\figref{fig:tablet}) by iteratively predicting the motion of an input device (such as a pen) and adjusting the position and strength of an underlying dynamic electromagnetic actuator. In addition to the control strategy, we introduced a prototypical haptic feedback device that allowed us to evaluate our approach (\figref{fig:hardware}). Our user study showed that users were more accurate with our method ($6.2mm \pm 0.8$) compared to open-loop ($38.0mm \pm 56.9$) or time-dependent closed-loop methods ($45.0mm \pm 50.8$), as illustrated in \figref{fig:single_user_control} and \tabref{tab:strategy_results}. \magpen demonstrates that an explicit predictive user model and system dynamics with a receding horizon cost function enable the control of a haptic system that trades off user autonomy with system automation. However, the reliance on known system dynamics posed a challenge, especially in embedding user behavior.

\subsection*{4. \marluiTitle}
In \chapref{ch:control:multi}, we introduce a model-free reinforcement learning approach to overcome the need for explicit system dynamics. We also proposed adaptive user interfaces that automatically adjust based on user context and previous inputs. In our MARL formulation, a user agent simulates real user interactions, while an interface agent learns to optimize the UI based on the user agent's performance. This approach significantly reduced the number of actions users needed ($3.34$) compared to supervised-learning ($3.87$) and static baselines ($5.73$), as illustrated in \figref{fig:userstudy} and \tabref{tab:desc}. Unique to our approach is that it requires minimal tuning between tasks and scenarios. To show this, we introduced four additional use cases: Interface Selection (\figref{fig:price}), Block Building (\figref{fig:building}), Out-of-reach Item Grabbing (\figref{fig:objectgrab}), and Photo Editing (\figref{fig:hierarchical_good}). \marlui demonstrates that it is possible to learn implicit models of human behavior and tasks by treating interaction as a multi-agent game. These implicit models can then be integrated into intelligent control strategies that take human decision-making into account.

\subsection*{Implications}
We demonstrate how to integrate models of human behavior explicitly and implicitly into optimal control strategies for intelligent systems. This approach allows systems to consider future states and actions and optimize inputs accordingly. In \magpen, we embed a user model explicitly in the control strategy, considering user behavior over a horizon, though mathematically formulating this behavior can be challenging. Conversely, in \marlui, we take a multi-agent RL approach where user behavior is implicitly learned, eliminating the need for explicit behavioral models or system dynamics descriptions, and allowing the method to generalize across interfaces and tasks. Both approaches successfully balance user agency and automation, assisting users in their tasks.

\section*{Conclusion}
We started with the question: \emph{``How can we algorithmically control intelligent systems with shared variables to balance user agency and system automation?''} To answer this question, we provide four distinct contributions in two parts: the design and control of shared variable interfaces. We demonstrate that integrating models of human behavior into control strategies, either explicitly or implicitly, enables intelligent systems to take user agency better into account. Furthermore, we show that user agency versus system automation is not just an algorithmic problem, and that this trade-off must also be taken into account during the engineering of physical devices and the design of interfaces. Hence, we advocate for an integrated end-to-end approach for the interaction with intelligent systems that takes an algorithmic, engineering, and design perspective.
\chapter{Outlook and Future Work}
\label{ch:outlook}
This dissertation has demonstrated the potential of shared control for interaction with intelligent systems that have shared variables, but further exploration is needed. We identify three important directions for future research: improving the system's understanding of human behavior, enabling systems to reason from a human perspective, and enhancing control strategies to incorporate models of human behavior.

\section{Understanding Humans}
\subsection{Human Sensing \& Inference}
The shift towards contextual intelligent systems will integrate the world at large—including the users themselves—into the interface. For instance, in a scenario where a robot and a human collaborate to clean an apartment, the robot should infer and predict the user's intent from the state of the world and the user's actions without explicit instructions.

Accurate human state estimation is essential for this, including physical states captured through computer vision and, more challengingly, latent states such as expertise, fatigue, and intent. Advances in machine learning, particularly in areas such as human latent state inference \cite{li2017inferring}, are needed to tackle these challenges. Additionally, the system should be aware of its environment and capable of extrapolating and predicting future states. Physics-based machine learning holds promise for understanding physical world dynamics \cite{christen2023learning}, while behavioral models are necessary for predicting human behavior. A first project could be to learn a classifier that predicts the extent to which a user is an expert in complex software such as CAD based on their cursor movement.

\subsection{Human Behavioral Models}
Sensing the human state alone is insufficient. To apply current state estimations to intelligent control, we need to predict future human states. This requires behavior models that capture the complex dynamics of human behavior. Challenges include stochastic system dynamics, changing user optimization goals, and the impact of system actions on users.

Data-driven approaches such as imitation learning and inverse reinforcement learning, combined with cognitive models, can help tackle these problems \cite{kwon2020inverse}. Reinforcement learning, assuming computational rationality, can approximate optimal human-like behavior. By incorporating data-driven priors, we can model more complex behaviors \cite{Langerak:2024:rile}. Integrating these approaches will enhance our understanding of human behavior and improve interface design and interaction paradigms. An interesting direction could be to use text-conditioned models that generate realistic user behavior given a prompt (e.g., "A novice user browses the website to look for new sneakers").

\subsection{Non-Stationary and Abstract Goals}
Our research has assumed specific, concrete goals, but this may not translate well to more abstract tasks such as creative activities or web browsing. Goals can also change dynamically during an activity.

Future research should focus on hierarchically structuring goals and dynamically adjusting their probabilities. Recognizing interactions as stochastic decisions at various levels, a hierarchical goal model could better adapt to uncertainty. Additionally, understanding the necessary accuracy of higher-level goals is crucial. In \magpen, we showed that a low-level polynomial approximation of a sketch was sufficient when updated iteratively.

\section{Understanding what Humans Understand}
The systems and humans might observe the world differently. Whether this is due to occlusions, quality of sensors, or other reasons, the system should be able to infer i) what the user has observed and ii) what the user believes about the world state. These capabilities might allow the system to predict future user actions and expectations more accurately.

\subsection{Human-Like Scene Understanding}
Humans do not have perfect scene understanding. They might miss something outside their field of view or something that does not match their prior knowledge. Current user models often assume complete scene awareness, which can be problematic for human-like user models.
Future studies should integrate insights from human cognitive perception with advanced computer vision techniques. Research questions include determining what a person can see from a monocular RGB image or video and estimating their field of view given a map and location. Extending gaze estimation techniques to real-world scenarios could provide valuable insights \cite{akinyelu2020convolutional}.

\subsection{Theory-of-Mind User Models}
Understanding user beliefs is crucial, but integrating these beliefs into control systems remains a challenge. Future research could explore incorporating probabilistic belief inferences into control frameworks. This requires precise belief inference, representation, and dynamics, along with optimization criteria that meaningfully utilize these beliefs.

Bayesian models offer a tractable path, keeping track of belief states at each inference step \cite{baker2011bayesian}. However, this requires knowing probability matrices. Recurrent networks such as LSTM can approximate complex dynamics, but the exact solution to the belief model problem remains an open question. A first project could predict human-belief states in simple games in which the user cooperates with an agent to solve a task (e.g., finding the exit of a maze), where the belief is limited to the agent state (e.g., position) and end-goal (the maze exit).

\section{Cooperative Control}
Finally, we need to integrate human sensing, inference, and behavioral models into intelligent control strategies, paving the way for more intuitive and seamless interactions between humans and machines that respect and enhance human decision-making processes.

\subsection{Embedding User Models in System Control}
Core to useful cooperative control strategies is embedding models of human behavior into the system control. However, how to embed human behavioral models is not entirely clear. We could embed these human models hierarchically into the system control strategy. However, this would be computationally expensive as we need to solve multiple control problems per timestep. Alternatively, we can learn implicit user models either from data or in simulation. While this might be a more scalable approach, how to learn these models implicitly remains an open question. A first project would be to directly compare explicit versus implicit embedded user models in a simple task to investigate the trade-offs involved.

\subsection{Combine Model-based and Model-free approaches}
Model-based methods like MPC and model-free methods like RL each have unique advantages. MPC can adjust to new and different objectives without retraining, while RL excels at developing policies for unpredictable system behaviors. Combining these approaches could greatly improve computational assistance for users.

Future developments might involve using MPC to offer initial direction in novel situations based on user goals, with RL continuously refining and customizing the system. Learning complex environmental dynamics and integrating them into model-predictive optimization can compute policies without retraining for new objectives. Alternatively, inverse reinforcement learning might offer cost functions to use in MPC when system dynamics are known.

\cleardoublepage

\manualmark%
\markboth{\spacedlowsmallcaps{\bibname}}{\spacedlowsmallcaps{\bibname}} 
\refstepcounter{dummy}
\addtocontents{toc}{\protect\vspace{\beforebibskip}} 
\addcontentsline{toc}{chapter}{\tocEntry{\bibname}}
\label{app:bibliography}
{%
  \emergencystretch=1em%
  \printbibliography%
}

\newcommand{\bibstyleheader}[1]{%
  \section*{\normalsize #1}  
  \markright{#1} 
  }
\bibstyleheader{Generative AI Used}
The following tools have been used during the writing of this dissertation. Prompts were, or in similar spirit to, "Correct the grammar and syntax [paragraph]", or "Give feedback on [paragraph]." Furthermore, they have been used to translate the English abstract into German. 
\begin{enumerate}
    \item \fullcite{openai2023chatgpt}
    \item \fullcite{perplexityai}
\end{enumerate}

\cleardoublepage%
\part{Appendix}
\cleardoublepage%
\appendix

\def\dir{chapters/appendix}

\chapter{Optimal Control for Electromagnetic Haptic Guidance Systems}
\label{app:dipoledipole}
\section{Dipole-Dipole Model}
\subsection{In-Plane}
    In this section we describe the derivation of the dipole-dipole model for the in-plane actuation force, as well as the case of considering a pen tilt $\angt$ of the pen. Please refer to the schematic Figure \ref{fig:dipole_dipole} for vector notations we use in this section. 
    
    The coordinate system is given by,
    \begin{eqnarray}
     \ed &=& \frac{\posm - \posp}{||\posm - \posp||} \\
     \ez &=& [0,0,1]^T \\
     \et &=& \ed \times \ez
    \end{eqnarray}
    \noindent with $\ed$ the in-paper-plane distance from the pen contact point to the electromagnet center projection, $\ez$ the vertical out-of-plane direction and $\et$ the orthogonal vector to the former two.
    
    The dipole-dipole expression for the force acting on $\mpBold$ due to $\mmBold$ and separated by $\RmagtopenBold$ is:
    \begin{multline}
       \mathbf{F_p} = {\dfrac  {3\mu _{0}}{4\pi \Rmagtopen^{5}}}
       \left [ \left(\langle\mpBold,\RmagtopenBold\rangle \right) \mmBold + 
       \left(\langle\mmBold,\RmagtopenBold\rangle\right) \mpBold \right . +
       \\
       \left(\langle\mpBold,\mmBold\rangle\right) \RmagtopenBold - 
        \left . {\dfrac{5\left(\langle\mpBold,\RmagtopenBold\rangle\right)
        \left(\langle\mmBold,\RmagtopenBold\rangle\right)}{\Rmagtopen^{2}}} \RmagtopenBold \right ] \ , \label{eq:ap.F21-dip}
    \end{multline}
    
    The two dipoles and the vector distance between them can be expressed in the proposed coordinate system as,
    \begin{eqnarray}
     \mmBold &=& \alpha \ m_m \ \ez \label{ap.mm}\\
     \mpBold &=& - (m_p \stheta \cphi) \ \ed \nonumber \\
              && + (m_p \stheta \sphi) \ \et \nonumber \\
              && + (m_p \ctheta) \ \ez \label{ap.mp} \\
     \RmagtopenBold &=& - (d+h_p \stheta \cphi) \ \ed \nonumber \\
              && + (h_p \stheta \sphi) \ \et \nonumber \\
              && + (h - (1-\ctheta) h_p) \ \ez \label{ap.rmp}
    \end{eqnarray}
    
    \noindent and the three scalar products of equation \ref{eq:ap.F21-dip},
    \begin{eqnarray}
     \langle\mmBold,\RmagtopenBold\rangle&=& \alpha \ m_m [h-(1-\ctheta)h_p] \label{eq:ap.mm.r.1}\\
     \langle\mpBold,\RmagtopenBold\rangle&=&mp \ [-\stheta \cphi (d + h_p \stheta \cphi) \nonumber\\
     &&+\stheta^2 \sphi^2 h_p +  \nonumber\\
     &&\ctheta (h - h_p(1-\ctheta))] \ \  \label{eq:ap.mp.r.1} \\
     \langle\mmBold,\mpBold\rangle&=& \alpha m_m m_p \ctheta \label{ap:mm.mp.1}
    \end{eqnarray}

\subsubsection{Position-aware dipole-dipole model}
    \label{sc:ap.position-dipole}
    We first derive the position-aware dipole-dipole model\del{ (3 DOF)}, before continuing to the full \del{6 DOF }\add{position-aware and angle-aware} model.
    We rewrite Eq. \ref{eq:ap.F21-dip} with an equivalent pen dipole $\mpBoldt$, obtained by applying the small tilting angle approximation ($\ctheta \simeq 1$ and $\stheta \simeq 0$) to Eq. \ref{ap.mp}, 
    \begin{equation}
        \mpBoldt = m_p \ \mathbf{e_z} \label{eq:m1t} \ ,
    \end{equation} 
    \noindent where the scalar magnetization is given by $m_p = B_r V/\mu_0$. $B_r$ is the residual magnetization of the permanent magnet and $V$ its volume and  $\mathbf{e_z}$ is the $z$-unit vector. This approximation removes the requirement for tracking the pen tilt. More importantly it drastically simplifies the force equation since both dipoles now only have a $z$ component and thus the actuation only depends on the distance $d$ between pen and magnet (not on $\angt$ nor $\angp$).
    This provides a simplified version of the 3D distance vector,
    \begin{equation}
        \RmagtopenBoldt = - d  \ \ed + h  \ \mathbf{e_z} , \label{eq:r21b} 
    \end{equation}
    \noindent where the vertical distance, $h = h_m + h_p$, is constant. Note that the in-plane distance $d = \norm{\posp - \posm}$ is one of the variables we seek to control, given the projections of the pen position ($\posm$) and the electromagnet position ($\posp$) onto the sketching plane. 
    
    The electromagnet dipole ($\mmBold$) is mounted in a fixed upright position. Therefore it can be expressed via Eq. \ref{ap.mm}, without incurring any approximation error.
    The magnetization value of the full-strength dipole $m_m$, which approximates the electromagnet, can be derived experimentally. For this purpose we scan the magnetic field generated by the electromagnet, setting $\alpha = 1$ and using a hall sensor and adjust the parameters of EM field equation to give a good fit, as explained below in section \textit{Electromagnet dipole equivalent}. Table \ref{tab:ap:em_model} reports the values of $m_m$, $m_p$ and $h$ that were used in our experiments.
    
    \begin{table}[tb]
      \caption{List of electromagnet model and hardware parameters}
      \label{tab:ap:em_model}
      \begin{tabular}{lll}
        \toprule
        Name&Value&Description\\
        \midrule
        $\mu_0$ & $4\pi \ 10^{-7}$ [H/m] & Vacuum permeability \\
        $B_r$ & 1.3 [T] & Pen magnet type (NIB N42) \\
        $V$ & 0.66 [cm$^3$] & Pen magnet volume \\
        $m_p$ & 0.683 [A m$^2$]& pen dipole ($=B_r V / \mu_0$)\\
        $m_m$ & 1.286 [A m$^2$]& electromagnet dipole\\
        $h$ & 2.71 [cm] & z-distance $\mmBold$ to $\mpBold$ \\
        $h_p$ & 1.40 [cm] & height pen-tip to magnet \\
        $h_m$ & 1.31 [cm] & z-distance from plane to $\mmBold$.\\  
        $F_0$ & 0.488 [N] & force factor in Eq. \ref{eq:Fa} \\
      \bottomrule
    \end{tabular}
    \end{table}
    
    The total force acting on the pen (Eq. \ref{eq:ap.F21-dip}) can now be decomposed into the in-plane and vertical force components: 
    \begin{equation}
        \mathbf{F_p} = F_a \ \mathbf{e_d} + F_z  \ \mathbf{e_z} \ . \label{eq:ap:Fp_decomp}
    \end{equation}
    
    \noindent Here $\mathbf{F_a} = F_a \ \mathbf{e_d}$ represents the quantity we seek to control. 
    By substituting the results form Eq. \ref{ap.mm}, \ref{eq:m1t} and \ref{eq:r21b} into Eq. \ref{eq:ap.F21-dip} and maintaining only the in-plane contributions ($\mathbf{e_d}$ direction), we obtain the expression for the actuation force as function of pen-magnet separation:
    \begin{equation}
        \mathbf{F_a} = \alpha \ F_0 \ \left( \frac{d \left(4 - \frac{d^2}{h^2}\right)}{h \left(1 + \frac{d^2}{h^2}\right)^\frac{7}{2}} \right)  \ \mathbf{e_d} , \label{eq:Fa}
    \end{equation}
    where $F_0$ is a constant force parameter given by the expression,
    \begin{equation}
     F_0 = \frac{3 \ \mu_0 \ m_p \ m_m}{4 \ \pi \ h^4} \ . \label{eq:F0}
    \end{equation}
    
\begin{figure}[!t]
        \centering
        \includegraphics[width=0.6\columnwidth]{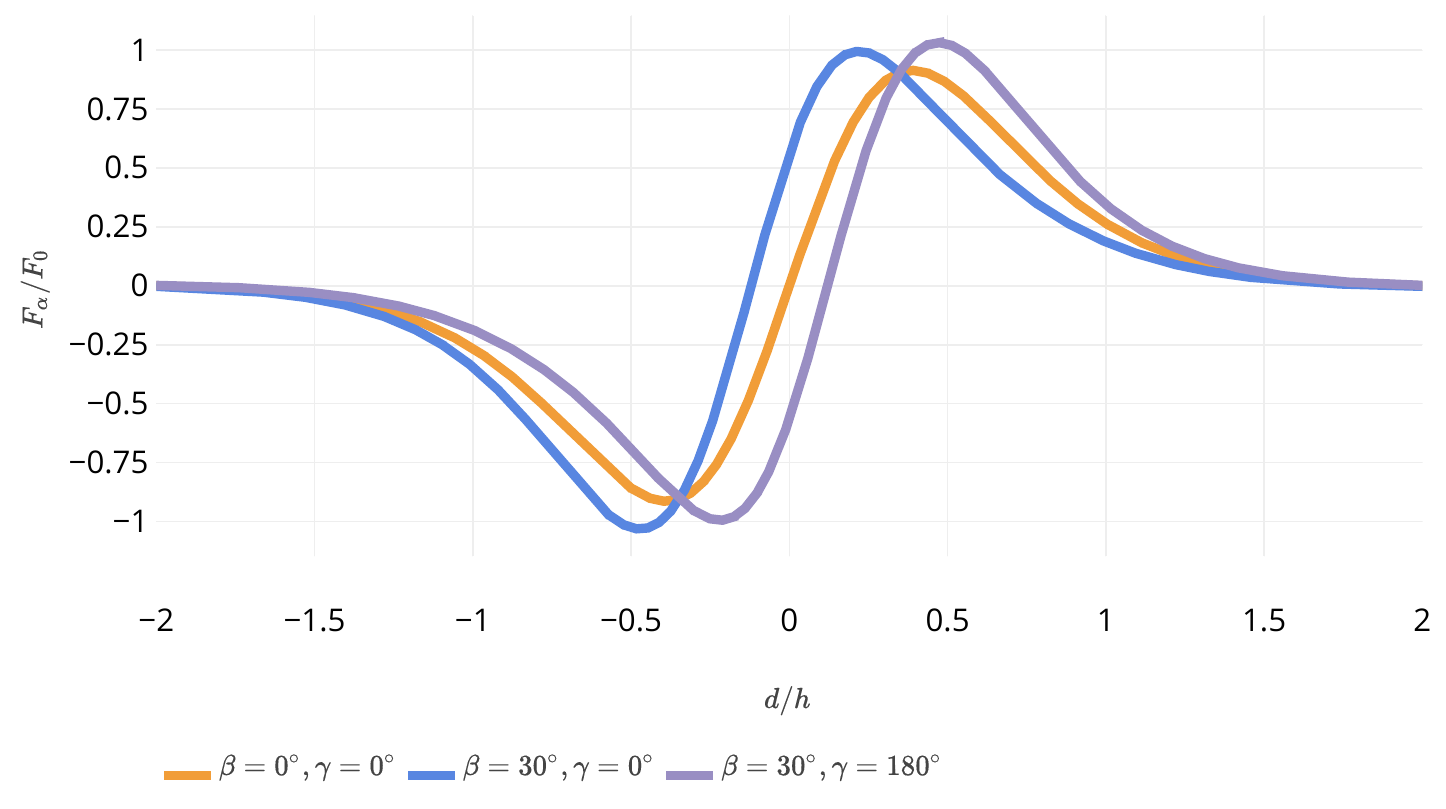} \\
        \caption{In-plane magnetic force as function of position. The horizontal displacement between curves (each denoting a different pen-tilt) is the approximation error induced by the upright pen (purple) assumption (angles defined in \protect\figref{fig:dipole_dipole}).}
        \label{fig:ap:approx_error}
    \end{figure}

    Fig. \ref{fig:ap:approx_error} illustrates how the dimensionless ratio within parentheses in Eq. \ref{eq:Fa} governs the force strength as function of distance $d=\norm{\mathbf{r_d}}$. 
    The actuation force $F_a$ is zero if the two magnets are aligned with one another ($d=0$), it has a maximum $F_a^{max} = 0.9 F_0$ at $d=0.39h$, and we can assume there is no more attraction for distances $d>2h$. In Table \ref{tab:ap:em_model} we report the value of $F_0$ we obtained for our prototype.
    
    Note that these simplifications lead only to a small approximation error.
    Compared to an angle dependent formulation, a tilt of up to $\angt = 30^{\circ}$ leads to a max error in our model (Eq. \ref{eq:Fa}) equivalent to shifting the distance $d$ by $\pm 3$ [mm] (Figure \ref{fig:ap:approx_error}). 
    
\subsubsection{Angle-aware dipole-dipole model}
\label{sc:ap.angle-dipole}
    In this section, we derive the \del{angle-aware EM model, that can be used for 6 DOF actuation}\add{complete EM model, using both, the pen position and its tilting angle as free variables}.
    We continue the deduction of $\mathbf{F_p}$ by substituting Eq. \ref{ap.mm}---\ref{ap:mm.mp.1} into the main expression Eq. \ref{eq:ap.F21-dip}. However, by following that path we wouldn't necessarily attain information on how strong the actuation force depends on the tilting angles $\angt$ and $\angp$. Here we take a different path. Based on the geometry of our system, we consider the cases where the pen is tilted by only a small angle\del{ ($\angt < 30 ^o$)}. We introduce this small-angle approximation by keeping only the first order terms in $\angt$,
    \begin{eqnarray}
     \stheta &\approx& \angt \ \ \ \ \text{(with} \ \angt \ \text{in \ radians)} \label{ap.appsin} \\
     \ctheta &\approx& 1 \label{ap.appcos}
    \end{eqnarray}
    \noindent As an indication of what this approximation means, for an angle $\angt = 30 ^\circ$, the difference between using $\stheta$ or $\ctheta$ or their approximations forms (Eq. \ref{ap.appsin} and \ref{ap.appcos}) is 5\% and 15 \%, respectively. Under the small-$\angt$ approximation, the dipoles' vectors are,
    \begin{eqnarray}
     \mmBold &=& \alpha \ m_m \ \ez \label{ap.mm2}\\
     \mpBold &\simeq& -m_p \angt \cphi \ed + m_p \angt \sphi \et + m_p \ \ez \label{ap.mp2}
    \end{eqnarray}
    \noindent and the distance between dipoles,
    \begin{equation}
    \RmagtopenBold \simeq -(d+h_p \angt \cphi) \ \ed + h_p \angt \sphi \ \et + h \ \ez
    \end{equation}
    \noindent with the length of that distance, at first order on $\angt$,
    \begin{equation}
        \Rmagtopen \simeq d^2 + h^2 + 2 d h_p \angt \cphi \label{eq:ap.rapp}
    \end{equation}
    
    In turn, the scalar products (Eq. \ref{eq:ap.mm.r.1}---\ref{ap:mm.mp.1}) can be written as,
    \begin{eqnarray}
     \langle\mmBold,\RmagtopenBold\rangle &\simeq& \alpha \ m_m h \label{eq:ap.mm.r.2}\\
     \langle\mpBold,\RmagtopenBold\rangle &\simeq& mp \ [-\angt \cphi d + h] \label{eq:ap.mp.r.2} \\
     \langle\mmBold,\mpBold\rangle &\simeq&\alpha \ m_m m_p \label{ap:mm.mp.2}
    \end{eqnarray}
    
    We can now substitute these expressions into the main force equation \ref{eq:ap.F21-dip}. As we do in \del{Section \nameref{sc:em_model}}\add{previous section}, we consider only the terms that contribute to the component $\ed$ of the force. Keeping only these terms that contain $\angt$ up to the first order,
     \begin{eqnarray}
     \mathbf{F_{p}^{(d)}} =&& \frac{3\mu _{0} \alpha m_m m_p}{4\pi \Rmagtopen^{5}} \left[-d + \frac{5 d h^2}{\Rmagtopen^2} - h \angt \cphi - h_p \angt \cphi + \right.\nonumber \\
     && \left. + \frac{5 h^2 h_p \angt \cphi}{\Rmagtopen^2} - \frac{5 h d^2 \angt \cphi}{\Rmagtopen^2} \right] \ \ed \\
     =&& \frac{3\mu _{0} \alpha m_m m_p}{4\pi (h^2+d^2)^{5/2}} \left[ \frac{-d (d^2+h^2) +5 d h^2}{(h^2+d^2)} + \right. \nonumber \\
     && \left. \angt \cphi \left( -h -h_p + \frac{5(h^2 h_p - h d^2)}{(h^2+d^2)} - \frac{5 d^2 h^2 h_p}{(h^2+d^2)^2}  \right) \right] \ed \nonumber \\
     && \label{eq:ap.Fp2} \\
      \mathbf{F_{p}^{(d)}} =&& \alpha \ F_0 \ \left[ \ f_0(d) \ + \ \angt \cphi \ f_1(d) \ \right] \ \ed \label{eq:ap.Fd}
     \end{eqnarray}
     \noindent where we define, 
    \begin{eqnarray}
     F_0 &=&  \frac{3 \ \mu_0 \ m_p \ m_em}{4 \ \pi \ h^4} \ . \label{eq:ap.F0}\\
     f_0(d) &=& \frac{d \left(4 - \frac{d^2}{h^2}\right)}{h \left(1 + \frac{d^2}{h^2}\right)^\frac{7}{2}} \label{ap.f_0}\\
     f_1(d) &=& \frac{1 + \frac{h_p}{h}}{\left(1 + \frac{d^2}{h^2} \right)^\frac{5}{2}} 
      + \frac{5 \left(\frac{h_p}{h} + \frac{d^2}{h^2}\right)}{\left(1 + \frac{d^2}{h^2} \right)^\frac{7}{2}} 
      - \frac{5 \left(\frac{h_p}{h}\right) \left(\frac{d^2}{h^2}\right)}{\left(1 + \frac{d^2}{h^2} \right)^\frac{9}{2}} 
    \end{eqnarray}
    
    Note that by considering the case $\angt = 0$ in Eq. \ref{eq:ap.Fd}, we recover what we obtain \del{in the Sec. \nameref{sc:em_model}} for $\mathbf{F_a}$ \add{ as calculated in Eq. \ref{eq:Fa}}. 
    That means that the equation for $\mathbf{F_{p}^{(d)}}$\del{ we obtained in this Appendix} subsumes the cases of the pen being tilted by a small angles, and it can be used in future EM actuated systems which may be able to track $\angt$ and $\angp$.

\chapter{MARLUI: Multi-Agent Reinforcement Learning for Adaptive UIs}
\def\dir{chapters/05_shared_control/rl}
\section{Learned Lower Level}
\label{app:learned}
We detail how to learn the low-level motor policy. The low-level policy needs to learn i) the coordinates and dimensions of menu slots, ii) an optimal speed-accuracy trade-off given a target slot, and its current position. 

To prevent the low-level motor control policy from correcting wrong high-level decisions and to increase general performance, we limit the state space $\StatePerPolicy_M$ to strictly necessary elements with respect to the motor control task \cite{christen2021hide}: 

\begin{equation}
    \StatePerPolicy_M = \left (\pos, \target \right ),
\end{equation}
with the current position $\pos \in I^2$, the target slot $\target \in \mathbb{Z}_2^{\nslots}$. 

The action space $\ActionPerPolicy_M$ is defined as follows:

\begin{equation}
    \ActionPerPolicy_M = \left(\mu_{\pos}, \sigma_{\pos} \right).
\end{equation}
It consists of $\mu_{\pos} \in I^2$ and $\sigma_{\pos} \in I$, i.e., the mean and standard deviation which describes the endpoint distribution in the unit interval. We scale the standard deviation linearly between a min and max value where the minimum value is the size of normalized pixel width and the max value is empirically chosen to be 15\% of the screen width. Once an action is taken, we sample a new end-effector position from a normal distribution: $\pos \sim \mathcal{N}\left(\mu_{\pos}, \sigma_{\pos}\right)$.

Given the predicted actions, we compute the expected movement time via the WHo model \cite{guiard2015mathematical}, similar to our non-learned low-level motor control policy in the main paper. 

The reward for the low-level motor control policy is based on the \emph{motoric} speed-accuracy trade-off. Specifically, we penalize: i) missing the target supplied by the high-level $(\miss)$, and ii) the movement time ($\mt$). Furthermore,  we add a penalty iii)  which amounts to the squared Euclidean distance between the center of the target $\target$ and $\mu_\pos$. This incentivizes the policy to hit the desired target in the correct location. Since the penalty only considers the desired point $\mu_\pos$, it will not impact the speed-accuracy trade-off (which is a function of $\sigma_\pos$). The total reward is defined as follows:

\begin{equation}
    \RewardPerPolicy_M = \underbrace{\satweight (\miss)}_{i)} - \underbrace{(1-\satweight) \mt}_{ii)} - \underbrace{\beta ||\mu_{\pos} - \mu_\target||_2^2}_{iii)},
\end{equation}
where $\miss$ equals $0$ when the target button is hit and $-1$ on a miss. A hit occurs when the newly sampled user position $\pos$ is within the target $\target$, while a miss happens if the user position is outside of the target. $\satweight$ is a speed-accuracy trade-off weight and $\beta$ is a small scalar weight to help with learning.

\end{document}